\documentclass[twocolumn]{aastex63}
\pdfoutput=1
\usepackage{enumerate, enumitem}
\usepackage{amssymb, amsmath, amsfonts}
\usepackage{gensymb}
\usepackage{mathtools}
\usepackage{hyperref}
\usepackage{xspace}
\usepackage{comment}
\usepackage{xcolor}
\usepackage{ulem}
\usepackage{lipsum} 
\usepackage{soul}

\newcommand{\Hipparcos}{\textsl{Hipparcos}\xspace}
\newcommand{\hipparcos}{\textsl{Hipparcos}\xspace}
\newcommand{\Gaia}{\textsl{Gaia}\xspace}
\newcommand{\gaia}{\textsl{Gaia}\xspace}

\newcommand{\orvara}{{\tt orvara}}
\newcommand{\htof}{{\tt htof}}
\newcommand{\ptemcee}{{\tt ptemcee}}

\newcommand{\Msun}{\mbox{$M_{\sun}$}}

\newcommand{\Lsun}{\mbox{$L_{\sun}$}}

\newcommand{\Mjup}{\mbox{$M_{\rm Jup}$}}

\newcommand{\addition}[1]{#1}
\newcommand{\changes}[1]{#1}
\newcommand{\subtractions}[1]{}

\begin{document}

\title{Precise Masses and Orbits for Nine Radial Velocity Exoplanets}

\author[0000-0002-6845-9702]{Yiting Li}
\affiliation{Department of Physics, University of California, Santa Barbara, Santa Barbara, CA 93106, USA}
\author[0000-0003-2630-8073]{Timothy D.~Brandt}
\affiliation{Department of Physics, University of California, Santa Barbara, Santa Barbara, CA 93106, USA}
\author[0000-0003-0168-3010]{G.~Mirek Brandt}
\altaffiliation{NSF Graduate Research Fellow}\affiliation{Department of Physics, University of California, Santa Barbara, Santa Barbara, CA 93106, USA}
\author[0000-0001-9823-1445]{Trent J.~Dupuy}
\affiliation{Institute for Astronomy, University of Edinburgh, Royal Observatory, Blackford Hill, Edinburgh, EH9 3HJ, UK}
\author[0000-0002-7618-6556]{Daniel Michalik}
\affiliation{European Space Agency (ESA), European Space Research and Technology Centre (ESTEC), Keplerlaan 1, 2201 AZ Noordwijk, The Netherlands}
\author[0000-0003-0054-2953]{Rebecca Jensen-Clem}
\affiliation{Astronomy $\&$ Astrophysics Department, University of California, Santa Cruz, CA 95064, USA}
\author[0000-0003-4594-4331]{Yunlin Zeng}
\affiliation{School of Physics, Georgia Institute of Technology, 837 State Street, Atlanta, Georgia, USA}
\author[0000-0001-6251-0573]{Jacqueline Faherty}
\affiliation{American Museum of Natural History, New York, NY}
\author[0000-0001-8345-6449]{Elena L.~Mitra}
\affiliation{CUNY Hunter College, New York, NY 10065, USA}

\begin{abstract}
Radial velocity (RV) surveys have discovered hundreds of exoplanetary systems but suffer from a fundamental degeneracy between planet mass $M_p$ and orbital inclination $i$. In this paper we break this degeneracy by combining RVs with complementary absolute astrometry taken from the \Gaia EDR3 version of the cross-calibrated \Hipparcos-\Gaia Catalog of Accelerations (HGCA). We use the Markov Chain Monte Carlo orbit code $\orvara$ to simultaneously fit literature RVs and absolute astrometry from the HGCA. We constrain the orbits, masses, and inclinations of nine single and massive RV companions orbiting nearby G and K stars. We confirm the planetary nature of six companions: HD~29021~b ($4.47_{-0.65}^{+0.67}\Mjup$), HD~81040~b ($7.24_{-0.37}^{+1.0}\Mjup$), HD~87883~b \changes{($6.31_{-0.32}^{+0.31}\Mjup$)}, HD~98649~b ($9.7_{-1.9}^{+2.3}\Mjup$), HD~106252~b ($10.00_{-0.73}^{+0.78}\Mjup$), and HD~171238~b ($8.8_{-1.3}^{+3.6}\Mjup$). We place one companion\changes{,} HD~196067~b \changes{($12.5_{-1.8}^{+2.5}\Mjup$)} on the planet-brown dwarf boundary, and two companions in the low mass brown dwarf regime: HD~106515~Ab (\changes{$18.9_{-1.4}^{+1.5}\Mjup$)}, and HD~221420~b \changes{(${20.6}_{-1.6}^{+2.0}\Mjup$)}. The brown dwarf HD~221420~b, with a semi-major axis of \changes{${9.99}_{-0.70}^{+0.74}$ AU, a period of ${27.7}_{-2.5}^{+3.0}$ years, and an eccentricity of $0.162_{-0.030}^{+0.035}$} represents a promising target for high-contrast imaging. The RV orbits of HD~87883~b, HD~98649~b, HD~171238~b, and HD~196067~b are not fully constrained yet because of insufficient RV data. We find two possible inclinations for each of these orbits due to difficulty in separating prograde from retrograde orbits, but we expect this will change decisively with future Gaia data releases. 
\end{abstract}


\section{Introduction} \label{sec:intro}
The radial velocity (RV) technique is one of the earliest and most commonly used indirect methods in the detection of exoplanets in both nearby and distant solar systems \citep{Hatzes_2016, Wei_2018, Wright_2018}. 51 Pegasi b, the first exoplanet around a Sun-like star, was discovered via RV in 1995 \citep{Mayor_1995}. Since then, RV instruments have overcome many design and performance challenges to reach extreme m/s precisions, at once revealing the richness and diversity of the population of exoplanets \citep[e.g.][]{Udry_2007,Akeson_2013,Rice_2014}. 

The RV method measures the Doppler shift in the spectrum of a star perturbed by one or multiple unseen orbiting companions \citep{WALKER_2012,plavchan2015radial,Fischer_2016,Wright_2018}. The RV semi-amplitude due to a companion scales with the companion's mass, the sine of the orbital inclination ($\sin{i}$), and the inverse of the square root of the semi-major axis. This means that massive and close-in RV companions are usually more easily and precisely characterized owing to complete orbital coverage. Much like 51 Pegasi b, these are often short-period Jupiter analogs orbiting solar-type stars \citep{Wright_2003}. 

Thanks to the growing temporal baselines of RV surveys, detection sensitivity is now being extended beyond the ice line \changes{($\gtrsim$ 3 AU)} \citep{Ford12616,Wittenmyer_2016}. In this wider orbital regime, long-period giant planets are frequently being revealed \citep{Perrier_2003,Marmier_2013,Rickman_2019}. Small planets or Uranus/Neptune-like ice giants on such wide orbits however are very challenging to detect due to long orbital periods (80-165 years), low transit probability, and the high contrast barrier. Still longer-period companions can induce measurable RV trends or accelerations \citep{Eisner_2001,Cumming_2004,Knutson_2014}. \subtractions{Sumi et al. 2016}.

Not all stars are accessible to RV searches for planets. RV surveys often target old and inactive main sequence FGK stars and M dwarfs because their lower RV jitter and abundance of narrow absorption lines allow very precise velocity measurements \citep[e.g.][]{Howard_2012,Marchwinski_2014,Gaidos_2016,Hsu_2019,Borgniet_2019,He_2020,Bauer_2020,Jin_2021}. \subtractions{Gould et al. 2010}
In addition, there are several limitations to using the RV method alone to characterize long period companions. Firstly, \addition{orbital phase coverage drives estimations of orbital periods, so the RVs must cover a significant fraction of an orbit which requires long term time commitments from RV-surveys.} \subtractions{so long-term time commitments from RV-surveys are required.} Secondly, RVs only constrain $M_p \sin i$, providing a minimum mass that is estimated assuming an edge-on system with $\sin i = 1$. This $\sin i$ degeneracy permits only a statistical estimate of a planet's true mass and limits our ability to understand an individual exoplanetary system \citep[e.g.,][]{Batalha_2019,Gaudi_2021}. 

\changes{One way to break the degeneracy in RV-only orbital fits is to combine RVs with precision astrometry. Astrometry measures the sky-projected Keplerian motion of accelerating stars. Together, radial velocity accelerations and astrometric accelerations probe orthogonal components of the 3D acceleration of a star through space, enabling us to map out the complete stellar reflex motion of a star under the influence of any unseen companion(s) \citep{Lindegren_2003}. }

\changes{Efforts to measure the reflex motion of stars astrometrically date back to the Fine Guidance Sensor (FGS) on the Hubble Space Telescope (HST). FGS/HST has achieved sub-milliarcsecond single-star optical astrometry precision \citep{Benedict_2008}. Pioneering works incorporated HST/FGS observations in ground-based astrometric and radial velocity fits to derive true masses for exoplanetary systems \citep[e.g.,][]{Benedict_2002,
Benedict_2006,McArthur_2010}.}
\changes{Beyond FGS/HST, the recent advent of Gaia astrometry and long-baseline interferometry open up the possibility of identifying and characterizing accelerating systems more systematically and on a larger scale. In particular,}  the combination of proper motion data from \Hipparcos and \Gaia provides a powerful and unique opportunity to realize this.

\Gaia \citep{GaiaCollaboration_2020,Lindegren+Klioner+Hernandez+etal_2020} and its predecessor \Hipparcos \citep{Hipparcos_Tycho_cat_1997,vanLeeuwen_2007} were launched by the European Space Agency (ESA) to map the absolute positions of stars across the celestial sphere. The data from both satellites are processed as least-squares solutions to model sky paths; the catalogs list the best-fit astrometric parameters such as proper motions and parallaxes along with their uncertainties and covariances. \subtractions{Differences in the proper motions between Hipparcos and Gaia indicate accelerations in an inertial frame. Together, RV and absolute astrometry measure orthogonal components of the true space velocity of the stars in inertial frames} \citep{Lindegren_2003}, \subtractions{enabling us to map out the stellar reflex motion of a star under the influence of any unseen companion(s).} \addition{\hipparcos and \gaia took data $\sim$25 years apart, and the positional difference between \hipparcos and \gaia divided by this temporal baseline offers a new and more precise proper motion. Comparing this third proper motion with either the \hipparcos or \gaia measurement indicates acceleration in an inertial frame. A cross-calibration between the \hipparcos and \gaia catalogs must first correct error underestimation and frame rotation. To achieve this,} \citet{Brandt_2018} and \citet{HGCAeDR3_2021} have placed the \hipparcos and \gaia DR2 and EDR3 catalogs, respectively, in a common reference frame presented in the cross-calibrated \Hipparcos-\Gaia catalog of accelerations (HGCA). HGCA astrometry has the potential to break the $\sin i$ degeneracy in RV-detected companions. 

Previous works have explored the idea of using \Hipparcos epoch astrometry in conjunction with RVs or relative astrometry to obtain precise dynamical masses \citep[e.g.,][]{Han_2001,Sozzetti_2010,Reffert_2011,Sahlmann_2011}. More recently, with the release of \gaia astrometry, a number of works have achieved improved astrometric/RV mass approximations by combining different data types: RV+relative astrometry \citep[e.g.,][]{Brown_2011,Boffin_2016}, RV+absolute astrometry \citep[e.g.,][]{Feng_2019,Kervella_2019,Venner_2021}, relative astrometry+absolute astrometry (e.g., Chen$\&$Li et al., in prep), or RV+relative astrometry+absolute astrometry \citep[e.g.,]{Snellen_2018,Brandt_2019,Grandjean_2019,Brandt_2020,Nielsen_2020,Mirek_2020,Kiefer_2021,Brandt+Michalik+Brandt+etal_2021}.

In this paper, we derive orbital solutions for nine RV-detected exoplanet candidates by jointly analyzing RVs and complementary absolute astrometry data from the HGCA (Gaia EDR3 version): HD~29021, HD~81040, HD~87883, HD~98649, HD~106252, HD~106515~A, HD~171238, HD~196067, and HD~221420. We use the MCMC orbit-fitting package \orvara\ \citep{Brandt+Dupuy+Li+etal_2021} that is capable of fitting one or more Keplerian orbits to any combination of RVs, direct imaging astrometry and absolute astrometry from the HGCA. Because the epoch astrometry from Gaia is not yet available, \orvara\ uses the intermediate astrometry fitting package \htof\ \citep{Brandt+Michalik+Brandt+etal_2021} to provide a workaround for this. \htof\ fetches the intermediate astrometry from Hipparcos and Gaia, and reproduces positions and proper motions from synthetic Hipparcos or Gaia epoch astrometry using five or higher degree parameter fits. We focus our work on systems with known single RV companions. Two host stars in our sample, HD~106515~A and HD~196067, have known wide stellar companions that we solve as 3-body systems with $\orvara$.

The paper is structured as follows. Section~\ref{sec:stellarcharact} presents our re-examination of the properties of the host stars, including a systematic re-analysis of the host stars' ages and masses based on an activity-age Bayesian model and PARSEC isochrones. Section~\ref{sec:rv} describes the literature radial velocity data we use to fit orbits. Section~\ref{sec:absoluteastrometry} discusses the stellar astrometry derived from the cross-calibrated \Hipparcos-\Gaia Catalog of Accelerations. In Section~\ref{sec:orbitfitting}, we demonstrate the method we use to fit orbits to our sample of nine RV companions, followed by Section~\ref{sec:udpatedparameters} where we show full orbital solutions from the orbital fits and discuss the implications of our work. We provide updated parameters, in particular, precise masses and inclinations for these planets in the same section. We summarize our results and findings in Section~\ref{sec:discussion}.

\section{Stellar characteristics} \label{sec:stellarcharact}

\movetabledown=2in
\begin{rotatetable*}
\begin{deluxetable*}{lccccccccccccccc}
\setlength{\tabcolsep}{0.050in}
\tabletypesize{\scriptsize}
\tablewidth{0pt}
\tablecaption{
Adopted stellar parameters for stars investigated in this work\tablenotemark{a}.\label{tab:stellar_params}}
\tablehead{
\colhead{Star (HD \#)}  &\colhead{29021}     &\colhead{81040}   &\colhead{87883}      &\colhead{98649}       
&\colhead{106252}       &\colhead{106515A}   & \colhead{106515B \tablenotemark{*}}    &\colhead{171238}       
&\colhead{196067}       &\colhead{196068 \tablenotemark{*}}     &\colhead{221420}    }
\startdata
HIP ID &21571	        &46076	&49699	&55409	&59610	&59743A	&59743B	&91085	&102125	&102128	&116250\\
$\bar{\omega}$ (mas)    &32.385	&29.063	&54.668	&23.721	&26.248	&29.315	&29.391	&22.481	&25.033	&25.038	&32.102\\
$\sigma[\bar{\omega}]$  &0.024	&0.041	&0.030	&0.022	&0.027	&0.030	&0.029	&0.032	&0.021	&0.017	&0.033\\
$V_{T}$ (mag)           &7.842	&7.791	&7.660	&8.066	&7.479	&8.063	&8.298	&8.737	&6.510	&7.070	&5.884\\
$\sigma[V_{T}]$         &0.011	&0.012	&0.012	&0.011	&0.011	&0.014	&0.015	&0.020	&0.010	&0.010	&0.009\\
$B_{T}$ (mag)           &8.612	&8.527	&8.800	&8.793	&8.158	&8.994	&9.284	&9.589	&7.193	&7.790	&6.643\\
$\sigma[B_{T}]$         &0.016	&0.017	&0.018	&0.017	&0.017	&0.021	&0.022	&0.029	&0.015	&0.015	&0.014\\
$K_{s}$ (mag)           &6.082	&6.159	&5.314	&6.419	&5.929	&6.151	&6.267	&6.831	&5.080	&5.713	&4.306\\
$\sigma[K_{s}]$         &0.020	&0.021	&0.020	&0.020	&0.026	&0.026	&0.017	&0.017	&0.016	&0.018	&0.036\\
\addition{$J$ (mag)               &6.518  &6.505  &5.839  &6.811  &6.302  &6.585  &6.746  &7.244  &5.417  &6.061  &4.997}\\
\addition{$\sigma[J]$             &0.029  &0.019  &0.020  &0.019  &0.024  &0.024  &0.030  &0.019  &0.021  &0.027  &0.252\\
$R_X$                   &$<-$5.14	&$-$5.27	&$<-$5.25	&$<-$4.87	&$<-$5.20	&$<-$4.84	&$<-$4.77	&$<-$4.83	&$<-$5.08	&$<-$4.87	&$<-$5.55}\\
$R'_{\rm HK}$               &\ldots	&$-$4.71	&$-$4.98	&$-$4.96	&$-$4.85	&$-$5.10	&$-$5.15	&$-$4.96	&$-$5.06	&$-$5.02	&$-$5.05\\
$P_{\rm rot}$ (days)        &\ldots	&$15.98$	&\ldots	&26.441	&\ldots	&\ldots	&\ldots	&\ldots	&\ldots	&\ldots	&\ldots \\
$\rm [Fe/H] (dex)$\tablenotemark{b}      &$-0.30\pm0.10$	&$-0.04\pm0.05$	&$0.11\pm0.10$	&$-0.06\pm0.04$	&$-0.06\pm0.04$	&$0.03\pm0.10$	&$0.03\pm0.10$	&$0.20\pm0.10$	&$0.23\pm0.07$	&$0.24\pm0.10$	&$0.34\pm0.07$\\
\changes{$M_{*,\rm model} (M_{\odot})$\tablenotemark{c}  &$0.86\pm0.02$	&$0.99\pm0.02$	&$0.80\pm0.02$	&$0.97\pm0.02$	&$1.05\pm0.02$	&$0.89\pm0.03$	&$0.86\pm0.03$	&$0.92\pm0.03$	&$1.26\pm0.07$	&$1.12\pm0.06$	&$1.28\pm0.08$}\\
\changes{$L_{*} (L_{\odot})$\tablenotemark{d}            &$0.659\pm0.017$	&$0.838\pm0.018$	&$0.338\pm0.008$	&$0.968\pm0.019$	&$1.328\pm0.030$	&$0.686\pm0.018$	&$0.567\pm0.017$	&$0.627\pm0.018$	&$3.435\pm0.068$	&$2.007\pm0.045$	&$3.850\pm0.072$}\\
Spectral Type           &G5	&G2/G3	&K0	&G3/G5 V	&G0	&G5 V	&G8 V	&G8 V	&G5 III	&G5 V	&G2 V\\
Age (Gyr)\tablenotemark{e}               &$5.5_{-1.0}^{+1.2}$	&$1.79_{-0.26}^{+0.30}$	&$7.6_{-1.8}^{+2.8}$	&$4.44_{-0.58}^{+0.68}$	&$3.00_{-0.60}^{+0.80}$	&$8.2_{-2.1}^{+2.7}$	&$8.4_{-2.1}^{+2.6}$	&$7.0_{-1.6}^{+6.0}$	&$5.8_{-2.2}^{+4.3}$	&$6.2_{-2.2}^{+4.1}$	&$6.6_{-2.0}^{+3.7}$\\
\enddata

\tablenotetext{*}{Secondary stellar companion}
\tablenotetext{a}{\textbf{References} -- $\bar{\omega}$ from Gaia eDR3 \citep{gaiacollaboration2020gaia}; $\rm V_{T}$ and $\rm B_{T}$ from Tycho-2 \citep{Hog_2000}, $K_{s}$ from 2MASS \citep{Cutri_2003}; $R_X$ from ROSAT \citep{Voges_1999}; Rotation period for HD 81040 from \citet{Reinhold_2020}, alternatively 15.2579 days from \citet{Oelkers_2018}; $\rm R'_{HK}$ from \citep{Pace_2013}; Rotation period for HD 98649 from \citet{Oelkers_2018}; Spectral types from \citet{Wright_2003} and \citet{Skiff_2014}.}
\tablenotetext{b}{We estimate $\rm [Fe/H]$ from the median of all entries in \citep{Soubiran_2016} with a default uncertainty of 0.1 dex. The references for each star are: \citet{Kim_2016} for HD~29021; \citet{Sousa_2006,Gonzalez_2010,Kang_2011,Luck_2017,Rich_2017,Sousa_2018} for HD~81040; \citet{Valenti_2005,Kotoneva_2006,Mishenina_2008,Maldonado_2012,Brewer_2016,Luck_2017} for HD~87883; \citet{Datson_2012,Santos_2013,Porto_2014,Ramirez_2014,Rich_2017} for HD~98649; \citet{Sadakane_2002,Heiter_2003,Laws_2003,Santos_2003,Santos_2004,Santos_2005,Huang_2005,Valenti_2005,Luck_2006,Takeda_2007,Fuhrmann_2008,Gonzalez_2010,Kang_2011,Datson_2015,Spina_2016,Luck_2017,Sousa_2018} for HD~106252; \citet{Santos_2013} for HD~106515A and HD~106515B; \citet{Brewer_2016} for HD~171238; \citet{Valenti_2005,Bond_2006,Santos_2013} for HD~196067 and HD~196068; and \citet{Valenti_2005,Bond_2006,Sousa_2006,Sousa_2008,Tsantaki_2013,Jofre_2015,Maldonado_2016,Soto_2018} for HD~221420.}
\tablenotetext{c}{The stellar masses are estimated from PARSEC isochrones (see text).} 
\tablenotetext{d}{The stellar luminosities are computed based on \changes{the absolute calibration of effective temperature scales performed by \citep{Casagrande_2010} (see text).} }
\tablenotetext{e}{The stellar ages and their $1\sigma$ error bars are approximated using the Bayesian technique of \citet{Brandt_2014}.}
\movetabledown=22mm
\end{deluxetable*}
\end{rotatetable*}

Our sample of targets comprises nine dwarf stars of G and early K spectral types, all of which have long term RV monitoring and precision spectroscopy. These stars have precise RV time series because their stable atmospheres and rich absorption spectra make them ideal targets for RV surveys: G and K stars constitute the majority of the host stars of RV-detected planets \citep{Lineweaver_2003}. In this Section, we provide a general overview on the photometric and atmospheric properties of the host stars using values from the literature.  We then derive our own constraints on the ages and masses of our host stars.  We need stellar masses in particular to infer planet masses from RV time series: the RV semi-amplitude is a function of both companion mass and total system mass.

To derive stellar masses, we first perform a uniform re-analysis on the stellar ages with an activity-based Bayesian method employing the stars' rotation and chromospheric activity \citep{Brandt_2014}. Next, we estimate bolometric luminosities from Tycho-2 and 2MASS photometry \changes{using the absolutely calibrated effective temperature scale in \citet{Casagrande_2010}}, and use metallicities, bolometric luminosities and the inferred stellar ages to infer masses using the PARSEC isochrones 
\citep{Bressan_2012}. We describe our approaches for stellar ages and masses in more detail in the following, and discuss the results for each star individually.  Our mass posteriors will later serve as the input priors in our joint orbit analysis of RV and absolute astrometry.

\subsection{Activity-based Age Analysis}

Stellar magnetic activity is a term that encompasses a range of phenomena observed of a star, including flares, star spots, and any chromospheric and coronal activity. Mid-F type and cooler main-sequence stars experience a decline in magnetic activity levels and rotational velocity as they age. The well-accepted explanation for the weaker activity and the slower rotation rates of old stars is the dynamo theory developed by \citet{Schatzman_1962,schatzman_1990,Parker_1955,Parker_1979}: late-type stars develop subsurface convection zones that support a magnetic dynamo. The dynamo provides the energy to heat the stellar chromosphere and corona \cite[][and references therein]{Kraft_1967,Hall_2008,Testa_2015}. The chromospheres eject winds in the form of jets or flares. These rotate with the stellar angular velocity out to the Alfv\'en radius, carrying away angular momentum which in turn causes the star to spin down over time. This effect is known as magnetic braking through the magnetic dynamo process \citep{Noyes_1984}. The declining rotational velocity due to magnetic braking results in a reduction in the magnetic field generated by the stellar dynamo which, in turn, diminishes the star's \ion{Ca}{2}~HK and X-ray emissions. \ion{Ca}{2}~HK and X-ray emission weaken fastest at young ages, thus age dating for young stars is easier than for their old field star counterparts \citep{Soderblom_2010}.

\ion{Ca}{2}~HK emission, via the chromospheric activity index $R'_{\rm HK}$, has been widely used as an age indicator for Solar-type (G and K) stars. Age dating of Sun-like stars with $R'_{\rm HK}$ was first done by \citet{Wilson_1968} at the Mount Wilson Observatory. The age-activity relationship has since been developed and calibrated across spectral types later than mid-F \citep{Skumanich_1972,Noyes_1984,Soderblom_1991,Barnes_2007,Mamajek_2008,David_2015,Angus_2015,VanSaders_2016,Lorenzo-Oliveira_2016}. Conventionally, $R'_{\rm HK}$ is calculated from the \ion{Ca}{2} $S$-index, a measurement of the strength of the emission. \cite{Mamajek_2008} calibrated the activity relations to an activity level of $\log R'_{\rm HK} \ge -5.0$\,dex or a Rossby number (ratio of rotation period to convective overturn time) of 2.2, roughly corresponding to the Solar activity and rotation.
    
X-ray activity traces magnetic heating of the stellar corona, providing another indirect probe of rotation \citep{Golub_1996,Mewe_1996,Jardine_2002,Manuel_2004,Zhuleku_2020}.  
\citet{Brandt_2014} combine $R'_{\rm HK}$ and X-ray activity to infer a Rossby number and, from this, an age via gyrochronology using the \cite{Mamajek_2008} calibration. \cite{Brandt_2014} further allows for a distribution of times, dependent on spectral type, on the C-sequence where the dynamo is saturated and magnetic braking is inefficient \citep{Barnes_2003}. Their model sets a floor for the Rossby number at $>2.2$, inflates the error in the activity and rotation by 0.06 dex to account for systematic uncertainties.  

A photometric rotation period provides a more direct measure of the Rossby number and a tighter constraint on the stellar age. For stars that have directly measured rotation periods, \cite{Brandt_2014} combine this measurement with the indirect probes of coronal and chromospheric emission to derive an age posterior. We refer the reader to that paper for a more detailed discussion.

We adopt the Bayesian inference model described in \citet{Brandt_2014} to estimate stellar ages. We take the \ion{Ca}{2}~HK $S$-indices from the catalog of \citet{Pace_2013} and references therein and convert them to Mt. Wilson $R'_{\rm HK}$ using the relations described in \citep{Noyes_1984}.  We extract X-ray activities $R_X$ from the ROSAT all-sky survey bright and faint source catalogs \citep{Voges_1999,Voges+Aschenbach+Boller+etal_2000}. For stars that are not in either ROSAT catalog, we take the nearest detection in the faint source catalog and use five times its uncertainty as our upper limit on X-ray flux.
    
\begin{figure*}
\centering
\includegraphics[width=0.45\linewidth]{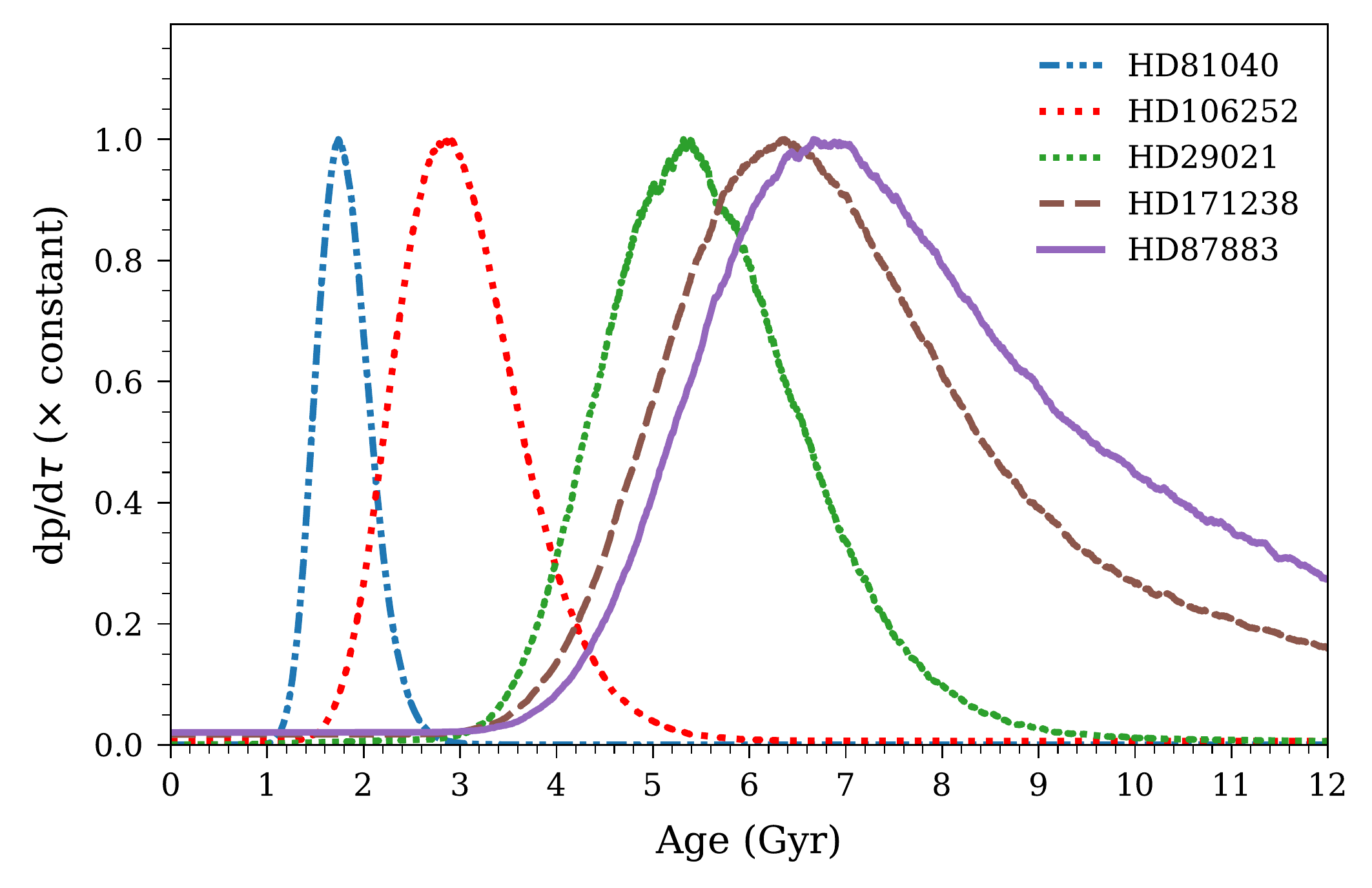} \quad
\includegraphics[width=0.45\linewidth]{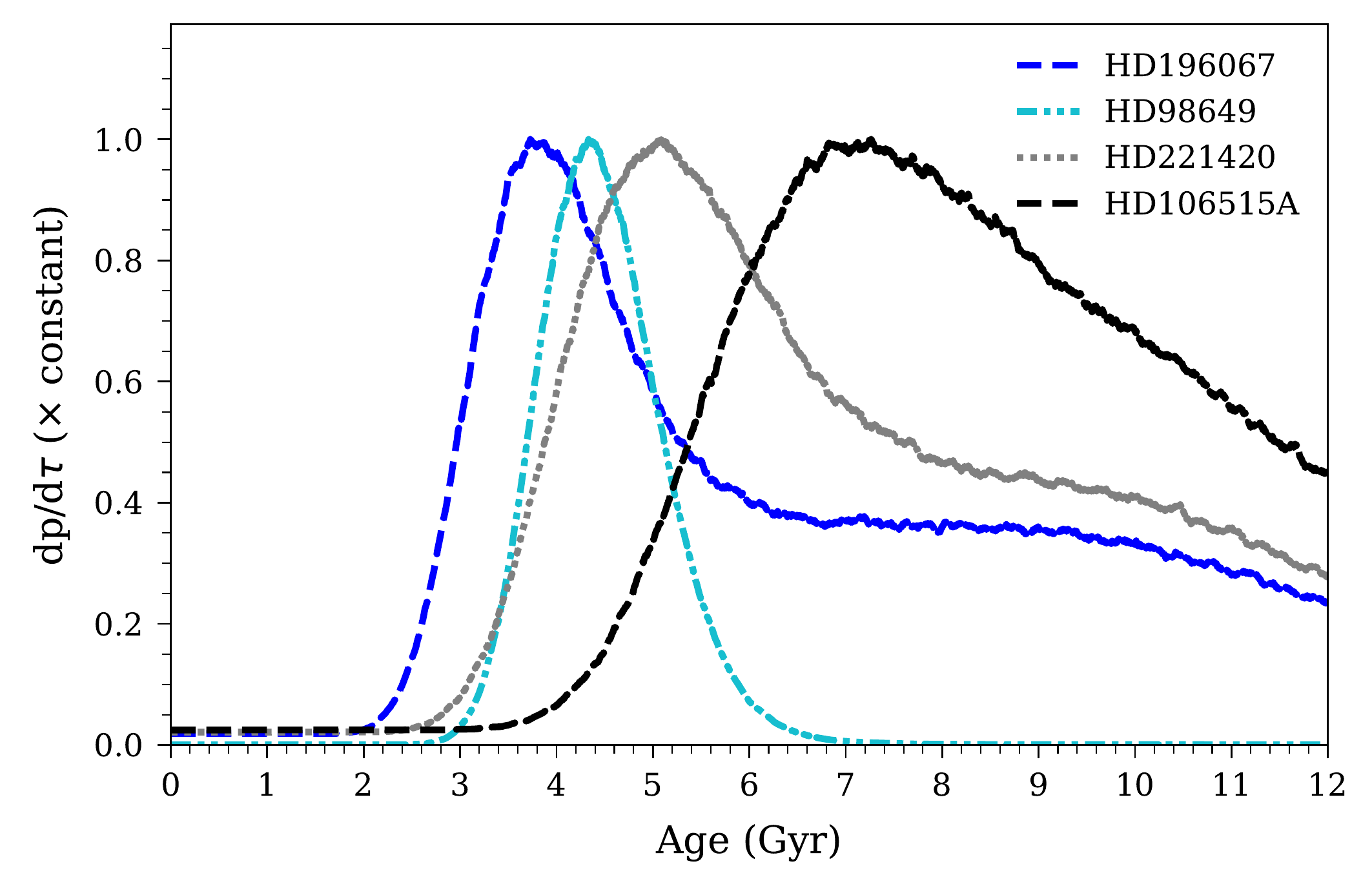}
\caption{Normalized age posteriors for the nine stars presented in this work following the Bayesian technique of \citet{Brandt_2014}. This method combines the X-ray ($R_{X}$) and chromospheric activity ($R'_{\rm HK}$) indicators with an optional rotation period to constrain the Rossby number. It is then transformed into an age diagnostic for a specific star based on the calibration of \citet{Mamajek_2008}. }
\label{fig:ageactivity}
\end{figure*}
    
\subsection{Stellar Luminosity Analysis} \label{subsec:lum_mass}

\changes{Here, we describe the method we use to estimate the bolometric luminosity for our sample of host stars. We adopt the absolutely calibrated effective temperature scales in \citet{Casagrande_2010} for FGK stars across a wide range of metallicities. These authors use the InfraRed Flux Method (IRFM) to derive model independent effective temperatures $T_{\rm eff}$ in stars with different spectral types and metallicities. The homogeneous, all-sky Tycho-2 $(B_T, V_T)$ optical photometry \citep{Hog_2000} was combined with the 2MASS infrared photometry in the JH$K_s$ bands \citep{Skrutskie_2006} to calibrate reliable $T_{\rm eff}$ for a sample of solar analogs via IRFM. }

\changes{All of our stars are late G and K sun-like dwarfs with optical and infrared photometry from Tycho-2 $(BV)_T$ and 2MASS $JHK_s$, thus we use Table 5 of \cite{Casagrande_2010} to derive their bolometric integrated fluxes. In Table 5, the set of Tycho-2 and 2MASS flux calibrations apply to a $[{\rm Fe/H}]$ range between $-$2.7 and 0.4, and a color range between 0.19 and 3.29. The standard deviation $\sigma_{\phi_{\xi}}$ in the Tycho-2 and 2MASS flux calibrations is the smallest for the $m_{\xi} = V_T$ flux calibrations with color indices $V_T - J$ (0.7$\%$) and $V_T - K_s$ (0.9$\%$). We use the $V_T - J$ colors to calibrate $V_T$ magnitudes for our sample of stars with the exception of HD~221420, which has a large uncertainty in its 2MASS $J$ band magnitude due to saturation ($4.997\pm0.252$ mag). Therefore, for HD~221420, we use the $V_T - K_s$ index instead to calculate the flux calibration in its $V_T$.}

\changes{The final bolometric magnitudes can be estimated in terms of the $V_T$ magnitude, the distance d, and the the integrated bolometric fluxes $\mathcal{F}_{\rm Bol}(\rm Earth)$:}
\begin{equation}
    L_{Bol} = 4 \pi d^2 10^{-0.4 V_{T}} \mathcal{F}_{\rm Bol}(\rm Earth)
\end{equation}
\changes{We express $L_{\rm Bol}$ in units of solar luminosity, adopting a value of $\Lsun = 3.828\times 10^{33} {\rm erg/s}$. Our parallaxes all have fractional uncertainties $\sim$10$^{-3}$; they contribute negligibly to our luminosity uncertainties.}

\changes{We propagate the errors using standard error propagation techniques. We estimate the final uncertainty in the luminosities as}
\begin{align}
    \frac{\sigma^2(L_{\rm Bol})}{L_{\rm Bol}^2} =  &\left(\frac{\sigma (\mathcal{F}_{Bol})}{\mathcal{F}_{Bol}}\right)^2 + \left(\frac{2\sigma(\varpi)}{\varpi}\right)^2 \nonumber \\
    &+ (0.4\ln(10) \sigma(V_T))^2 + (\sigma_{\phi_{\xi}})^2 
\end{align}
\changes{where the uncertainty in the flux calibration is $\sigma_{\phi_{\xi}}$ is 0.9$\%$ for HD~221420 and 0.7$\%$ for the rest of the sample. We summarize in Table~\ref{tab:stellar_params} the Tycho-2 $(BV)_T$ and the 2MASS $JHK_s$ photometry and our derived luminosity values for the host stars that we study.}

\subsection{Stellar Masses and Metallicities} \label{subsec:mass_metallicities}

We infer masses for all of our stars using the PARSEC isochrones \citep{Bressan_2012} by matching the observed luminosity at the spectroscopic metallicity and activity-based age.  

We take spectroscopic metallicities from the PASTEL catalog \citep[][and references therein]{Soubiran_2016}. The PASTEL catalog is a large compilation of high-dispersion spectroscopic measurements from the literature. For the metallicities, we take the median of all entries in the PASTEL catalog. We estimate the uncertainty in the metallicity as the standard deviation of the measurements in the PASTEL catalog, multiplied by $\sqrt{n/(n-1)}$ to correct for the bias in the estimator. Three targets---HD~171238, HD~29021, and HD~106515~A---have only single entries in the PASTEL catalog. For these stars we use the single measurement and assume a conservative uncertainty of 0.1 dex.

Once we have a probability distribution for metallicity, we take a fine grid of stellar models and, for each, compute a likelihood using 
\begin{align}
    {\cal L} = p(\tau_{\rm model}) \times \exp \bigg[ -&\frac{\left( L_\star - L_{\rm model}\right)^2}{2\sigma^2_L} \nonumber \\
    &- \frac{\left( [{\rm Fe/H}]_\star - [{\rm Fe/H}]_{\rm model}\right)^2}{2\sigma^2_{[\rm Fe/H]}} \bigg].
    \label{eq:like_stellarmass}
\end{align}
In Equation \eqref{eq:like_stellarmass}, $p(\tau_{\rm model})$ is our activity-based age posterior at the model age.  We then marginalize over metallicity and age to derive a likelihood as a function of stellar mass.  We report the mean and standard deviation of this likelihood in Table \ref{tab:stellar_params} as $M_\star (\rm model)$ and adopt it as our (Gaussian) prior for orbital fits.  The variances of these distributions are all small; weighting by an initial mass function has a negligible effect.  

\subsection{Results on individual stars} \label{subsec:results_stars}

Here, we describe the results of the activity-based age analysis and the background for each individual star in our sample. The stellar distances we refer to in this section are inferred from \Gaia EDR3 parallaxes. The fractional uncertainties on parallax are all $\lesssim$0.1\%, so we simply take distance as the inverse of the parallax.  The adopted stellar parameters in this paper, including Tycho-2 \citep{Hog_2000} and 2MASS \citep{Cutri_2003} photometry, $R'_{\rm HK}$, and $R_X$, are summarized in Table \ref{tab:stellar_params}. 

We combine the chromospheric index $R'_{\rm HK}$, the X-ray activity index $R_X$, and in some cases a photometric rotation period, to derive an age posterior for each star. Figure~\ref{fig:ageactivity} shows the activity-based age posteriors for the whole sample of stars. 

\subsubsection{HD~29021 (HIP~21571)}
HD~29021 is a G5 star located at a distance of 30.9\,pc. \citet{Rey_2017} estimated the stellar activity level of the star using high-resolution spectra obtained with the SOPHIE spectrograph at the 1.93\,m telescope of the Haute-Provence Observatory, and no significant variability of the star was found from the $H\alpha$ activity indicator. The age of HD~29021 is determined to be $\approx$6 Gyr from our activity-based analysis, slightly lower than 7.4 Gyr from \citet{Winter_2020}, the only age reported for this star in the literature. We obtain a luminosity of \changes{$0.659 \pm 0.008$\,\changes{$\Lsun$}}, consistent with the luminosity of $\approx$0.7\,$\Lsun$ from \citet{Anderson_2012,GaiaDR2_2018}. The mass estimate of the star is between 0.85-1.0 $\Msun$ from multiple sources \citep{Chandler_2016,Mints_2017,Goda_2019}. Combining our luminosity and age with our adopted metallicity, we infer a mass of $0.86\pm0.02$\,$\Msun$. 

\subsubsection{HD~81040 (HIP~46076)}
HD~81040 is a bright, nearby, early G-dwarf star, commonly classified as a very young Galactic disc star \citep{Montes_2001}. \citet{Sozzetti_2006} used \ion{Ca}{2}~HK lines in the Keck/HIRES spectra to determine a chromospheric age of $0.73\pm 0.1$\,Gyr. A significant Lithium (Li) abundance of $\log \epsilon(Li) = 1.91 \pm 0.07$ also suggests youth \citep{Sozzetti_2006,Gonzalez_2010}. \citet{Oelkers_2018} and \citet{Reinhold_2020} measured rotation periods of 15.26 and 15.98 days, respectively, about half the Solar period. Our activity-based analysis produces a narrow age posterior centered just under 2 Gyr, slightly older than the \cite{Sozzetti_2006} age. The mass of the star is consistently inferred to be between 0.93\,$\Msun$ and 1.01\,$\Msun$ \citep{Sousa_2006,Butler_2006,Goda_2019}. We obtain a mass of $0.99\pm0.02$\,$\Msun$ from PARSEC isochrones. 

\subsubsection{HD~87883 (HIP~49699)}
HD~87883 is a K star at a distance of 18.3 pc. \citet{Fischer_2009} found modest chromospheric activity with $\log R'_{\rm HK} = -4.86$\,dex, a stellar luminosity of $0.318\pm0.018\,\Lsun$, and a stellar jitter of 4.5\,m\,s$^{-1}$. We have adopted a slightly lower $\log R'_{\rm HK}$ value of $-4.98$\,dex computed using $S$-indices from \citet{Strassmeier_2000,Gray_2003,Isaacson_2010}.
We derive a slightly higher luminosity of \changes{$0.338\pm0.004\,\Lsun$}.
Literature estimates of the stellar mass range from 0.78--0.85\,$\Msun$ \citep{Valenti_2005,Howard_2010,Luck_2017,Santos_2017,Anders_2019,Maldonado_2019,Stassun_2019}. We find a stellar mass from PARSEC isochrones of $0.80\pm0.02\,\Msun$, in good agreement with previous results. Several studies have constrained the age of this star to $\sim$5--10\,Gyr \citep{Valenti_2005,Takeda_2007,Bonfanti_2015,Brewer_2016,Bonfanti_2016,Yee_2017,Luck_2017,Winter_2020}. Our adopted parameters combined produce an activity-based age peak at $\approx$7\,Gyr which falls within that age range. The combined properties paint a picture of an old aged main-sequence K star with low to modest chromospheric and coronal activity levels.

\subsubsection{HD~98649 (HIP~55409)}
HD~98649, located at a distance of 42.2 pc from the Sun, is a G3/G5V dwarf. \citet{Porto_2014} label it as a solar analog candidate that closely matches the Sun in colors, absolute magnitude, chromospheric activity as measured from the $H\alpha$ indicator, and atmospheric characteristics. Our luminosity from Tycho-2 \changes{and 2MASS photometry yield $L_{*} = 0.968\pm0.012\,\Lsun$}. \citet{Pace_2013} and \citet{Gaspar_2016} have determined the age of the star to be between 4-5 Gyr, in agreement with our activity-based age of $4.436_{-0.576}^{+0.677}$ Gyr. 

We adopt an activity index of $R'_{\rm HK} = -4.96$ using $S$-indices from \citet{Arriagada_2011,Jenkins_2011,Pace_2013}, consistent with the values published in many sources \citep{Jenkins_2008,Jenkins_2011,Herrero_2012,Marmier_2013,Gondoin_2020}. The spectroscopic mass of the star is between 0.96-1.03\,$\Msun$ \citep{Allende_1999,Mortier_2013,Santos_2013,Mints_2017,Goda_2019}. Our PARSEC isochrone fitting provides a value of $0.97\pm0.02$\,$\Msun$ that is consistent with the literature values. The stellar rotation period is 26.44 days \citep{Oelkers_2018}, again a close match to the Solar period. Based on the $\approx$4\,Gyr peak from our activity-based analysis, HD~98649 is a chromospherically inactive G dwarf and a close Solar analog. 
    
\subsubsection{HD~106252 (HIP~59610)}
HD~106252, at a distance of 38.1 pc, is a Sun-like G type star \citep{Gonzalez-Hernandez_2010,Datson_2012}. Its chromospheric index is comparable with that of the Sun at the minimum of the solar cycle \citep{Lanza_2018}. A weak lithium absorption feature \citep{Perrier_2003} and a $R'_{\rm HK} = -4.85$ from \citet{Pace_2013} indicate low stellar activity and an old age for HD~106252. Literature values agree on an age of $\sim$4.5 - 7.5 Gyr \citep{Marsakov_1995,Gonzalez_2010,Casagrande_2011,Gontcharov_2012,Pace_2013,Bonfanti_2016,Aguilera_2018}. Our activity-based age indicators however favor a somewhat younger age, with our distribution peaking at $\sim$3 Gyr; this is driven by a single relatively active measurement by \cite{White+Gabor+Hillenbrand_2007}. The stellar luminosity we infer \subtractions{from the Tycho brightness} is higher than that of the Sun. The mass of HD~106252 derived from the PARSEC isochrones model is $1.05\pm0.02\,\Msun$, in agreement with independent determinations \citep{Fischer_2005,Butler_2006,Marchi_2007,Goda_2019}.

\subsubsection{HD~106515A (HIP~59743A)}
HD~106515 is a high common proper motion binary system consisting of the G-type stars HD~106515 A and B, with $V = 7.96$ and $V = 8.22$ mag, respectively. HD~106515~AB is a gravitationally bound system; the presence of the stellar companion HD~106515~B perturbs the orbit of the planet orbiting around HD~106515~A \citep{rica_2017}. HD~106515~B has an estimated spectroscopic mass of $0.925\,\Msun$ and an effective temperature of 5425\,K \citep{Mugrauer_2019}. 
\cite{Dommanget_2002} list a third visual stellar companion $98.\!\!^{\prime\prime}7$ from HD~106515~A with a magnitude of $V=10.3$ (BD-06 3533, or TYC 4946-202-1). \gaia EDR3 places this star at a distance of nearly 1\,kpc, conclusively identifying it as a chance alignment. 

HD~106515~A has a slow rotational velocity and low levels of chromospheric activity, with $R'_{\rm HK} = -5.10$ \citep{Schroder_2009,Meunier_2019}, suggesting an old star. Our activity-based age analysis results in a broad posterior peaking at an age just under 8 Gyr. Previous mass measurements of HD~106515~A range between 0.87-0.95 $\Msun$ \citep{Allende_1999,Santos_2017,Anders_2019,Goda_2019,Queiroz_2020,Gomes_2021}. Our PARSEC isochrone fitting yields a mass of $0.90\pm0.03$\,$\Msun$. We use the same method to derive a mass of \changes{$0.89 \pm 0.03$\,$\Msun$} for HD~106515~B.
 
\subsubsection{HD~171238 (HIP~91085)}
HD~171238 is a G8 dwarf at a distance of 44.5\,pc. With theoretical isochrones, \citet{Segransan_2010} obtain a mass of $M_{*} = 0.94 \pm 0.03$\,$M_\odot$ and an age of $4.9 \pm 4.1$\,Gyr. \citet{Bonfanti_2016} studied the age consistency between exoplanet hosts and field stars with the PARSEC evolutionary code, and found a stellar age of $4 \pm 1.2$ Gyr. Our activity-based age analysis yields a broad posterior spread of 3-12 Gyr with a peak at 6 Gyr, in mild tension with above mentioned values. The discrepancies in the ages and a broad posterior suggest that the age of the star is poorly constrained. \citet{Bonfanti_2016} \subtractions{also derive a stellar luminosity of $0.774\pm0.003$\,$\Lsun$, $24\%$ higher than our inferred value of $0.627\pm0.012$\,$\Lsun$.}
\subtractions{We obtain a similar value of $0.648\pm0.016$\,$\Lsun$ using $V-Ks$ instead of $B-V$ to interpolate the bolometric correction; this could not account for the discrepancy with} \citet{Bonfanti_2016}.

\subsubsection{HD~196067 (HIP~102125)}
HD~196067 is in a bright visual binary gravitationally bound to the G-dwarf star HD~196068 (=HIP~102128) \citep{Gould_2004}. The binary pair is located 39.94\,pc from the Sun. HD~196067 appears to be slightly evolved on the color-magnitude diagram. \citet{Marmier_2013} derived an age of $3.3\pm 0.6$\,Gyr and a mass $M_{*} = 1.29\pm 0.08 \Msun$, with ${\rm [Fe/H]}=0.34\pm 0.04$. Our activity-based age for HD~196067 peaks at $\approx$4 Gyr, consistent with literature findings that most strongly favor an age of about 3 Gyr \citep{Casagrande_2011,Delgado_2015,Winter_2020}. These same authors as well as \citet{Valenti_2005,Mortier_2013,Santos_2017,Goda_2019,Maldonado_2019} infer masses between 1.23-1.32 $\Msun$. The PARSEC isochrones yield a mass of \changes{$1.26\pm$0.07\,$\Msun$}, which agrees well with literature values. We derive a mass of \changes{$1.12 \pm 0.06$\,$\Msun$} for its companion, HD~196068.

\subsubsection{HD~221420 (HIP~116250)}
HD~221420 is a slightly evolved G dwarf located 31.2\,pc from the Sun. \citet{Holmberg_2009} derive a photometric metallicity of $\rm [Fe/H]=0.19$ and an age of $\sim$5 Gyr. Spectroscopic measurements suggest an even higher metallicity \citep[e.g.][]{Valenti_2005,Sousa_2006,Tsantaki_2013}. Other authors have inferred ages between 4 and 6\,Gyr \citep{Valenti_2005,Baumann_2010,Casagrande_2011,Pace_2013,Tsantaki_2013,Aguilera_2018}. Our activity-based age analysis yields an age posterior centered between 4-6 Gyr that closely matches the literature measurements. 
Unlike most of our other stars, HD~221420 has a luminosity and surface gravity that place it above the main sequence.  We therefore verify the activity-based age with an analysis using only the observed luminosity combined with the PARSEC isochrones \citep{Bressan_2012}.  We derive an age of 3 to 5\,Gyr with this approach, consistent with our activity-based age but excluding the tail to old ages. 

We derive a mass of \changes{$1.28 \pm 0.08$\,$M_\odot$} using an activity-based age prior. This increases slightly to $1.35 \pm 0.09$\,$M_\odot$ if we use a uniform age prior. We adopt \changes{$1.28 \pm 0.08$\,$M_\odot$} for our analysis here.  Using the higher mass would result in a slightly longer inferred orbital period and higher mass of the substellar companion.

\section{Radial Velocity data} \label{sec:rv}

The RV time series for the stars come from decades of monitoring by several long-term RV surveys. Four of the targets, HD~171238~b, HD~98649~b, HD~196067~b and HD~106515~Ab, are long period and massive planets from the CORALIE survey \citep{Marmier_2013}. HD~221420~b is a massive companion reported in \citet{Kane_2019} with radial velocities from the 3.9-meter Anglo-Australian Telescope (AAT) in the course of the Anglo-Australian Planet Search \citep[AAPS; ][]{diego_1990}. It was also observed with the High Accuracy Radial velocity Planet Searcher (HARPS) spectrograph at the ESO 3.6-m telescope in La Silla \citep{Trifonov_2020}. The rest of the sample have multi-decade RV data acquired with the HIRES instrument on the Keck~I Telescope, the Hamilton spectrograph at Lick Observatory, and the ELODIE and SOPHIE spectrographs at Haute-Provence Observatory. In this section, we briefly summarize the long-term RV history for each star and the data we use for orbital fits. 

\subsection{HD~29021~b}
HD~29021 was observed with SOPHIE+ survey for giant planets with 66 RV measurements on a 4.5\,yr time baseline \citep{Rey_2017}. \subtractions{No abnormal radial velocity dispersion was detected.} The resulting orbital solution yielded a mass of $M\sin i = 2.4\,\Mjup$ and an orbital period of 3.7 years, placing the planet just outside the outer limit of its star's habitable zone. \addition{The dispersion of the residuals from this fit is within the expected levels for SOPHIE+ and no long term drift was observed by these authors.} We use all of the 66 RVs from the SOPHIE search for northern exoplanets. 

\subsection{HD~81040~b}
\citet{Sozzetti_2006} reported the detection of a massive planetary companion orbiting the young disc star HD~81040 based on five years of precise RV measurements with the HIRES spectrograph on the 10-m Keck telescope and the ELODIE fiber-fed echelle spectrograph on the 1.93\,m telescope at the Observatoire de Haute-Provence in France. HIRES/Keck monitored this star \addition{for 8.5 months} in 1991, obtaining a total of 3 RV measurements, and 23 follow-up RV data were taken with the ELODIE spectrograph from 2002-2005. The orbital fit to all of the data unveiled a massive planet with period $P = 1001.7\pm7.0$\,days, $e=0.526\pm0.042$ and $M\sin i = 6.86\pm0.71\,\Mjup$ \citep{Sozzetti_2006}. We utilize the combined data from ELODIE and HIRES/Keck for our orbit analysis.
      
\subsection{HD~87883~b}
The RV monitoring of HD~87883 began in 1998 December at Lick \citep{Fischer_2013}. A total of 44 RV measurements with Lick have a median uncertainty of 5.3\,m\,s$^{-1}$. An additional 25 higher-quality RV measurements with the HIRES instrument in 2009 on the Keck telescope have a median precision of 4.0\,m\,s$^{-1}$ \citep{Fischer_2009}. A Keplerian fit to RV data from both Lick and Keck revealed a single long-period RV companion with a period of $7.55\pm0.24$ years, an eccentricity of $0.53\pm0.12$ and a minimum mass of $M\sin i = 1.78\pm0.13\,\Mjup$ \citep{Fischer_2009}. \citet{Butler_2017} has published an additional 46 RVs with a median uncertainty of 1.35\,m\,s$^{-1}$ \citep{Butler_2017} that extends the total time baseline to more than 20 years. We use all of the above measurements in our orbit analysis.
           
\subsection{HD~98649~b} 
For HD~98649, a total of 11 radial velocity measurements were published from CORALIE-98 with a median measurement uncertainty of 4.4\,m\,s$^{-1}$. Then, 37 data were published from CORALIE-07 with a median error of 3.3\,m\,s$^{-1}$. Using the combined data, \citet{Marmier_2013} uncovered a companion with $M\sin i = 6.8\,\Mjup$, a period of $13.6^{+1.6}_{-1.3}$~years on an eccentric orbit with $e = 0.85$. CORALIE measured 15 additional RVs in 2014. We make use of all of the 70 RV measurements acquired with the CORALIE spectrograph in the 16-year time span. 

\subsection{HD~106252~b}
HD~106252 was observed with the ELODIE high-precision echelle spectrograph at Haute-Provence Observatory as part of the ELODIE survey for northern extra-solar planets \citep{Perrier_2003}. A total of 40 high-precision RV measurements with a median uncertainty of 10.5 m\,s$^{-1}$ were obtained since March 1997 \citep{Perrier_2003}. Using these data, \citet{Perrier_2003} performed orbit analysis for the companion and found a well-constrained solution with minimum mass of $M\sin i \approx 7.56\,\Mjup$, an eccentricity of $0.471\pm0.028$, and a period of $4.38\pm0.05$ years. \citet{Fischer_2002} independently confirmed the planet using 15 RV measurements (median uncertainty of 11 m\,s$^{-1}$) from Hamilton/Lick, but with poorer constrains due to shorter temporal coverage \citet{Perrier_2003}. We use both the 40 RVs from ELODIE and 15 measurements from Hamilton. 

HD~106252 was observed again with both ELODIE and Hamilton after the initial publications in 2002/2003, providing an additional 15 RVs from ELODIE with a median uncertainty of 11.1 m\,s$^{-1}$, and 54 RVs from Hamilton with median errors of 10.5 m\,s$^{-1}$ in 2006 \citep{Butler_2006}. We also use the most recent RV measurements in 2009 that include 12 RVs from CDES-TS2 (median uncertainty 10.35 m\,s$^{-1}$) and 43 RVs (median uncertainty 9.3 m\,s$^{-1}$) from the High Resolution Spectrograph (HRS) instrument mounted on the 9.2 m Hobby-Eberly Telescope (HET) at the McDonald Observatory \citep{Wittenmyer_2016}. Altogether, we use 179 RV data points spanning 12 years.

\subsection{HD~106515~Ab} 
In the same CORALIE survey, a massive planet HD~106515~Ab was discovered around one star in the binary system HD~106515 \citep{Marmier_2013}. A total of 19 and 24 RV measurements have been collected with CORALIE-98 and CORALIE-07 with median measurement uncertainties of 6.3\,m\,s$^{-1}$ and 4.1\,m\,s$^{-1}$, respectively \citep{Marmier_2013}. Its orbital parameters are well constrained, with $M \sin i = 9.61\pm 0.14\,\Mjup$, $P = 3630\pm12$\,days, $e=0.572\pm0.011$, and a semi-major axis of $4.590\pm0.010$\,AU. Similar to HD~196067, with the presence of a stellar companion, the Kozai-Lidov pumping mechanism \citep{Kozai_1962,Lidov_1962} may be a viable explanation for the high eccentricity. We use all the CORALIE RVs.

\subsection{HD~171238~b}
The RV monitoring of HD~171238 started in October 2002 using the CORALIE spectrograph mounted on the 1.2 Euler Swiss telescope at La Silla Observatory in Chile \citep{Segransan_2010}. Those authors published 32 RV measurements from CORALIE-98 with a median measurement uncertainty of 7.3\,m\,s$^{-1}$. An additional 65 RV measurements from CORALIE-07 show a median measurement error of 3.7\,m\,s$^{-1}$. Using these radial velocities, \citet{Segransan_2010} reported a massive companion with $M\sin i = 2.60\,\Mjup$, a period of 4.17\,yrs, a semi-major axis of 2.5\,AU, and an eccentricity of $e = 0.40$. They also speculate that the presence of spots on the stellar surface and active chromospheric activity might be the cause of the large RV jitter (10 m\,$s^{-1}$) from their single planet Keplerian model fitting. We adopt the 32 RVs from CORALIE-98, 65 RVs from CORALIE-07 and an additional 9 RVs published in 2017 with the HIRES spectrograph on the Keck Telescope \citep{Butler_2017}. 
    
\subsection{HD~196067~b} 
Like HD~98649~b, HD~196067~b was also discovered in the CORALIE survey for southern exoplanets. A total of 82 Doppler measurements have been obtained since September 1999: 30 were from CORALIE-98 with a median measurement uncertainty of 7.5\,m\,s$^{-1}$ and 52 from CORALIE-07 with a median uncertainty of 5.9\,m\,s$^{-1}$ \citep{Marmier_2013}. The data for C98 are sparsely sampled due to poor coverage near periastron \citep{Marmier_2013}. With the caveats from this poor coverage, the companion's $M \sin i$ is constrained between $5.8-10.8 \Mjup$, the period between 9.5-10.6 yr, and the eccentricity between 0.57-0.84. We use the RVs from both CORALIE-98 and CORALIE-07.
    
\subsection{HD~221420~b} 
The companion to HD~221420 was discovered by \citet{Kane_2019} using 18 years of RV data acquired with the Anglo-Australian Telescope (AAT) \citep[AAPS; ][]{diego_1990}. The median uncertainty of the 88 measurements is 1.33\,m\,s$^{-1}$. \citet{Kane_2019} obtain an orbital period of $61.55_{-11.23}^{+11.50}$ years, a RV semi-amplitude of $54.7_{-3.6}^{+4.2}$, an eccentricity of $0.42_{-0.07}^{+0.05}$, and a $M\sin i$ of $9.7_{-1.0}^{+1.1}\,\Mjup$; they infer a semi-major axis of $18.5\pm2.3$\,AU.
The High Accuracy Radial velocity Planet Searcher (HARPS) spectrograph (ESO) also observed the target from 2003-2015, offering 74 higher precision measurements with a median uncertainty of 0.94\,m\,s$^{-1}$. Both AAT+HARPS data are recently utilized by \citet{Venner_2021} to derive a precise dynamical mass of $22.9\pm2.2 \Mjup$. We include all the RV measurements from AAT and HARPS for this target. 

\begin{deluxetable*}{lcrcrcccccccrrc}
\tablecaption{Summary of Hipparcos and Gaia (EDR3) Astrometry from HGCA.\tablenotemark{a} \label{tab:absAST}}
\tabletypesize{\scriptsize}
\tablewidth{0pt}
\tablehead{ \\[-0.5em]
Star & Data 
&\multicolumn{1}{c}{$\mu_{x,\alpha*}$}
&$\sigma[\mu_{\alpha*}]$
&\multicolumn{1}{c}{$\mu_{x,\delta}$}
&$\sigma[\mu_{\delta}]$
&Correlation
&Epoch, $\alpha$
&Epoch, $\delta$ 
&\multicolumn{1}{c}{$\Delta \mu_{x-HG,\alpha*}$}
&\multicolumn{1}{c}{$\Delta \mu_{x-HG,\delta}$}
&Significance\\
&Source &  \multicolumn{2}{c}{mas\,yr$^{-1}$} &  \multicolumn{2}{c}{mas\,yr$^{-1}$} & Coefficient & \multicolumn{2}{c}{year}
& \multicolumn{2}{c}{mas\,yr$^{-1}$}
& 
& 
}
\startdata
%
HD~29021  &Hip  &60.884      &0.674   &22.917      &0.717  &\phantom{$-$}0.160   &1991.11  &1990.92     &$-1.004\pm0.674$               &$-0.261\pm0.717$   &\\ 
          &HG   &61.888      &0.020   &23.178      &0.020  &\phantom{$-$}0.020\\
          &Gaia &62.078      &0.025   &23.430      &0.025  &\phantom{$-$}0.266   &2016.41  &2016.09     &\phantom{$-$}$0.190\pm0.032$   &\phantom{$-$}$0.252\pm0.032$   &8.89$\sigma$\\
HD~81040  &Hip  &$-$151.520  &0.835   &35.954      &0.525  &$-$0.320             &1991.66  &1991.34     &$-0.369\pm0.836$               &\phantom{$-$}$0.190\pm0.525$   &\\ 
          &HG   &$-$151.151  &0.031   &35.764      &0.019  &$-$0.265\\
          &Gaia &$-$151.265  &0.061   &35.708      &0.049  &$-$0.404             &2016.11  &2016.32     &$-0.114\pm0.068$               &$-0.056\pm0.053$               &1.98$\sigma$\\
HD~87883  &Hip  &$-$64.197   &0.665   &$-$60.478   &0.466  &$-$0.006             &1991.50  &1991.09     &$-0.391\pm0.665$               &\phantom{$-$}$0.113\pm0.466$   &\\ 
          &HG   &$-$63.806   &0.024   &$-$60.591   &0.017  &$-$0.062\\
          &Gaia &$-$64.293   &0.037   &$-$61.438   &0.035  &$-$0.214             &2016.13  &2016.33     &$-0.487\pm0.044$               &$-0.847\pm0.039$               &26.57$\sigma$\\
HD~98649  &Hip  &$-$199.571  &0.663   &$-$177.725  &0.575  &$-$0.432             &1991.18  &1991.51     &$-0.019\pm0.663$               &$-0.187\pm0.575$               &\\ 
          &HG   &$-$199.552  &0.024   &$-$177.538  &0.022  &$-$0.478\\
          &Gaia &$-$199.735  &0.031   &$-$177.620  &0.023  &$-$0.047             &2015.81  &2015.84     &$-0.183\pm0.039$               &$-0.082\pm0.032$               &5.68$\sigma$\\
HD~106252 &Hip  &23.781      &0.886   &$-$279.571  &0.444  &$-$0.068             &1991.28  &1991.42     &\phantom{$-$}$0.179\pm0.886$   &\phantom{$-$}$0.602\pm0.444$   &\\ 
          &HG   &23.602      &0.029   &$-$280.173  &0.017  &$-$0.144\\
          &Gaia &23.315      &0.044   &$-$279.896  &0.028  &$-$0.409             &2015.94  &2015.49     &$-0.287\pm0.053$               &\phantom{$-$}$0.277\pm0.033$   &8.65$\sigma$\\
HD~106515A&Hip  &$-$249.256  &1.074   &$-$53.466   &0.875  &$-$0.372             &1991.39  &1991.10     &\phantom{$-$}$1.650\pm1.075$   &$-1.749\pm0.876$               &\\ 
          &HG   &$-$250.906  &0.048   &$-$51.717   &0.035  &$-$0.318\\
          &Gaia &$-$251.469  &0.043   &$-$51.330   &0.030  &$-$0.317             &2016.10  &2016.10     &$-0.563\pm0.064$               &\phantom{$-$}$0.387\pm0.046$   &10.27$\sigma$\\
HD~171238 &Hip  &$-$30.814   &1.549   &$-$110.327  &1.042  &\phantom{$-$}0.069   &1991.04  &1991.22     &$-1.776\pm1.550$               &$-0.713\pm1.042$               &\\ 
          &HG   &$-$29.038   &0.043   &$-$109.614  &0.030  &$-$0.046\\
          &Gaia &$-$29.539   &0.046   &$-$109.580  &0.038  &\phantom{$-$}0.258   &2016.25  &2016.55     &$-0.501\pm0.063$               &\phantom{$-$}$0.034\pm0.048$   &7.87$\sigma$\\
HD~196067 &Hip  &150.240     &2.186   &$-$159.708  &2.343  &$-$0.477             &1991.31  &1991.16     &$-5.894\pm2.187$               &\phantom{$-$}$1.833\pm2.344$   &\\ 
          &HG   &156.134     &0.074   &$-$161.541  &0.079  &$-$0.277\\
          &Gaia &156.404     &0.026   &$-$162.214  &0.031  &$-$0.298             &2015.97  &2016.24     &\phantom{$-$}$0.270\pm0.078$   &$-0.673\pm0.085$               &7.76$\sigma$\\
HD~221420 &Hip  &15.884      &0.437   &1.648       &0.385  &\phantom{$-$}0.159   &1991.14  &1991.30     &\phantom{$-$}$0.911\pm0.437$   &\phantom{$-$}$1.193\pm0.385$   &\\ 
          &HG   &14.973      &0.014   &0.455       &0.012  &\phantom{$-$}0.042\\
          &Gaia &16.306      &0.056   &0.736       &0.058  &\phantom{$-$}0.110   &2016.06  &2015.98     &\phantom{$-$}$1.333\pm0.058$   &\phantom{$-$}$0.281\pm0.059$   &23.11$\sigma$\\
\hline
\enddata
\tablenotetext{a}{from \citet{HGCAeDR3_2021}} 
\end{deluxetable*}

\section{Absolute Stellar Astrometry} \label{sec:absoluteastrometry}

Absolute astrometry of each of our host stars provides a constraint orthogonal to that from radial velocities.
The ESA satellites \Hipparcos and \Gaia each measure a position and proper motion in an inertial reference frame (the International Celestial Reference Frame, or ICRF), but taken 25 years apart. 
\Hipparcos detected proper motions 
near epoch 1991.25 ($\mu_{Hip}$), and 
\Gaia near 2016.0 ($\mu_{Gaia,EDR3}$). Differences between the two proper motions probe a star's acceleration. An additional, more accurate tangential proper motion is given by the \Hipparcos-\Gaia positional difference divided by the $\sim$25 year time baseline ($\mu_{HG}$). We reference this scaled positional difference by subtracting it from both \Hipparcos and \Gaia proper motions ($\Delta \mu_{Hip-HG}$ or $\Delta \mu_{Gaia-HG}$), and we use them as astrometric constraints for the host stars' orbits. We report these variables in Table~\ref{tab:absAST}. The proper motion of a star in conjunction with its radial velocity define a three-dimensional vector that traces out the true space velocity of the star. 

Table~\ref{tab:absAST} summarizes the \Gaia EDR3 version HGCA catalog astrometry for the host stars in our sample. While \Gaia and \Hipparcos's measurements are independent, a slight covariance results from the use of \gaia parallaxes to improve the \Hipparcos astrometry \citep{Brandt_2018}.  We neglect this small covariance in our orbital fits.

All of our stars except for HD~81040 have acceleration $\chi^2$ values greater than 11.8 (i.e., 3$\sigma$ detections assuming Gaussian errors and two degrees of freedom), indicating significant accelerations between \Hipparcos and \Gaia. 

\section{Single-epoch relative astrometry for 3-body fitting} \label{sec:singleRelAst}

Among the stars in our sample, two systems have gravitationally bound, wide stellar companions measured independently in \Gaia EDR3. These include HD~196067's stellar companion HD~196068, and HD~106515~A's secondary stellar companion HD~106515~B as discussed in Section~\ref{sec:stellarcharact}. Table \ref{tab:relast_secondary} lists the single epoch relative astrometry data for both of these systems here. This is obtained using the single epoch measurements of position in R.A. and Dec. ($\alpha$, $\delta$) from \gaia EDR3 at epoch 2016.0 and converting them into position angle east of north (PA) and projected separation ($\rho$) between the host star and the secondary stellar companion.  

We adopt uncertainties of 10\,mas in separation and $0.\!\!^\circ1$ in PA.  These are much larger than the formal \gaia uncertainties, but still represent tiny fractional uncertainties.  Adopting the actual formal \gaia uncertainties gives measurements so precise that it makes the MCMC chains much slower to converge.

\gaia EDR3 also gives proper motions for both stellar companions.  We use these proper motions together with the relative astrometry to constrain the orbital fit.  Both stellar companions are of similar brightness to the primary stars that we fit.  As a result, the magnitude-dependent frame rotation seen by \cite{Cantat-Gaudin+Brandt_2021} will be shared by both components and will not affect our analysis.

\begin{deluxetable}{crr}
\tablewidth{0pt}
\tablecaption{Adopted single epoch relative astrometry derived from Gaia EDR3.\label{tab:relast_secondary}}
\tablehead{ \\[-0.5em]
\colhead{Companion}
&\colhead{HD~196068}
&\colhead{HD~106515~B}}
\startdata
Epoch                                           &2016.0     &2016.0\\
$\rho$ ($''$)                                   &16.62      &6.86\\
$\sigma_\rho$ ($''$)\tablenotemark{a}           &0.01       &0.01 \\
PA ($^\circ$)                                   &19.3       &85.9\\
$\sigma_{\rm PA}$ ($^\circ$)\tablenotemark{a}   &0.1        &0.1 \\
$\mu_{{\rm Gaia},\alpha*}$    (mas\,yr$^{-1}$)                        &163.531    &$-$244.603\\
$\sigma[\mu_{{\rm Gaia},\alpha*}]$    (mas\,yr$^{-1}$)                &0.016      &0.031\\
$\mu_{{\rm Gaia},\delta}$     (mas\,yr$^{-1}$)                        &$-$171.346 &$-$67.744\\
$\sigma[\mu_{{\rm Gaia},\delta}]$  (mas\,yr$^{-1}$)                   &0.018      &0.021\\
Corr Coefficient                                &$-$0.156   &$-$0.650 \\
$\Delta \mu_{{\rm comp}-{\rm host},\alpha*}$    (mas\,yr$^{-1}$)            &7.127      &6.866\\
$\sigma[\Delta \mu_{{\rm comp}-{\rm host},\alpha*}]$  (mas\,yr$^{-1}$)      &0.031      &0.053\\
$\Delta \mu_{{\rm comp}-{\rm host},\delta}$           (mas\,yr$^{-1}$)      &-9.132     &-16.414\\
$\sigma[\Delta \mu_{{\rm comp}-{\rm host},\delta}]$   (mas\,yr$^{-1}$)      &0.036      &0.037\\
\enddata
\tablenotetext{a}{Adopted errors are much larger than \gaia EDR3 values to aid MCMC convergence. }
\end{deluxetable}

We impose priors on the masses of the secondary star in both cases.  We derive these from the same stellar isochrone fitting described in Section \ref{subsec:lum_mass}. For HD~106515~B we find a mass of \changes{$0.86\pm0.03$\,\Msun}. This value is slightly lower than $0.925 \pm 0.05 \,\Msun$ from \citet{Mugrauer_2019}. For HD~196068, we adopt a prior of \changes{$1.18\pm0.06 \,\Msun$}.

\section{Orbital Fit} \label{sec:orbitfitting}

Precise masses and inclinations become possible for RV planets because RV and astrometry measure orthogonal components of the motion in inertial frames. We perform full orbital analyses for our sample of RV planets using the orbit fitting package $\orvara$ \citep{Brandt+Dupuy+Li+etal_2021}. $\orvara$ fits Keplerian orbits to an arbitrary combination of radial velocity, relative and/or absolute astrometry data. None of the RV-detected companions in our sample have previous relative astrometry data from direct imaging. Three of these systems, HD~87883, HD~106252 and HD~106515~AB, were previously observed with AO imaging that did not show signs of companions in the systems. HD~87883 was observed with the Calar Alto 2.2-m telescope with the lucky imaging camera AstraLux in \citep{Ginski_2012,Luck_2017}. HD~106252 was imaged with Palomar/Keck as part of the AO survey of young solar analogs in June, 2004 \citep{Metchev_2009}. Finally, the HD~106515~AB system was imaged with AdoPT@TNG \citep{Desidera_2012}. For the present analysis, we use published RV data from the literature described in Section 3 and absolute astrometry data from the HGCA described in Section 4. 

$\orvara$ uses the intermediate astrometry fitting package \htof\ \citep{Brandt+Michalik+Brandt+etal_2021}. \htof\ parses the intermediate \Hipparcos astrometric data by accounting for the scan angles and uncertainties to construct covariance matrices to solve for best-fit positions and proper motion relative to the barycenter. The \Gaia epoch astrometry is currently unavailable, so \changes{\htof} uses an approach to forward model \Gaia observations using the predicted scan angles and observation times, and fits a five-parameter astrometric model to these synthetic data. The \Gaia EDR3 predicted observation times and scan angles are publicly accessible via the \gaia Observation Forecast Tool \footnote{\url{https://gaia.esac.esa.int/gost/index.jsp}}. $\orvara$ then compares the resulting positions and proper motions to the values provided in the HGCA. 

$\orvara$ adopts the parallel-tempering Markov Chain Monte Carlo (MCMC) sampler with \ptemcee\ \citep{Foreman-Mackey_2013,Vousden+Farr+Mandel_2016} to explore the 9-dimensional parameter space comprised of the host star and companion masses $M_{*}$ and $M_{\rm sec}$, RV jitter, semi-major axis $a$, inclination $i$, PA of the ascending node $\Omega$, mean longitude at a particular reference epoch of 2455197.5 JD ($\lambda_{ref}$), the eccentricity $e$ and the argument of periastron ($\omega$) fitted as $\sqrt{e}\sin\omega$ and $\sqrt{e}\cos\omega$. For our three-body systems, there are six more orbital elements plus another mass; these become 16-parameter fits.  

In addition to the nine free parameters that we fit (sixteen for three-body systems), \orvara\ marginalizes out several nuisance parameters to reduce computational costs.  These include the systemic RV zero point (one per instrument), the parallax, and the barycenter proper motion.  We assume the \Gaia EDR3 parallax as our parallax prior, and use flat priors for the RV zero point(s) and barycenter proper motion. \orvara\ thus produces posterior distributions for all of these parameters as well.

\begin{deluxetable}{cc}
\tablewidth{0pt}
\tablecaption{Adopted Priors.
    \label{tab::priors}}
\tablehead{ \\[-0.5em]
Parameter & Prior}
\startdata
RV Jitter  $\sigma_{\rm jit}$     &  $1/\sigma_{\rm jit}$ (log-flat) \\
Primary Mass $M_{\rm *}$        &  $1/M$ (Gaussian) \\
Secondary Mass $M_{\rm sec}$      &  $1/M$ (log-flat) \\
Semimajor axis $a$                &  $1/a$ (log-flat) \\
$\sqrt{\varepsilon} \sin \omega$  &  uniform \\
$\sqrt{\varepsilon} \cos \omega$  &  uniform \\
Inclination $i$                   & $\sin(i)$, 0$^{\circ}<i<180 ^{\circ}$ (geometric)\\
Mean longitude at 2010.0 $\lambda_{\rm ref}$ &                         uniform \\
Ascending node $\Omega$           &  uniform \\
Parallax $\varpi$                 &  $\exp\left[-\frac{1}{2} (\varpi - \varpi_{\rm EDR3})^2/\sigma_\varpi^2 \right]$ 
\enddata
\end{deluxetable}

We perform orbital fits for each RV companion in our sample with \orvara. We chose informative priors on the host stars' masses according to stellar masses derived in Section 2 and listed in Table~\ref{tab:stellar_params}. These are Gaussian with the means and standard deviations given in Table \ref{tab:stellar_params}. We assume uninformative priors for our other fitted parameters: either log-uniform, uniform, or geometric (see Table~\ref{tab::priors}), except for parallax, which we have marginalized out from the fit. For each target, we use \ptemcee\ to fit for the nine parameters, employing 30 temperatures and 100 walkers over $5\times10^5$ steps per walker. In each case, the MCMC chains converge after no more than 15,000 steps, we thus discard the first 20,000 steps as burn-in and use the rest for inference. We post-process the MCMC chains with $\orvara$, and we discuss the results for each individual system in the following Section. 

\section{Updated orbital parameters}
\label{sec:udpatedparameters}

\setlength\extrarowheight{4pt}
\movetabledown=73mm
\begin{rotatetable*}
\begin{deluxetable*}{lcccccccccccc}
\tablewidth{0pt}
\tablefontsize{\scriptsize}
\tablecaption{Posteriors of single RV companions in our sample from orvara MCMC analysis.
\label{tab:planet_params}}
\tablehead{
Companion   &
HD~29021~b   &  
HD~81040~b   &  
\changes{HD~87883~b}   &  
HD~98649~b   &  
HD~106252~b  &  
\changes{HD~106515~Ab} &  
\changes{HD~106515~B}\tablenotemark{*} & 
HD~171238~b  &  
\changes{HD~196067~b}  &  
\changes{HD~196068}\tablenotemark{*}  & 
\changes{HD~221420~b}  &  
}
\startdata
    \multicolumn{11}{c}{Fitted parameters}\\
    \hline
    $\mathrm{RV\, Jitter\, \sigma\, (m\,s^{-1})}$
    &${3.66}_{-0.58}^{+0.60}$
    &${4.14}_{-4.1}^{+0.69}$
    &${4.92}_{-0.12}^{+0.06}$
    &${4.74}_{-0.34}^{+0.19}$
    &${2.9}_{-2.9}^{+1.7}$
    &${7.6}_{-1.3}^{+1.3}$
    &${7.6}_{-1.3}^{+1.3}$
    &${4.96}_{-0.06}^{+0.03}$
    &${7.7}_{-1.1}^{+1.2}$
    &${7.7}_{-1.1}^{+1.2}$
    &${3.75}_{-0.23}^{+0.25}$\\ 
    $\mathrm{Primary\, Mass\, (M_{\odot})}$ 
    &${0.86}\pm	0.02$
    &${0.97}\pm	0.02$
    &${0.80}\pm	0.02$
    &${0.97}\pm	0.02$
    &${1.05}\pm	0.02$
    &${0.90}\pm	0.03$
    &${0.90}\pm	0.03$
    &${0.92}\pm	0.03$
    &${1.34}\pm	0.06$
    &${1.34}\pm	0.06$
    &${1.30}\pm	0.08$\\
    $\mathrm{Secondary\, Mass\, (M_{\rm Jup})}$  
    &${4.47}_{-0.65}^{+0.67}$			
    &${7.24}_{-0.37}^{+1.0}$			
    &${6.31}_{-0.32}^{+0.31}$			
    &${9.7}_{-1.9}^{+2.3}$			
    &${10.00}_{-0.73}^{+0.78}$			
    &${18.9}_{-1.4}^{+1.5}$			
    &${904}_{-31}^{+31}$
    &${8.8}_{-1.3}^{+3.6}$			
    &${12.5}_{-1.8}^{+2.5}$
    &${1235}_{-59}^{+58}$ 
    &${20.6}_{-1.6}^{+2.0}$\\
    $\mathrm{Semimajor\, axis}\, a\, \mathrm{(AU)}$  
    &${2.294}_{-0.019}^{+0.019}$	
    &${1.946}_{-0.014}^{+0.014}$	
    &${3.77}_{-0.094}^{+0.12}$	
    &${5.97}_{-0.21}^{+0.24}$	
    &${2.655}_{-0.017}^{+0.017}$	
    &${4.48}_{-0.050}^{+0.050}$	
    &${335}_{-42}^{+96}$
    &${2.518}_{-0.033}^{+0.032}$	
    &${5.10}_{-0.17}^{+0.22}$
    &${1701}_{-186}^{+206}$
    &${9.99}_{-0.70}^{+0.74}$	\\
    ${\sqrt{e}\, \sin\, \omega}$       
    &${-0.007}_{-0.039}^{+0.038}$
    &${0.714}_{-0.020}^{+0.018}$
    &${-0.831}_{-0.019}^{+0.019}$
    &${-0.865}_{-0.012}^{+0.023}$
    &${-0.645}_{-0.010}^{+0.010}$
    &${0.625}_{-0.020}^{+0.019}$
    &${0.39}_{-0.28}^{+0.16}$
    &${0.437}_{-0.024}^{+0.024}$
    &${0.426}_{-0.066}^{+0.056}$
    &${0.839}_{-0.073}^{+0.028}$
    &${-0.254}_{-0.063}^{+0.062}$\\
    ${\sqrt{e}\, \cos\, \omega}$  
    &${-0.672}_{-0.010}^{+0.010}$
    &${0.114}_{-0.041}^{+0.041}$
    &${0.178}_{-0.055}^{+0.055}$
    &${-0.312}_{-0.10}^{+0.051}$
    &${0.253}_{-0.017}^{+0.016}$
    &${-0.424}_{-0.029}^{+0.029}$
    &${-0.54}_{-0.12}^{+0.41}$
    &${0.408}_{-0.040}^{+0.038}$
    &${-0.702}_{-0.13}^{+0.080}$
    &${-0.20}_{-0.18}^{+0.16}$
    &${-0.312}_{-0.071}^{+0.10}$\\
    $\mathrm{Inclination}~(i<90^\circ)~(^\circ)$ \tablenotemark{a}
    &${33.7}_{-4.9}^{+6.8}$
    &${73}_{-16}^{+12}$
    &${16.8}_{-1.4}^{+1.7}$
    &${43.7}_{-8.1}^{+13}$
    &${46.0}_{-4.1}^{+4.9}$
    &${29.2}_{-2.2}^{+2.4}$
    &${18.8}_{-8.3}^{+7.8}$
    &${19.1}_{-8.5}^{+7.9}$
    &${41.2}_{-9.1}^{+28}$
    &${10.6}_{-5.2}^{+6.1}$
    &${17.8}_{-2.8}^{+2.9}$\\
    $\mathrm{Inclination}~(i>90^\circ)~(^\circ)$ \tablenotemark{a}
    &${146.3}_{-6.8}^{+4.9}$
    &${107}_{-12}^{+16}$
    &${163.2}_{-1.7}^{+1.4}$
    &${136.3}_{-13}^{+8.1}$
    &${134.0}_{-4.9}^{+4.1}$
    &${150.8}_{-2.4}^{+2.2}$
    &--
    &${162.9}_{-3.1}^{+5.0}$
    &${138.8}_{-28}^{+9.1}$
    &--
    &${162.2}_{-2.9}^{+2.8}$\\
    Ascending node $\mathrm{\Omega\, (^{\circ})}$  
    &${40.8}_{-9.1}^{+7.1}$
    &${77}_{-23}^{+27}$
    &${109.9}_{-4.1}^{+4.0}$
    &${54}_{-26}^{+44}$
    &${105.3}_{-6.1}^{+10}$
    &${59.9}_{-4.1}^{+4.6}$
    &${62}_{-37}^{+45}$
    &${71}_{-13}^{+47}$
    &${101}_{-82}^{+59}$
    &${163}_{-158}^{+12}$
    &${66.6}_{-3.1}^{+5.4}$\\
    Mean longitude at $\mathrm{\lambda_{\rm ref}~(^{\circ})}$
    &${15.2}_{-2.7}^{+2.7}$
    &${148.3}_{-3.8}^{+3.7}$
    &${73.7}_{-7.4}^{+13}$
    &${75.8}_{-7.6}^{+4.1}$
    &${174.6}_{-1.6}^{+1.5}$
    &${100.5}_{-1.8}^{+1.8}$
    &${128}_{-110}^{+37}$
    &${59.0}_{-3.0}^{+3.0}$
    &${150}_{-15}^{+13}$
    &${93}_{-11}^{+12}$
    &${170.5}_{-2.2}^{+2.3}$\\
    \hline
    \multicolumn{11}{c}{Derived parameters}\\
    \hline
    Period (years)          
    &${3.737}_{-0.018}^{+0.018}$
    &${2.7452}_{-0.0093}^{+0.011}$
    &${8.23}_{-0.34}^{+0.32}$
    &${14.74}_{-0.75}^{+0.88}$
    &${4.202}_{-0.010}^{+0.011}$
    &${9.927}_{-0.032}^{+0.030}$
    &${4630}_{-850}^{+2150}$
    &${4.148}_{-0.046}^{+0.045}$
    &${9.88}_{-0.43}^{+0.63}$
    &${44300}_{-7600}^{+8700}$
    &${27.7}_{-2.5}^{+3.0}$\\
    Argument of periastron $\omega\, (^{\circ})$
    &${180.6}_{-3.2}^{+3.4}$
    &${81.0}_{-3.3}^{+3.3}$
    &${282.1}_{-3.7}^{+3.7}$
    &${250.2}_{-6.3}^{+3.2}$
    &${291.4}_{-1.5}^{+1.5}$
    &${124.1}_{-2.5}^{+2.6}$
    &${144}_{-42}^{+26}$
    &${46.9}_{-3.7}^{+3.9}$
    &${148.6}_{-5.5}^{+7.6}$
    &${103}_{-11}^{+13}$
    &${219}_{-12}^{+17}$\\
    Eccentricity $e$        
    &${0.453}_{-0.013}^{+0.014}$
    &${0.525}_{-0.026}^{+0.024}$
    &${0.720}_{-0.027}^{+0.038}$
    &${0.852}_{-0.022}^{+0.033}$
    &${0.480}_{-0.010}^{+0.010}$
    &${0.571}_{-0.012}^{+0.012}$
    &${0.399}_{-0.064}^{+0.12}$
    &${0.358}_{-0.026}^{+0.028}$
    &${0.70}_{-0.12}^{+0.14}$
    &${0.744}_{-0.027}^{+0.024}$
    &${0.162}_{-0.030}^{+0.035}$\\
    Semimajor axis (mas)             
    &${74.29}_{-0.62}^{+0.61}$
    &${56.55}_{-0.41}^{+0.41}$
    &${206.4}_{-5.2}^{+6.7}$
    &${141.6}_{-4.9}^{+5.7}$
    &${69.68}_{-0.46}^{+0.45}$
    &${131.2}_{-1.5}^{+1.5}$
    &${9813}_{-1227}^{+2815}$
    &${56.63}_{-0.74}^{+0.73}$
    &${127.6}_{-4.2}^{+5.5}$
    &${42600}_{-4700}^{+5200}$
    &${321}_{-23}^{+24}$\\
    Periastron time (${\rm JD}-2450000$)
    &${5824.8}_{-9.0}^{+9.4}$
    &${5511.1}_{-8.7}^{+10}$
    &${6913}_{-16}^{+17}$
    &${10500}_{-280}^{+330}$	
    &${6463.0}_{-8.6}^{+8.4}$	
    &${5435.8}_{-8.5}^{+8.7}$
    &${1570000}_{-330000}^{+810000}$	
    &${5905}_{-20}^{+20}$
    &${6560}_{-270}^{+350}$
    &${485600}_{-9700}^{+8500}$ 		
    &${6580}_{-280}^{+430}$		\\
    $M_{p} \sin i$ ($\Mjup$)  (this work) 
    &${2.483}_{-0.068}^{+0.070}$
    &${6.87}_{-0.19}^{+0.19}$
    &${1.82}_{-0.10}^{+0.12}$
    &${6.64}_{-0.23}^{+0.45}$
    &${7.20}_{-0.13}^{+0.13}$
    &${9.25}_{-0.25}^{+0.25}$
    &--
    &${2.599}_{-0.088}^{+0.090}$
    &${7.9}_{-1.5}^{+4.7}$
    & --
    &${6.24}_{-0.59}^{+0.65}$\\
    $M_{p}\sin i$ ($\Mjup$) (literature) 
    &$2.4_{-0.2}^{+0.2}$
    &$6.86_{-0.71}^{+0.71}$
    &$1.54_{-0.26}^{+0.26}$
    &$7.27_{-0.98}^{+0.98}$
    &$6.93_{-0.27}^{+0.27}$
    &$9.08_{-0.20}^{+0.20}$
    &--
    &$2.72_{-0.49}^{+0.49}$
    &$6.9_{-1.1}^{+3.9}$
    &--
    &$6.31_{-0.61}^{+0.60}$\\
    Literature mass reference\tablenotemark{b} 
    &	R17		
    &	S06 S17		
    &	S17	F09	
    &	R19	M13	
    &	S17	W09	L14
    &	S19	M13	
    &   --			
    &	M18	S10	
    &	M13		
    &	--		
    &	V21	K19	\\
    Difference in $M_{p}\sin i$ ($\%$)
    &3.3$\%$
    &0.1$\%$
    &15.4$\%$
    &9.5$\%$
    &3.8$\%$
    &1.8$\%$
    &--
    &4.7$\%$
    &12.7$\%$
    &--
    &1.1$\%$\\
    Difference in $M_{p}\sin i$ ($\sigma$)
    &0.39$\sigma$
    &0.01$\sigma$
    &1.01$\sigma$
    &0.63$\sigma$
    &0.90$\sigma$
    &0.53$\sigma$
    &--
    &0.24$\sigma$
    &0.54$\sigma$
    &--
    &0.08$\sigma$\\
\enddata

\tablenotetext{*}{Secondary stellar companion. \addition{Their \orvara\ masses in solar masses are ${0.86}\pm 0.03\,\Msun$ for HD~106515~B and ${1.18}\pm	0.06\,\Msun$ for HD~106515~B.}}
\tablenotetext{a}{The inclination distributions are usually bimodal, so we separately report the values for prograde and retrograde orbits.}
\tablenotetext{b}{References abbreviated as S06 \citep{Sozzetti_2006}; F09 \citep{Fischer_2009}; W09 \citep{Wittenmyer_2009}; S10 \citep{Segransan_2010}; M13 \citep{Marmier_2013}; L14 \citep{Liu_2014}; R17 \citep{Rey_2017}; S17 \citep{Stassun_2017}; M18 \citep{Ment_2018}; V21 \citep{Venner_2021}; K19 \citep{Kane_2019}; R19 \citep{Rickman_2019}; S19 \citep{Saffe_2019}. Tabulated masses are from the first reference for each star.  The other measurements are $1.78\pm0.34$ (F09) for HD 87883~b; $7.61\pm0.39$ (W09) and $6.92\pm0.16$ (L14) for HD 106252b; $2.60\pm0.15$ (S10) for HD 171238b; $7.27\pm0.98$ (S17) for HD 81040b; $6.8\pm0.5$ (M13) for HD 98649b; and $9.61_{-0.14}^{+0.14}$ (M13) for 106515~Ab.}
\end{deluxetable*}
\end{rotatetable*}

In this section, we summarize the results of our orbital fits to each target. \addition{The \orvara\ radial velocity fits and relative astrometric orbits are shown in Figure Set \ref{fig:HD29021b_RVAST}, and the corner plots and covariances of our orbital posteriors are showcased in Figure Set \ref{fig:HD29021b_corner}, both are presented in the appendix.}
Table~\ref{tab:planet_params} lists the nine basic Keplerian orbital elements \orvara\ fit, and the inferred parameters computed using the fitted parameters, as well as the $M\sin i$ values from the literature. We obtain tight constraints on the masses of companions except for HD~171238. Aside from HD~221420~b, whose precise mass was measured by \citet{Venner_2021}, we present the first precise dynamical mass and orbital inclination measurements for the RV companions in our sample. Our derived $M\sin i$ values agree with the RV-only literature values to within 1$\sigma$. 

For the inclinations (or equivalently the position angle of the ascending node $\Omega$) derived in this paper, there are two complementary values whose absolute values wrap around $180^{\circ}$. This is unsurprising since the orbital inclination is dependent on whether the planet is in prograde ($0\ge i_1 \le90$) or retrograde ($i_2 = 180^{\circ} - i_1$) motion \citep{Kervella_2020}, and there are limited ways to determine which orbit the companion is on without high contrast imaging. Unfortunately, besides the long-period brown dwarfs in our sample, none of the other companions are accessible to imaging in the near term.  

\begin{figure*}[htp]
    \vspace*{+10mm}
    \includegraphics[width=0.33\textwidth,height=0.2\textheight]{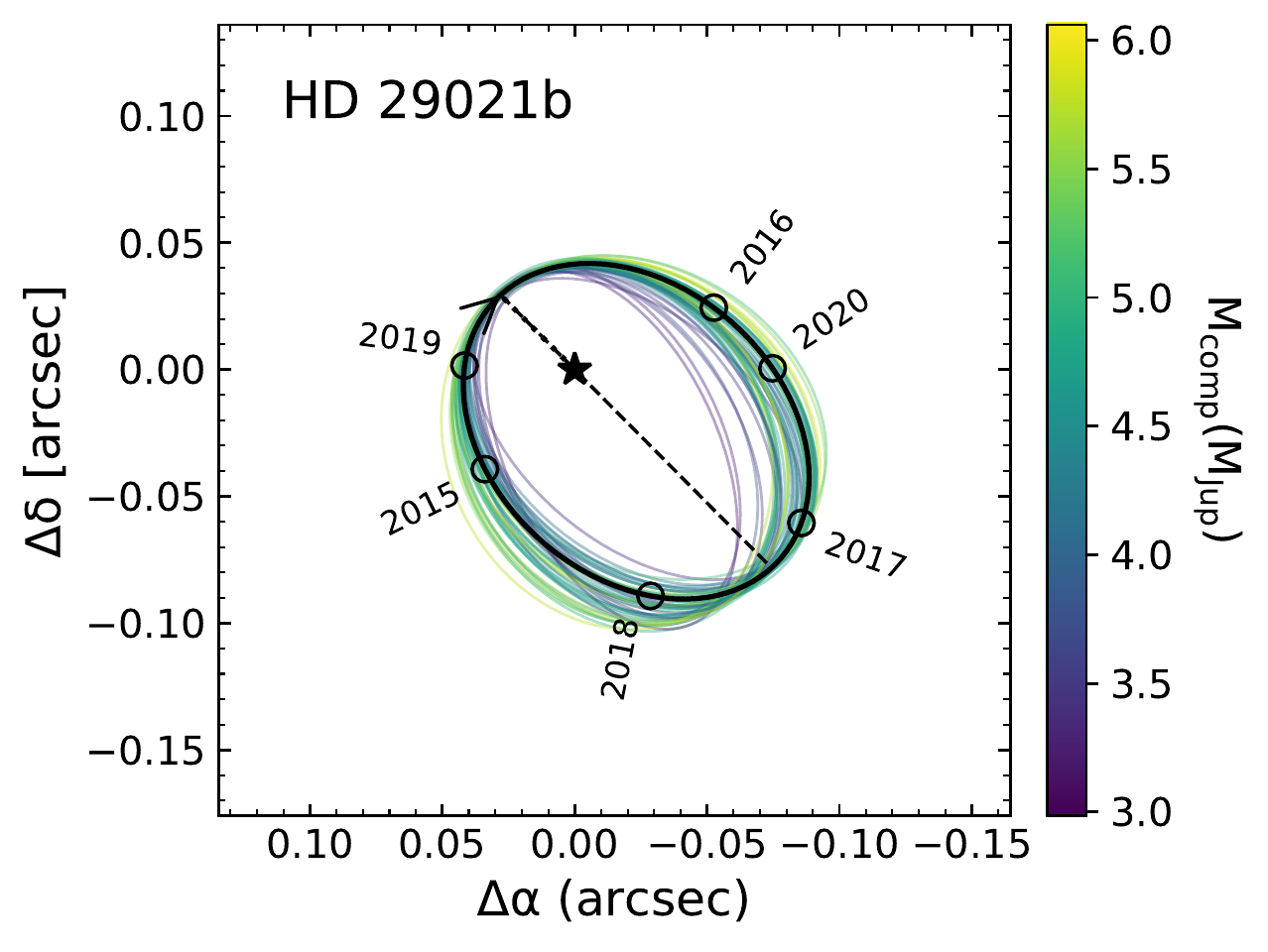}  \quad
    \includegraphics[width=0.33\textwidth,height=0.2\textheight]{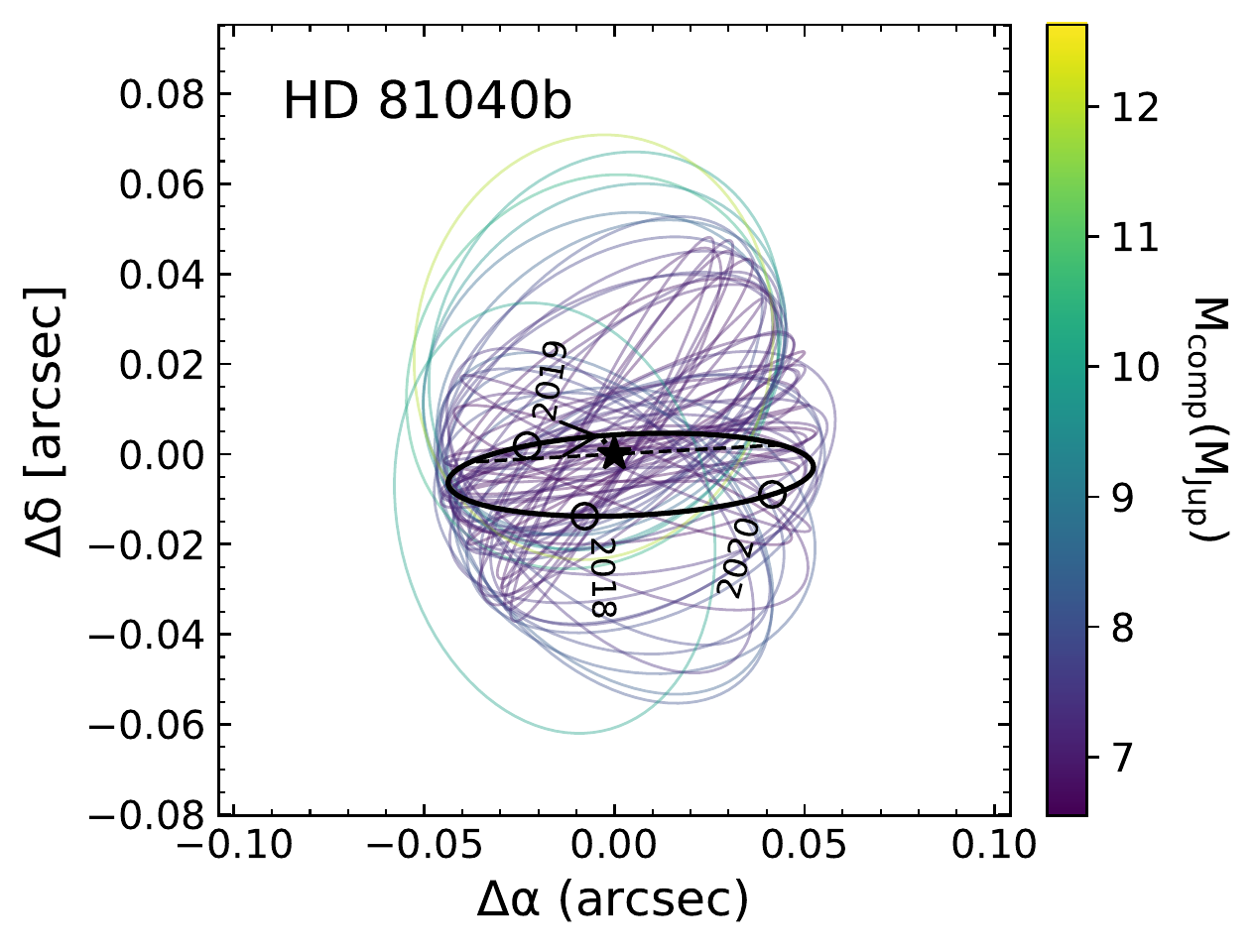}  \quad
    \includegraphics[width=0.33\textwidth,height=0.2\textheight]{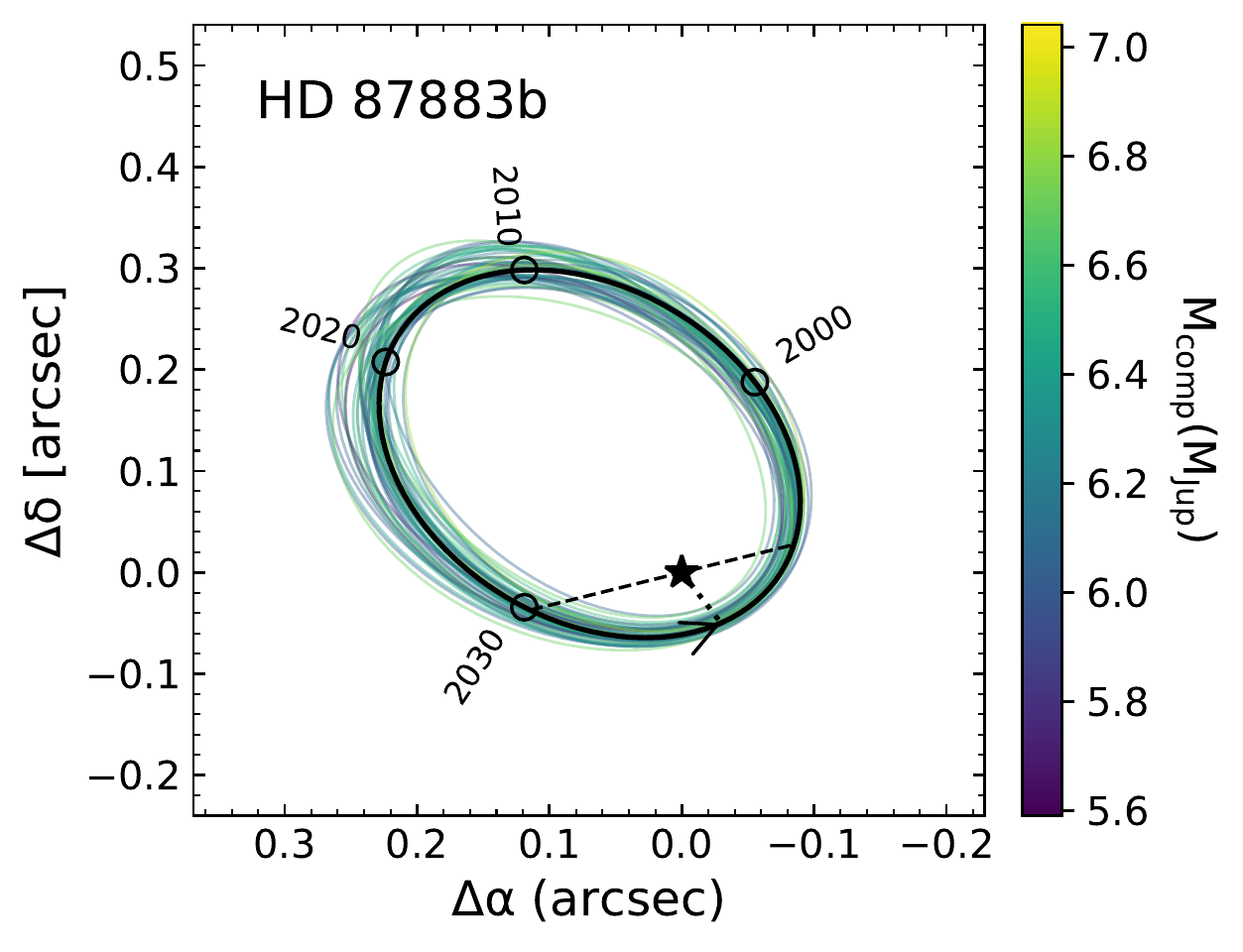}  \\[+5em]
    \includegraphics[width=0.33\textwidth,height=0.2\textheight]{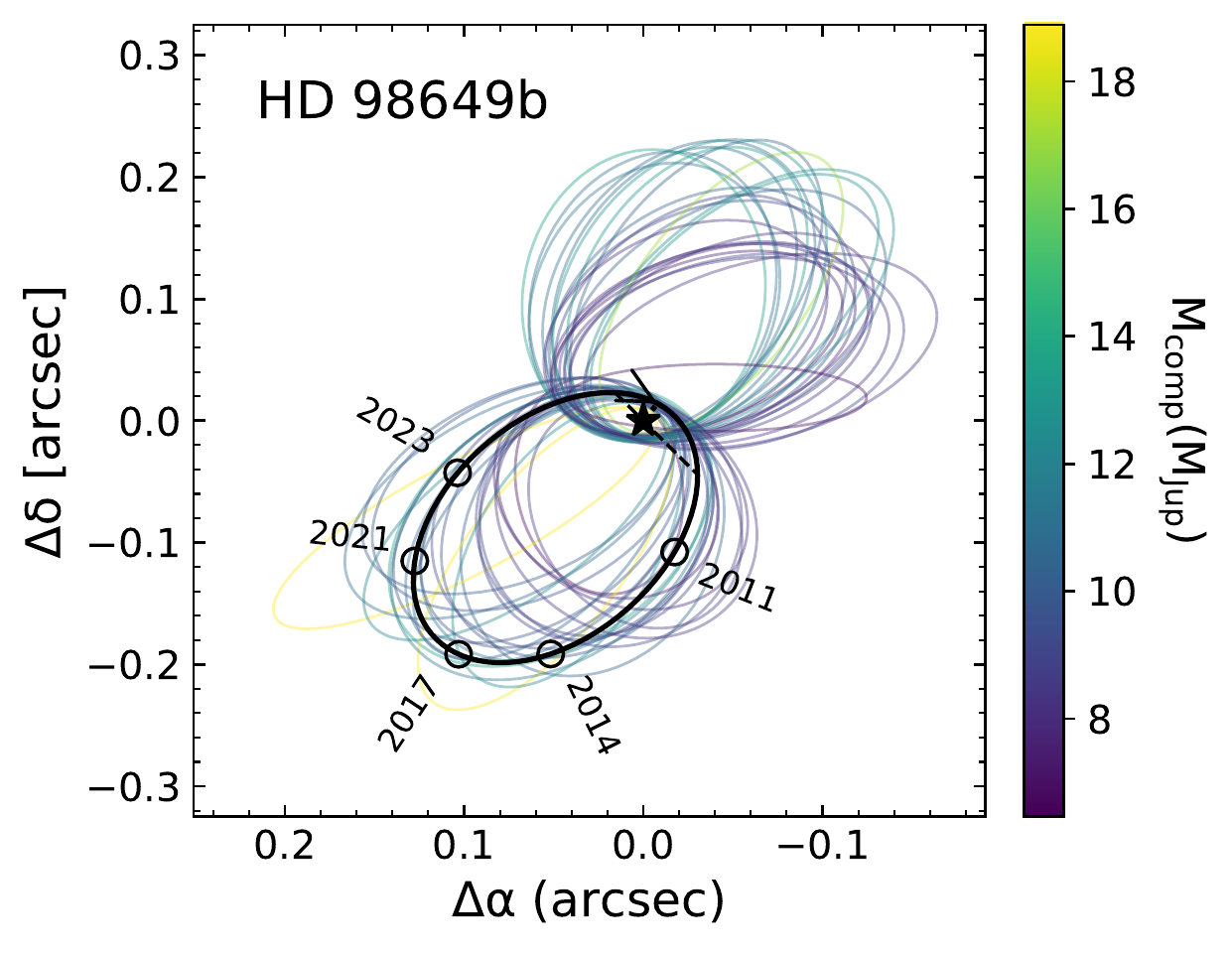} \quad
    \includegraphics[width=0.33\textwidth,height=0.2\textheight]{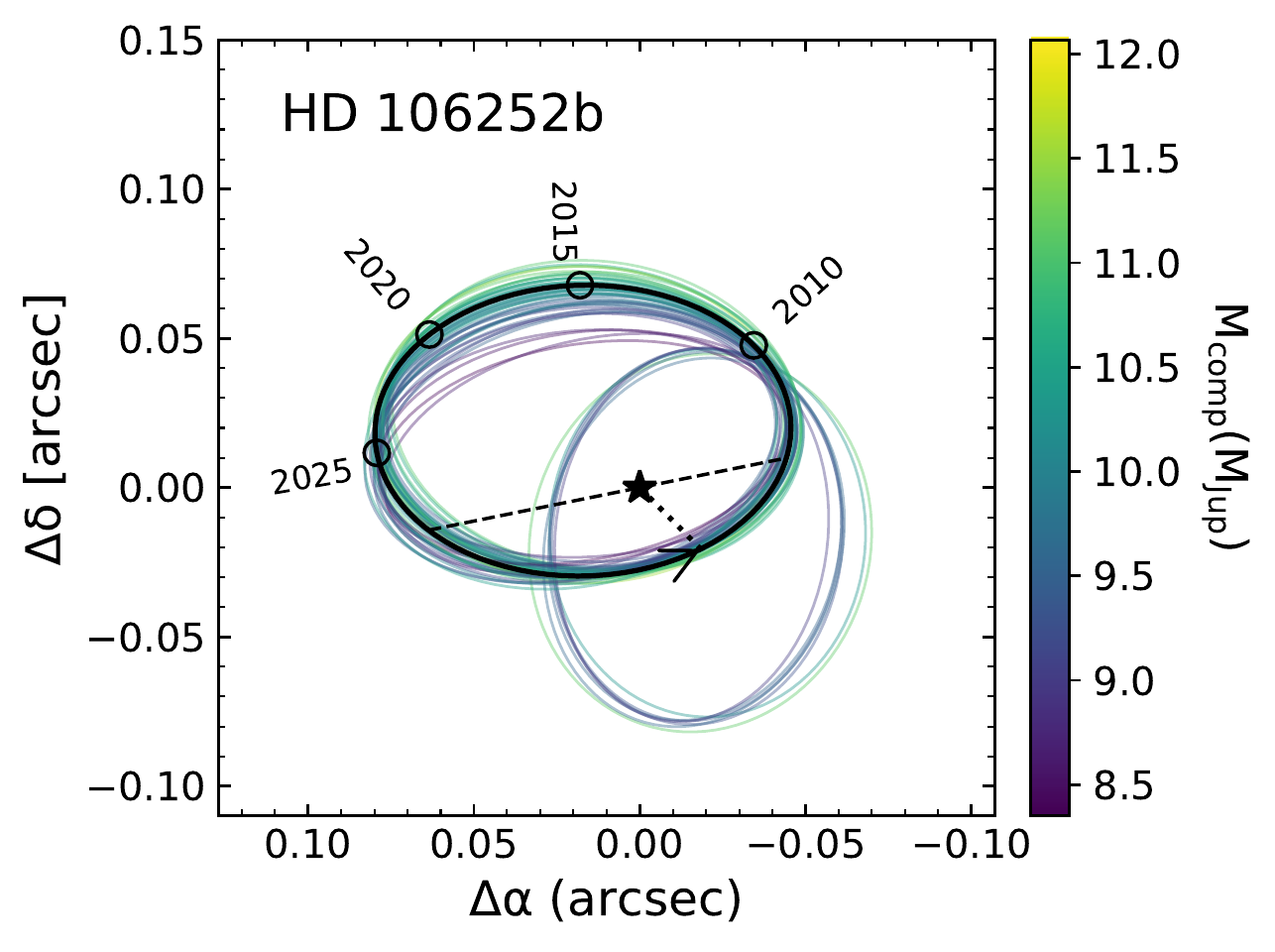} \quad
    \includegraphics[width=0.33\textwidth,height=0.2\textheight]{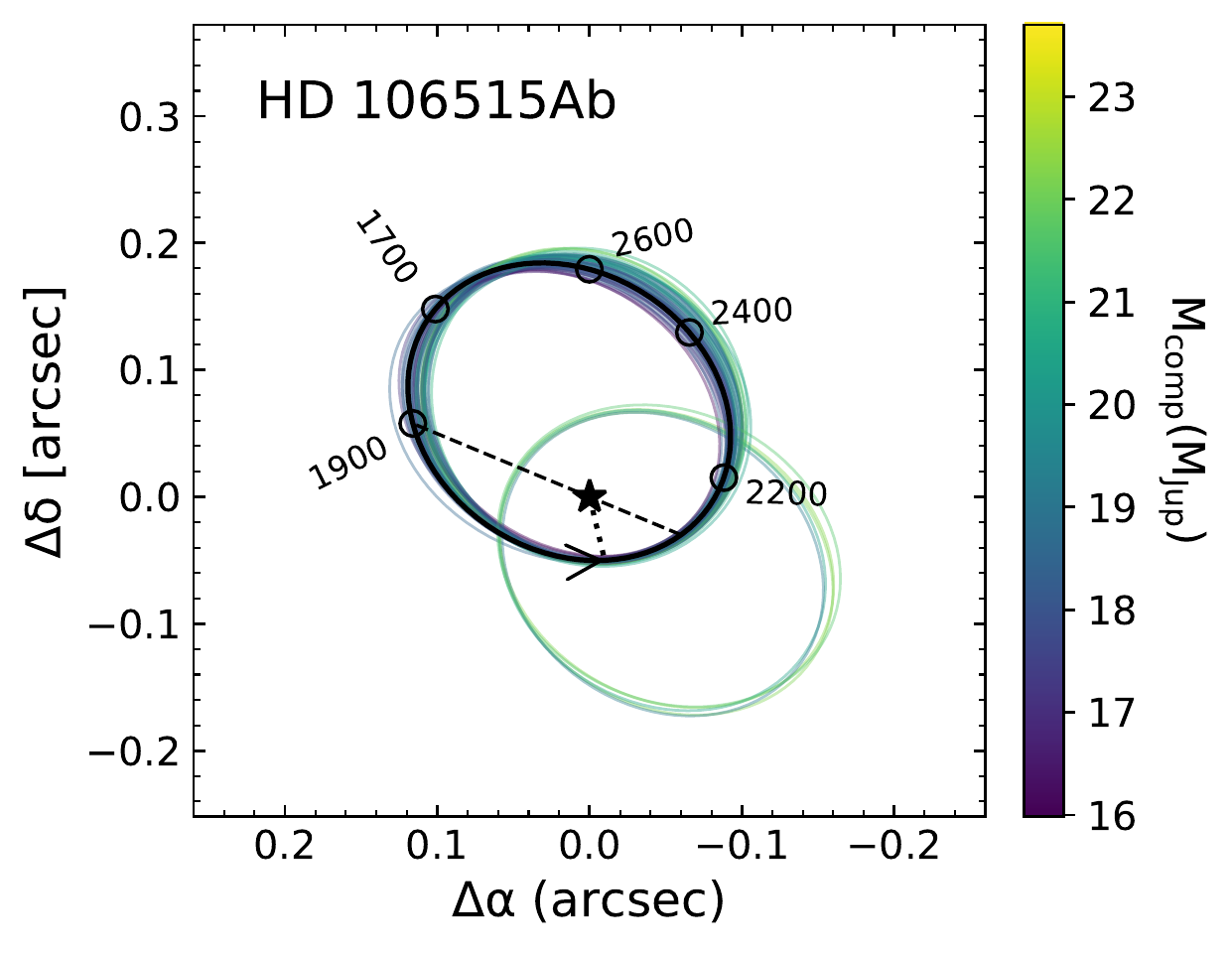} \\[+5em]
    \includegraphics[width=0.33\textwidth,height=0.2\textheight]{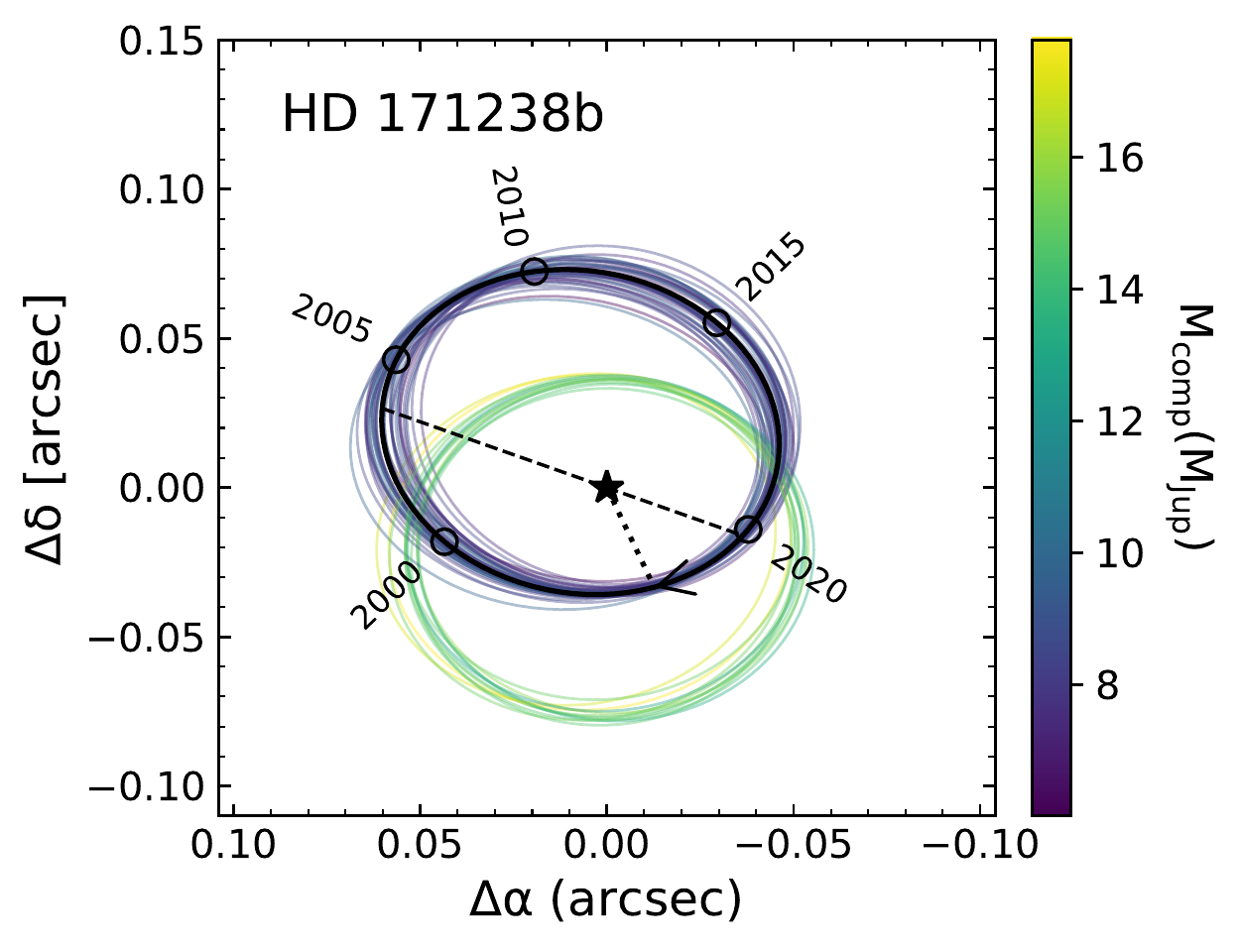} \quad
    \includegraphics[width=0.33\textwidth,height=0.2\textheight]{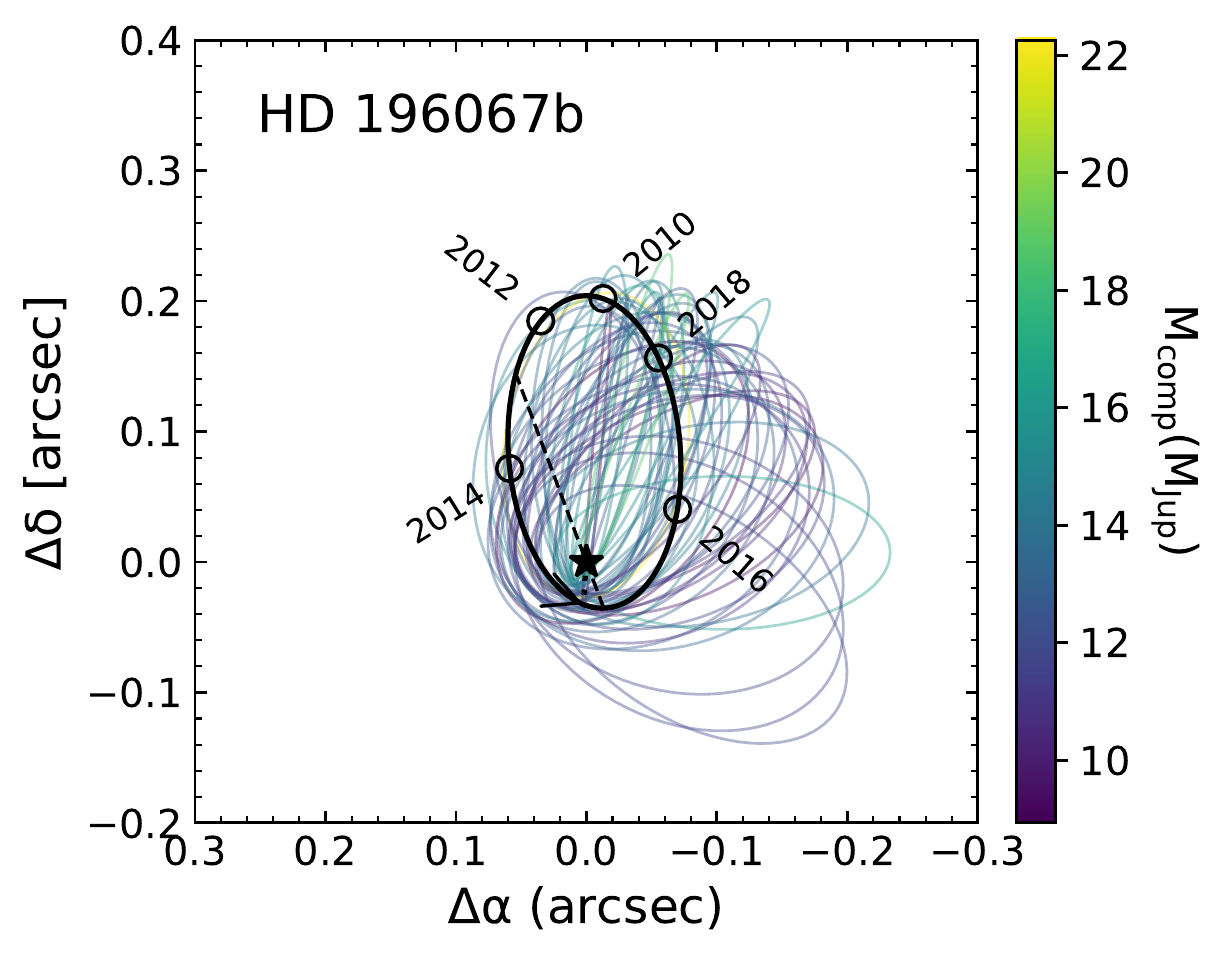} \quad
    \includegraphics[width=0.33\textwidth,height=0.2\textheight]{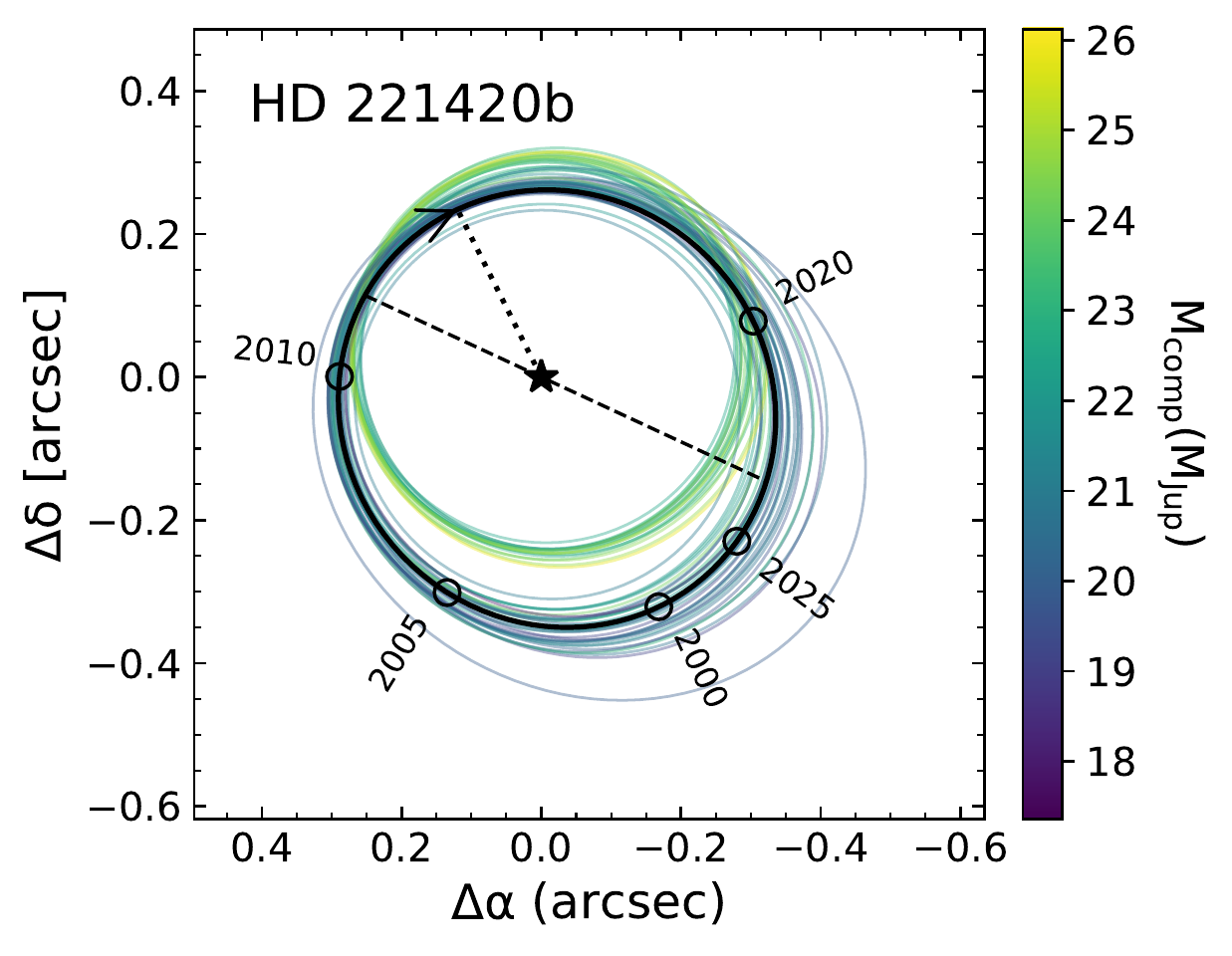}  \\[+5em]
    \caption{Relative astrometric orbits of the nine companions studied in this paper. The thick black lines indicate the best-fit orbit and the colorful thin lines are 50 orbits drawn randomly from corresponding posterior distribution, color-coded by the mass of the secondary companion from purple (low mass) to yellow (high mass). The dashed lines are the line of nodes connecting the host star to the companion's periastron passage. The arrows on the best-fit orbit indicate the direction of motion of the companion revolving around the host star. The hollow circled points label selected epochs along the orbit.
\label{fig:astrometric_plots}}
\end{figure*}

The relative astrometric orbits of every RV companion with respect to its host star are demonstrated in Fig.~\ref{fig:astrometric_plots}. As a result of bimodal inclinations, the prograde and retrograde orbits are clearly evident from the two families of orbits shown in the plots. 

The orbital fits to the RVs and absolute astrometry from HGCA are presented in Figure Set~\ref{fig:HD29021b_RVAST}. The host stars' astrometric reflex motions over the 25-year time baseline between Hipparcos and Gaia are clearly seen. These figures show that the astrometric reflex motion of a star oscillates with a fixed period over 25 years, short period companions induce more cycles of oscillation.  One of our companions, HD~221420~b, is a possible target for direct imaging.  Figure 12 shows its predicted locations at four future dates. Finally, for each system, we show corner plots in Figure Set~\ref{fig:HD29021b_corner} of five astrophysically interesting parameters, including the primary and companion masses $M_{\rm pri}$ and $M_{\rm sec}$, semi-major axis $a$, eccentricity $e$ and inclination $i$, and their covariances. 

\subsection{HD~29021~b}
Figure Set~\ref{fig:HD29021b_corner} shows the fitted posterior distributions for the orbital elements of HD~29021~b. All are Gaussian apart from the inclination, which is bi-modal with equal likelihoods for prograde and retrograde orbits: either $33.\!\!^\circ7_{-4.9}^{+6.8}$ or $146.\!\!^\circ3_{-6.8}^{+4.9}$. Although the RV data only cover one and a half periods, the RV orbit of HD~29021~b is well-constrained. For HD~29021~b, we obtain a dynamical mass of $4.47_{-0.65}^{+0.67} \Mjup$, an eccentricity of $0.453_{-0.013}^{+0.014}$, and a semi-major axis of $74.29_{-0.62}^{+0.61}$ AU. The $M\sin i$ inferred from our fit is $2.483_{-0.068}^{+0.070} \Mjup$. We compare our orbital solution with the only orbital fit for this system published by  \citet{Rey_2017} where $M\sin i = 2.4\pm0.2 \Mjup$, $e=0.459\pm0.008$, $a=2.28_{-0.08}^{+0.07}$, and $P=3.732_{-0.012}^{+0.013}$ years; all of our parameters are consistent with this orbital solution.

\subsection{HD~81040~b}
HD~81040~b has the shortest period in our sample; the RV orbit is well-constrained by the joint HIRES and ELODIE RV data (see Figure Set~\ref{fig:HD29021b_RVAST}). The posterior distributions of selected parameters for HD~81040 are displayed in Figure Set~\ref{fig:HD29021b_corner}. Our orbital inclination indicates that it presents itself as an edge-on system with an orbital inclination of either $73{^\circ}_{-16}^{+12}$ or $107{^\circ}_{-12}^{+16}$. Our solution reveals that the true mass of HD~81040~b is $7.24_{-0.37}^{+1.0}\,\Mjup$, slightly higher than these previous RV-only $M\sin i$ values. Out of the two $M\sin i$ values reported for this system, our inferred $M\sin i = 6.87\pm0.19\,\Mjup$ agrees better with a value of $6.86\pm0.71$\,$\Mjup$ from \citet{Sozzetti_2006} than a value of $7.27\pm0.98$ from \citet{Stassun_2017}. \addition{The orbit of HD~81040~b is consistent with being edge-on: we find a 0.9\% probability that the planet will transit.  If it does transit, we predict transit times of 2021-11-11 and 2024-10-08, each with an unfortunately large uncertainty of about 25 days.}

\subsection{HD~87883~b}
The posterior probability distributions of selected parameters and their covariances for HD~87883~b are shown in Figure Set~\ref{fig:HD29021b_corner}. The posterior for the semi-major axis is bi-modal, peaking at 3.7 AU and 3.9 AU, respectively. Both the prograde and retrograde orbital inclinations are relatively face-on with values of \changes{$16.\!\!^\circ 8_{-1.4}^{+1.7}$ and $163.\!\!^\circ 1_{-1.7}^{+1.4}$}. Figure Set~\ref{fig:HD29021b_RVAST} shows the HIRES and Hamilton RVs, and the corresponding best-fit Keplerian models and the residuals. Two families of orbits are possible; current RV data are insufficient to completely constrain the planet's semi-major axis and eccentricity.

HD~87883 was assessed by \citet{Fischer_2009} prior to the 2007 publication of the HIRES/Keck RVs. Their eccentricity was poorly constrained ($e\ge$0.4) due to the incomplete coverage of the orbital phase approaching periastron. Using the extra RV data, we find consistent orbital parameters with \citet{Fischer_2009} and \citet{Stassun_2017} except for $M\sin i$ where we find a higher value. HD~87883's orbit is face-on and eccentric from our analysis. Our most likely orbit yields \changes{$M = 6.31_{-0.32}^{+0.31}\,\Mjup$, a=$3.77_{-0.094}^{+0.12}$ AU, P=$8.23_{-0.34}^{+0.32}$ years, and $e=0.720_{-0.027}^{+0.038}$.} Still, further RV monitoring of target will be required to fully constrain its orbit. 

\subsection{HD~98649~b}
The posterior distributions for HD~98649~b are depicted in Figure Set~\ref{fig:HD29021b_corner}, and the fits to the RVs and astrometric accelerations are shown in Figure Set~\ref{fig:HD29021b_RVAST}. Our solution shows that HD~98649~b is a highly eccentric and massive planet with an eccentricity of $0.852_{-0.022}^{+0.033}$ and a true planetary mass of $9.7_{-1.9}^{+2.3}\,\Mjup$. There are two possible inclinations: either $136.\!\!^\circ3_{-13}^{+8.1}$ or $43.\!\!^\circ7_{-8.1}^{+13}$. The orbit of HD~98649~b was studied in \citet{Rickman_2019} and \citet{Marmier_2013}. Both studies agree on a $M\sin i$ value of around $6.8\pm0.5\,\Mjup$, and an eccentricity $e=0.85\pm0.05$ but digress on the semi-major axis and the period of the system. We obtain a value of $a=5.97^{+0.24}_{-0.21}$ AU, which agrees with a value of $a=5.6\pm0.4$ AU found by \citet{Marmier_2013}, and marginally with the value of $6.57_{-0.23}^{+0.31}$ AU derived by \citet{Rickman_2019}. 
\subsection{HD~106252~b}
The posterior probabilities for HD~106252~b are illustrated in Figure Set~\ref{fig:HD29021b_corner}. The posterior probabilities follow nearly Gaussian distributions, except for the orbital inclination. The inclination is slightly bimodal, depending on whether the companion is in retrograde or prograde orbital motion. Figure Set~\ref{fig:HD29021b_RVAST} demonstrates the agreement between the calibrated \Hipparcos-\Gaia proper motions from the HGCA. As shown in Figure Set~\ref{fig:HD29021b_RVAST}, the RV orbit of HD~106252~b is well-constrained thanks to full orbital phase coverage from 12 years of RV data. We obtain relatively tight constraints on the dynamical mass of the system, with \changes{$M = 10.0_{-0.73}^{+0.78}\,\Mjup$}, and an orbital inclination of $46.\!\!^\circ 0^{+4.9}_{-4.1}$ (or $134.\!\!^\circ 0^{-4.9}_{+4.1}$). Our derived $M\sin i$ of $7.20\pm0.13\,\Mjup$ corroborates the dynamical analyses of \citet{Wittenmyer_2009}, \citet{Stassun_2017} and \citet{Liu_2014} within 5$\%$. Other orbital parameters such as semi-major axis, period, and eccentricity are also in perfect agreement with results obtained by these authors.

\subsection{HD~106515~Ab} 
The orbits of both HD~106515~Ab and HD~106515~B in the three-body system HD~106515~Ab are fully constrained (see Figure Set~\ref{fig:HD29021b_RVAST}). The induced astrometric acceleration by HD~106515~Ab on HD~106515~A is the most significant in our list. The posterior distributions for HD~106515~Ab and HD~106515~B are shown in Figure Set~\ref{fig:HD29021b_corner}. Our 3-body fit reveals a brown dwarf companion to HD~106515~A with a precise dynamical mass of \changes{$18.9_{-1.4}^{+1.5}\,\Mjup$ on a $\approx$10-year period. Our $M\sin i$ value of $9.25\pm0.25\,\Mjup$} agrees with the $9.08\pm0.14\,\Mjup$ from \citet{Saffe_2019}'s analysis within 2$\%$. All other fitted parameters are also in agreement with \citet{Saffe_2019} and \citet{Marmier_2013}. 

Our three-body fit yields a semi-major axis of \changes{$335_{-42}^{+96}$ AU} for the HD~106515~A/B binary. This is consistent with a 201~AU projected separation given in \gaia EDR3 (see Table~\ref{tab:relast_secondary}) and a 329 AU semi-major axis value provided in \citet{Marmier_2013} based on the positions given by \citet{Gould_2004}. \citet{rica_2017} also found a semi-major axis of 345 AU and an eccentricity of about 0.42 that agrees with ours.

\subsection{HD~171238~b}
Our best-fit orbit for HD~171238~b reveals a dynamical mass of $8.8_{-1.3}^{+3.6}\,\Mjup$, an $M\sin i = 2.599_{-0.088}^{+0.090}\,\Mjup$, a semi-major axis of $a=2.519_{-0.033}^{+0.032}$ AU and an eccentricity $e=0.358_{-0.026}^{+0.028}$, all with tight $1\sigma$ errors. 
The MCMC posterior (Figure Set~\ref{fig:HD29021b_corner}) for the mass of HD~171238~b shows a secondary peak around 15\,$\Mjup$. The RV orbit presented in Figure Set~\ref{fig:HD29021b_RVAST} is constrained by three RV instruments. The inclination ($17.\!\!^{\circ}1_{-5.0}^{+3.1}$ or $162.\!\!^{\circ}9_{-3.1}^{+5.0}$) from our solution indicates a relatively face-on orbit for HD~171238~b.
The $M\sin i$, eccentricity and semi-major axis are more consistent with the orbit analysis of \citet{Segransan_2010} ($M\sin i = 2.60\pm0.15 \Mjup$, $a=2.54\pm0.06$ AU and $e=0.400^{+0.061}_{-0.065}$) than that of \citet{Ment_2018}. HD~171238~b's mass is bimodal, possibly either $\approx$9$\Mjup$ or $\approx$15$\Mjup$. Future \Gaia data releases will confirm the planet's orbit and mass. 

\subsection{HD~196067~b}
HD~196067~b is in a 3-body system with HD~196067 and HD~196068, HD~196067 being the host star and HD~196068 being the stellar companion to HD~196067. The wide orbit stellar companion HD~196068 does accelerate HD~196067, though the $>$1000\,AU separation results in a minimal contribution to the astrometric acceleration from HD~196068. Our MCMC posteriors from the 3-body fit with $\orvara$ are shown in Figure Set~\ref{fig:HD29021b_corner} for the inner and outer companions, respectively. 

The RV orbit of HD~196067~b (see Figure Set~\ref{fig:HD29021b_RVAST}) is not perfectly constrained. CORALIE-98 data did not sufficiently sample the sharp turnaround region near periastron. This lack of data results in a bimodal distribution for the eccentricity in the MCMC posteriors, one near $e=0.6$ and the other near $e=0.85$ (see Figure Set~\ref{fig:HD29021b_corner}). For HD~196067~b, our 3-body solution favors a dynamical mass of \changes{$12.5_{-1.8}^{+2.5}\,\Mjup$, a semi-major axis of $5.10_{-0.17}^{+0.22}$, and an orbital period of $9.88_{-0.43}^{+0.63}$ years.} The orbital period is consistent with the minimum period of 9.5 years found by \citet{Marmier_2013} who have studied the system using the same data as us. They also found an eccentricity of $e=0.66_{-0.09}^{+0.18}$, which is more consistent with the lower peak (e$\approx 0.60$) than the higher one in our bi-modal eccentricity posteriors. The inclination distribution is again, bi-modal: \changes{either $41\,\,\!\!^{\circ}2\,_{-9.1}^{+28}$ or $138\,\,\!\!^{\circ}8\,_{-28}^{+9.1}$.} We find that its true mass was underestimated by RV-only works. Our estimate of its true dynamical mass of $12.8_{-1.8}^{+2.6}\,\Mjup$ is nearly twice as much as the minimum mass found by \citet{Marmier_2013}. HD~196067~b is interesting as it lies extremely close to the deuterium burning limit \citep{Spiegel_2011} that divides planets and brown dwarfs. We expect that further RV data near the periastron passage of the companion will enable us to finalize its full orbit and confirm the dynamical nature of this companion. 

The posterior distributions for HD~196068 are shown in Figure Set~\ref{fig:HD29021b_corner}. The inclination of the stellar companion HD~196068, instead of being bi-modal, has a single value of \changes{$10.\!\!^{\circ}6_{-5.2}^{+6.1}$}, suggesting a nearly face-on orbit. This is because the single relative astrometric measurement from \gaia at epoch 2016.0 described in Section~\ref{sec:singleRelAst} successfully differentiated prograde from retrograde orbits of HD~196068. Our 3-body solution leads to a projected binary semi-major axis of $1631_{-213}^{+208}$ AU, significantly higher than a value of 932 AU in \citep{Marmier_2013}. This can be explained by the stellar companion being near apastron in an eccentric orbit where  $e=0.731_{-0.031}^{+0.026}$. 

\subsection{HD~221420~b}
\label{sec::hd221420_results}

The posterior distributions for selected parameters for HD~221420~b from the joint orbit fit are illustrated in Figure Set~\ref{fig:HD29021b_corner}. The RV orbits and astrometric proper motions are displayed in Figure Set~\ref{fig:HD29021b_RVAST}. The best fit curves agree with both the RV data and proper motions from the HGCA. 

The orbital solution for HD~221420~b was first derived by \citet{Kane_2019} using only the AAT RVs. This orbital solution presents a period of $61.55_{-11.23}^{+11.50}$ years, an eccentricity of $0.42_{-0.07}^{+0.05}$, a semi-major axis of $18.5_{-2.3}^{+2.3}$ AU, and a minimum mass of $9.7_{-1.0}^{+1.1}\Mjup$. 
This minimum mass is $36.5\%$ discrepant with ours. This companion has recently been revisited by \citet{Venner_2021} who use AAT and HARPS RVs, and Hipparcos-Gaia (DR2) astrometry to constrain the precise dynamical mass. They find an orbital inclination of $164.\!\! ^{\circ}0_{-2.6}^{+1.9}$, a precise dynamical mass of $22.9\pm2.2\,\Mjup$, a semi-major axis of $10.15_{-0.38}^{+0.59}$ AU, and a period of $27.62_{-1.54}^{+2.45}$ years. Using the AAT and HARPS RVs, and absolute astrometry from \Hipparcos and \Gaia EDR3, we obtain a companion mass of \changes{$20.6_{-1.6}^{+2.0}\Mjup$, an eccentricity of $0.162_{-0.030}^{+0.035}$, a semi-major axis of $9.99_{-0.70}^{+0.74}\,AU$, and a period of $27.7_{-2.5}^{+3.0}$.} Our EDR3 solution agrees well with that of \citet{Venner_2021}, with one distinction being our inclination is again bimodal with equal maximum likelihood: \changes{i = $17. \!\! 8^{\circ}5_{-2.8}^{+2.9}$ or $162.\!\! ^{\circ}2_{-2.9}^{+2.8}$}. We validate the findings by \citet{Venner_2021} that the companion is near the $25\Mjup$ upper limit for core accretion and disk instability to be considered plausible formation channels. \citet{Venner_2021} also identify a M-dwarf stellar companion HD~221420~B that may be bound to HD~221420~A, but at a separation of more than 20,000~AU, it contributes negligibly to HD~221420's acceleration. HD~221420~b has the longest period and highest companion mass among the RV companions we sample. As a result, it is also the most accessible of our substellar companions for future direct imaging. 

Direct imaging can probe the outer architecture of a system \citep{Chauvin_2005,Lafreniere_2008,Ireland_2010,Rameau_2013,Lagrange_2014,Bowler_2016}. \subtractions{Bowler et al 2017} However, only about a dozen wide-orbit giant planets have been discovered at separations of $\sim$10-150 AU of their host stars \citep[e.g.][]{Chauvin_2018,Marois_2008}. These surveys have mostly been blind. HD~221420b represents a rare case of a substellar companion whose mass and position we can determine before imaging.

\begin{figure*}[htp]
\label{fig:HD221420b_AstPredict}
\centering
    \includegraphics[width=0.4\textwidth]{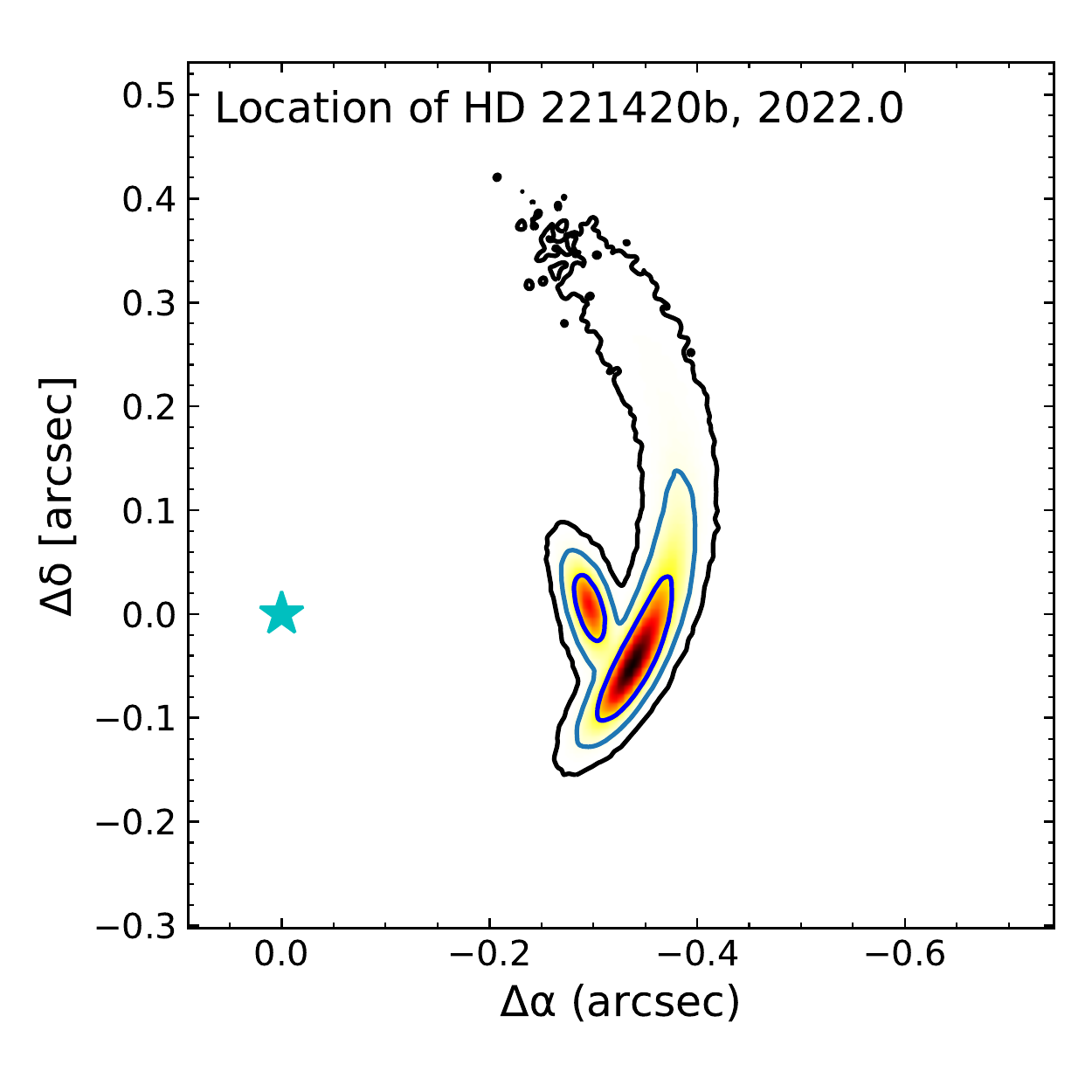}\quad
    \includegraphics[width=0.41\textwidth]{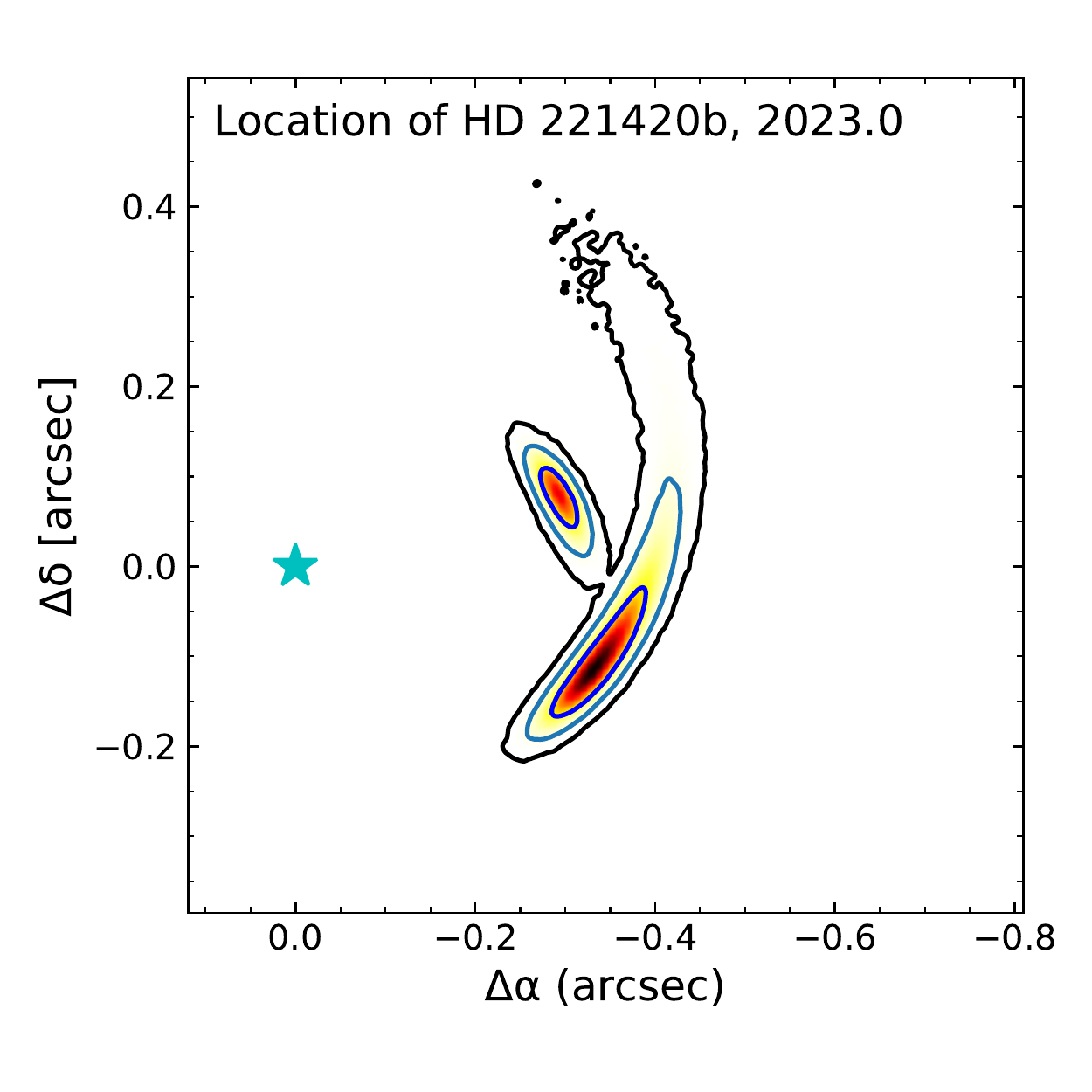}\quad
    \includegraphics[width=0.4\textwidth]{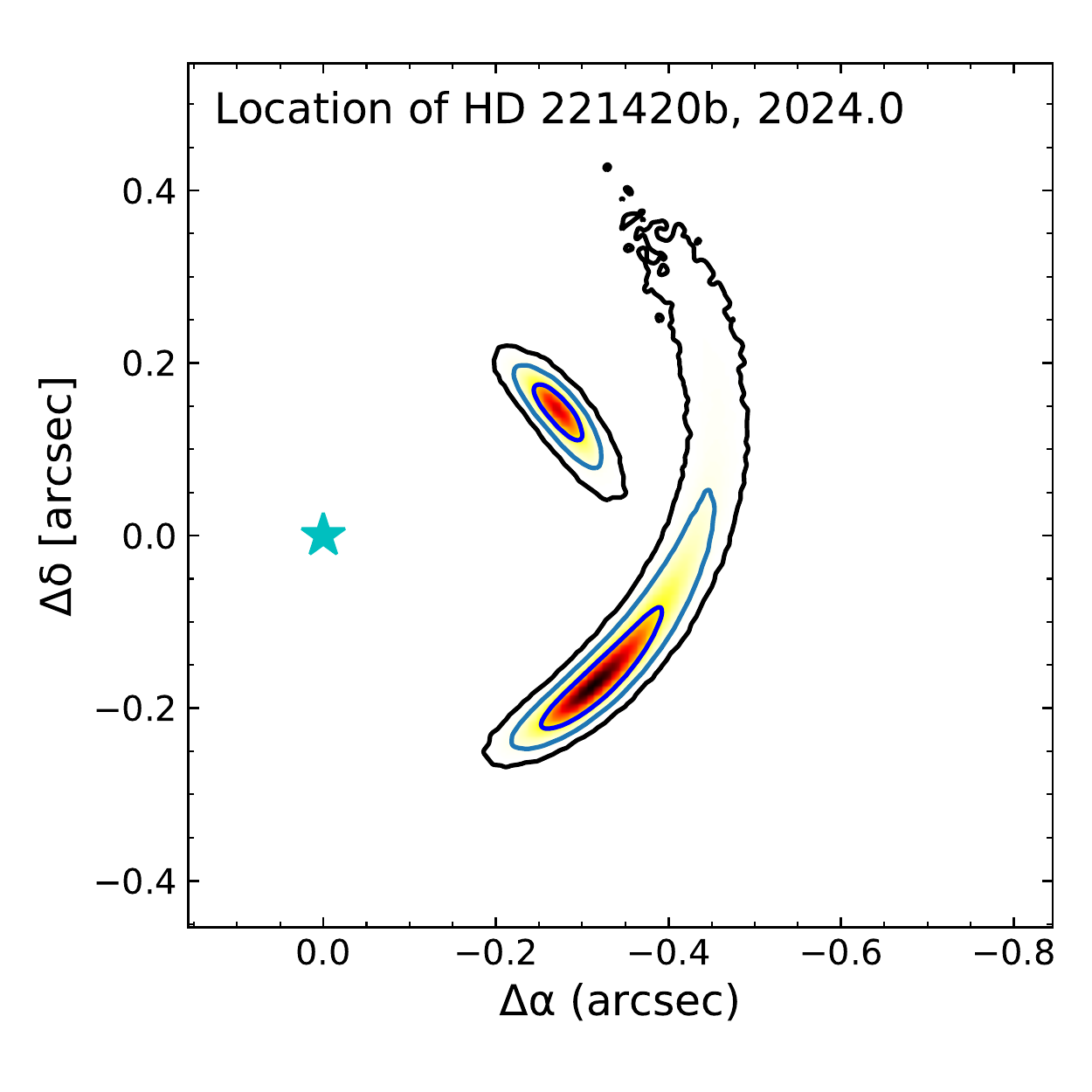}\quad
    \includegraphics[width=0.4\textwidth]{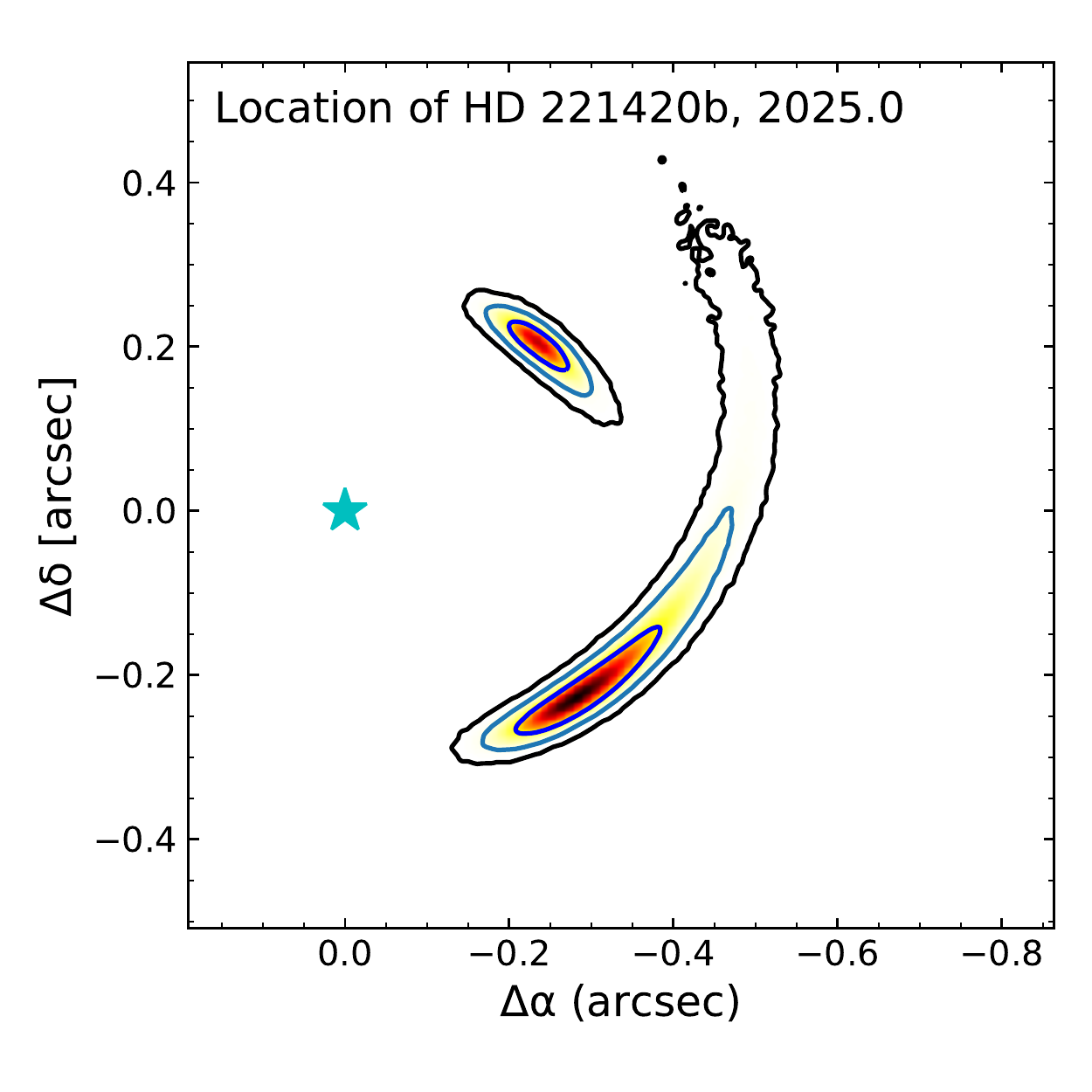}
    \caption{Predicted positions of HD~221420~b in epoch 2022, 2023, 2024, and 2025. The light blue stars denote the host star HD~221420. The contours indicate the predicted relative coordinates of the brown dwarf companion HD~221420~b with respect to the location of the host star. The contour lines enclose the $1\sigma$, $2\sigma$ and $3\sigma$ probabilities with descending normalized likelihood from inner to outer contours (black to red to light yellow). Two possible sets of contours correspond to prograde or retrograde orbits. If the companion can be resolved via high-contrast imaging, it will eventually tell the two orbital motions apart.
\label{fig:HIP116250_predictions}}
\end{figure*}

We use our orbital fit to predict the position of HD~221420~b at future epochs. In Fig.~\ref{fig:HIP116250_predictions}, we show the predicted locations of HD~221420~b at epochs 2022.0 and 2023.0, 2024.0, and 2025.0.  Despite the fact that HD~221420~b has never been imaged, our 68\% confidence interval places it within a box of about $100 \times 200$\,mas, almost sufficient to locate a fiber for the GRAVITY interferometer \citep{GRAVITY_2017}.  For the next couple of years the brown dwarf will be offset about $0.\!\!^{\prime\prime}4$ west of its host star in a slow orbit.  Unfortunately, we lack the data to establish whether this orbit is clockwise or counter-clockwise on the sky.

HD~221420~b is an unusually promising accelerating system for high-contrast imaging follow-up. It consists of a relatively low-mass brown dwarf orbiting a bright ($V=5.8$) and nearby ($d=31.2$~pc) host star.  \addition{Depending on the system age (Section \ref{subsec:results_stars}), the ATMO2020 evolutionary models \citep{Phillips+Tremblin+Baraffe+etal_2020} predict an effective temperature of 400-600\,K for HD~221420~b.  These low temperatures correspond to a late-T or early-Y spectral type.  The $H$-band contrast predicted from ATMO2020 ranges from 16 to 20 magnitudes depending on the brown dwarf's mass and the system age, while the predicted $L'$ contrast ranges from 13 to 15 magnitudes.  These exceed the typical performance of SPHERE \citep{Beuzit_2008} but may be achievable with long integrations and, as Figure \ref{fig:HIP116250_predictions} shows, with the knowledge of where to look.}  \subtractions{While the old age of its host star makes the brown dwarf faint, it may be detectable by SPHERE on the VLT} 

\clearpage

\section{Discussion and Conclusion} \label{sec:discussion}

\begin{figure*}
    \centering
    \includegraphics[width=0.8\linewidth]{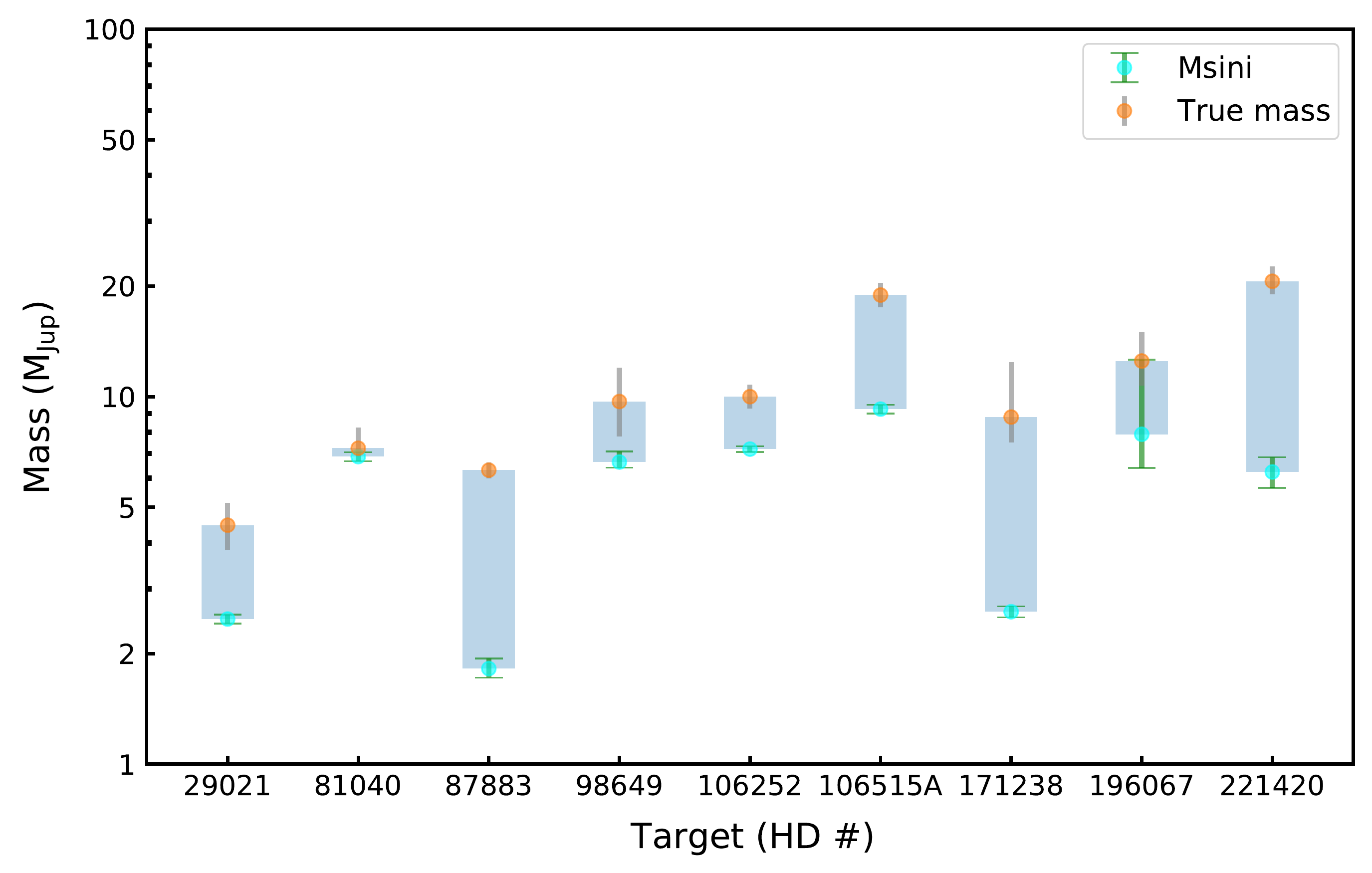}
    \caption{A summary of the differences in the final mass and Msini estimations for the nine RV companions studied in this work. The values for each companion are listed in Table~\ref{tab:planet_params}. The orange points and grey error bars are the true dynamical masses derived by this work, and the cyan dots plus green error bars are the Msini values inferred from the orbital parameters in this work. The blue regions highlight how much the true dynamical masses can be underestimated by RV-only fits as a consequence of uncertain inclinations. }
    \label{fig:mass_summary}
\end{figure*}

In this paper, we have refined orbital solutions for nine single RV companions orbiting G and K-type stars using the orbit-fitting package $\orvara$. For every target, we have obtained 1$\sigma$ error bars on the dynamical masses, and we find bi-modal orbital inclinations. Apart from HD~221420~b recently assessed by \citet{Venner_2021}, no past dynamical mass measurements were available for the RV companions we study. This motivates us to compare our $M \sin i$ with previous RV-only $M \sin i$ values. Our $M \sin i$ values agree with the literature reports within 1$\sigma$. \addition{Fig.~\ref{fig:mass_summary} compares the minimum mass and true mass of each companion from our orbital fits. The RV-only minimum companion masses were underestimated due to uncertain inclinations by at least $\approx$ 5 percent and at most $\approx$ 250 percent. The mass increase is greater for more face-on orbits such as that of HD~87883~b, HD~171238~b, and HD~221420~b, and less so for edge-on orbits like that of HD~81040b.}

Several of our companions have periods comparable to the 33-month \gaia EDR3 baseline: HD~29021~b ($3.737_{-0.018}^{+0.018}$ years), HD~81040~b ($2.7452_{-0.0093}^{+0.011}$ years), HD~106252~b ($4.202_{-0.010}^{+0.011}$ years), and HD~171238~b ($4.148_{-0.046}^{+0.045}$ years), with the shortest being HD~81040~b. The other five -- HD~87883~b, HD~98649~b, HD~106515~Ab, HD~196067~b, HD~221420~b -- have long periods $\ge$ 8 years. The Renormalised Unit Weight Error (RUWE) of the short-period planet HD~81040~b (=1.598) is slightly higher than the usual good astrometry solution \gaia recommends of $<$1.4 \citep{GaiaDR2_2018}. Since there is no evidence of its binarity in the literature, the anomaly in RUWE may be attributed to the perturbation effects from HD~81040~b, its $7.24_{-0.37}^{+1.0}\,\Mjup$ planetary companion. Furthermore, given that its orbital period is comparable to the 33-month baseline of \gaia EDR3, it is challenging to constrain its astrometric orbital motion without the \gaia intermediate astrometric data. In their absence, all epoch astrometry data are forward-modeled by \htof. \citet{Mirek_2020} has tested the fidelity of \htof\ on data integration against the $\tt REBOUND$ code for a 3-body \orvara\ fit to the short period planet $\beta$ Pictoris~c, and found that \htof\ and \orvara\ recover the $\tt REBOUND$ fit. 
Future \gaia data releases will include non-single star fits and epoch astrometry, enabling precise orbital fits to these shorter-period systems.

Interestingly, once masses and inclinations are separately constrained, the true dynamical masses of some of these companions are revealed to be much higher than the literature RV-only minimum masses. HD~196067~b, with a true mass of \changes{$12.5_{-1.8}^{+2.5}\,\Mjup$}, lies on the transition between giant planets and brown dwarfs according to the $13\,\Mjup$ upper mass limit for the ignition of deuterium \citet{Boss_2008,Spiegel_2011}. HD~106515~Ab and HD~221420~b are rare long period and low-mass brown dwarfs ($\approx 20\,\Mjup$) according to this deuterium-burning mass limit. HD~87883~b, HD~171238~b, and HD~221420~b have almost face-on orbits and, as a result, high masses. This places them in the population suggested by \cite{Schlaufman_2018} to be uncorrelated with host star metallicity, though spectroscopic measurements find each of these stars to be of super-Solar metallicity. Our finding of several companions with relatively face-on inclinations may be a selection effect: we have fit stars with signficant astrometric accelerations in the HGCA. Many of our inclinations have bimodal posteriors; these will be resolved with future \gaia data releases. 

Eccentricities of planets provide insights into their dynamical history and formation mechanisms. \citet{Bowler_2020} conducted a population-level study of the eccentricity distributions of directly imaged companions. They found that the eccentricity distribution peaks around $e=0.23$ for single long-period giant planets, around $e=0.5$ for brown dwarf on closer orbits (5-30 AU), and around $e=0.6$ for low-mass companions within 10 AU.

Three of our wide orbit companions---HD~87883~b, HD~98649~b, and HD~196067~b---have high eccentricities $>$0.7. HD~81040~b and HD~106515~Ab also have moderately high eccentricities of between 0.5 and 0.6. HD~221420~b has a surprisingly low eccentricity of \changes{$0.162_{-0.030}^{+0.035}$}, much lower than the mean eccentricity of $e=0.5$ found by \citet{Bowler_2020} for short-period brown dwarfs. HD~98649~b, with $e=0.822_{-0.024}^{+0.030}$ is the most eccentric planet known with a period longer than 600 days and one of the more eccentric planetary orbits ever discovered. 

HD~106515~Ab and HD~196067~b have outer stellar companions, raising the possibility that Kozai-Lidov \citep{Kozai_1962,Lidov_1962} oscillations brought them onto highly eccentric orbits \citep{Marmier_2013}. 
For HD~106515~Ab and HD~196067~b, the relative inclination angles between the orbital planes of the planet and the stellar companion are \changes{$10.\!\! ^{\circ}4^{+8.2}_{-8.6}$ and $30.\!\! ^{\circ}6^{+28.7}_{-10.5}$}, respectively. The Lidov–Kozai interactions \citet{Kozai_1962,Lidov_1962} require a relative inclination angle of $39^{\circ}$ in order to be invoked. This does not apply to HD~106515~Ab, but does apply to HD~196067~b whose relative inclination is subject to large uncertainties. The high orbital eccentricities of HD~87883~b and HD~98649~b may also be explained by the Kozi mechanism, but there's a lack of evidence of the existence of a third body in their systems. Precise \Hipparcos and \Gaia astrometry over a 25-year baseline limits the parameter space in which additional, outer companions could be hiding. 

HD~29021, HD~81040, and HD~106252 are rare examples of stars with giant planetary companions but with subsolar metallicity. This may support the finding by \citet{Adibekyan_2013} that metal-poor stars host planets with longer periods than metal-rich stars. HD~81040 and HD~106252 seem to fall in the high-mass category suggested by \cite{Schlaufman_2018} to be uncorrelated with stellar metallicity. Our dynamical masses and orbits, including our substantial corrections to $M \sin i$ for several of our planets, will enable stronger comparisons between the masses of planets around metal-poor and metal-rich stars.

From our RV fits, four of the targets HD~87883~b, HD~171238~b, HD~98649~b and HD~196067~b need at least one more period of RV monitoring in order to completely constrain their orbital solutions. \changes{Our companions have typical separations of $\lesssim 0.2''$ and distances within 50 pc. The high contrasts needed to image these planets are on the order of a few $10^{-7}$ for HD 221420~b in the thermal near-infrared, and slightly worse for HD 106515~Ab.  The remainder would have contrasts of $\sim$10$^{-9}$  in reflected light depending on orbital phase and albedo. The current generation of high contrast imagers like SPHERE for VLT \citep{Beuzit_2008} and GPI for Gemini \citep{Chilcote_2018} may marginally resolve HD 221420~b. Most of our stars are old and their companions are very cold, so the companions are light-reflecting planets/low-mass brown dwarfs with little thermal emission. These companions will provide comparisons to thermal spectra of outer Solar system planets and very old field brown dwarfs. }
\changes{The systems we study are not the most favorable for the Coronagraph Instrument (CGI) on the Nancy Grace Roman Space Telescope \citep{Kasdin_2020}: their $\sim$0.$\!\!''2$ angular separations are too close. However, HD~106515~Ab and HD~221420~b are excellent targets for the Giant Magellan Telescope (GMT), the European Extremely Large Telescope (ESO ELT) , and the Thirty Meter Telescope (TMT).}

\subtractions{Our orbital analysis also identifies HD 221420b as a promising (if challenging) candidate for current high contrast imaging surveys. Both HD 196067b and HD 106515b are also massive companions on wide orbits ($\approx$10 AU), but such separation pushes the limit of hitherto achievable high-contrast imaging. All of the stars are old; their companions will be very cold and will provide comparisons to
thermal spectra of outer Solar system planets and very old field brown dwarfs. Future generations of imaging
instruments like WFIRST (Zellem et al. 2020) might be
able to target these stars already knowing where to find
the planets and how massive they are.}

Absolute astrometry can break the mass-inclination degeneracy inherent in RV work, and can enable the prediction of a planet's location on the sky.  We have derived precise dynamical masses of nine planets in this work.  Many still have an uncertain orientation on the sky, with existing astrometry unable to distinguish prograde from retrograde orbits. This will change conclusively with future \Gaia data releases, as \Gaia achieves the temporal coverage needed to map out orbits on its own. The synergy between RV and absolute astrometry will then enable precise mass and orbit measurements for all of the nearby giant exoplanets.

\clearpage
\appendix

Figures \ref{fig:HD29021b_RVAST}-\ref{fig:HD221420b_corner} show the full results from the {\tt orvara} orbital fits to the 11 companions (nine planets and brown dwarfs, plus two stellar companions) in nine systems.  Figures \ref{fig:HD29021b_RVAST}-\ref{fig:HD221420b_RVAST} show the radial velocity measurements and fitted orbits in the left panels, and the absolute astrometry in the right panels.  Figures \ref{fig:HD29021b_corner}-\ref{fig:HD221420b_corner} show corner plots of the astrophysically interesting parameters: masses, semimajor axis, and eccentricity, along with inclination.

\renewcommand{\thefigure}{A\arabic{figure}}

\setcounter{figure}{0}

\begin{figure*}[htp]
\centering
    \includegraphics[width=\textwidth]{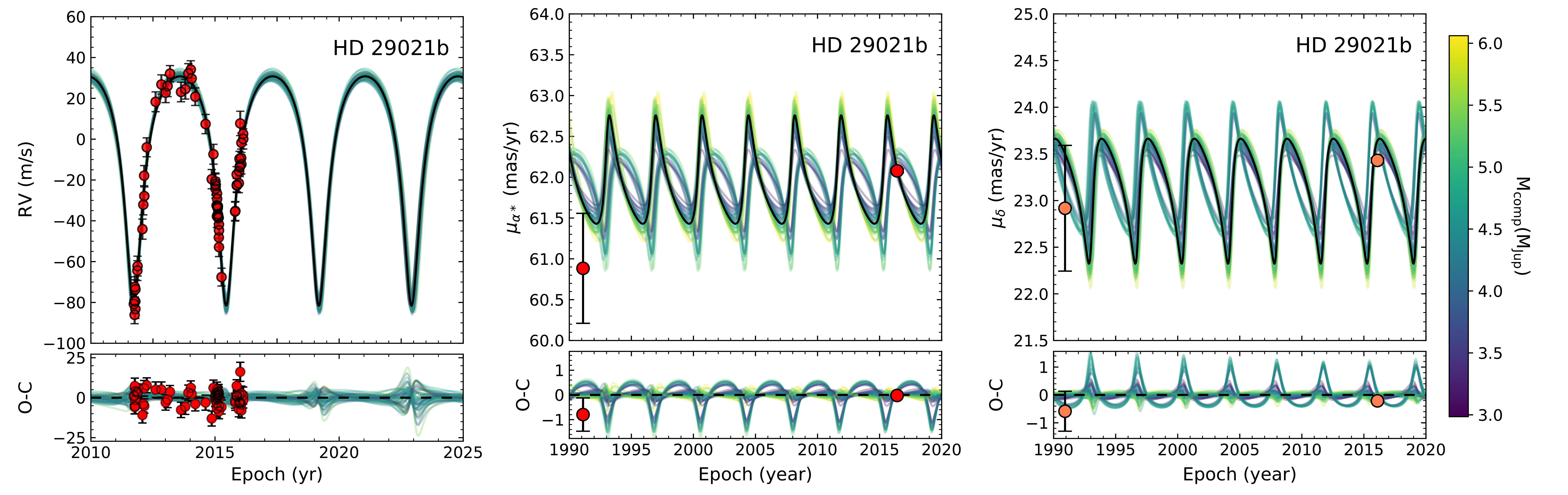}
    \caption{Left panel: Radial Velocity orbit for HD~29021~b. The RV data are from the OHP 1.93m SOPHIE spectrograph. Middle and right panels: astrometric acceleration in right ascension and declination. The red and orange points are proper motion measurements from \Hipparcos near epoch 1991 and \gaia near epoch 2016. The thick black lines indicate the best-fit orbits in the MCMC chain while the colored lines are solutions randomly drawn from the chain, color-coded by the dynamical mass of the companion.  
\label{fig:HD29021b_RVAST}}
\end{figure*}

\begin{figure*}[htp]
\centering
    \includegraphics[width=\textwidth]{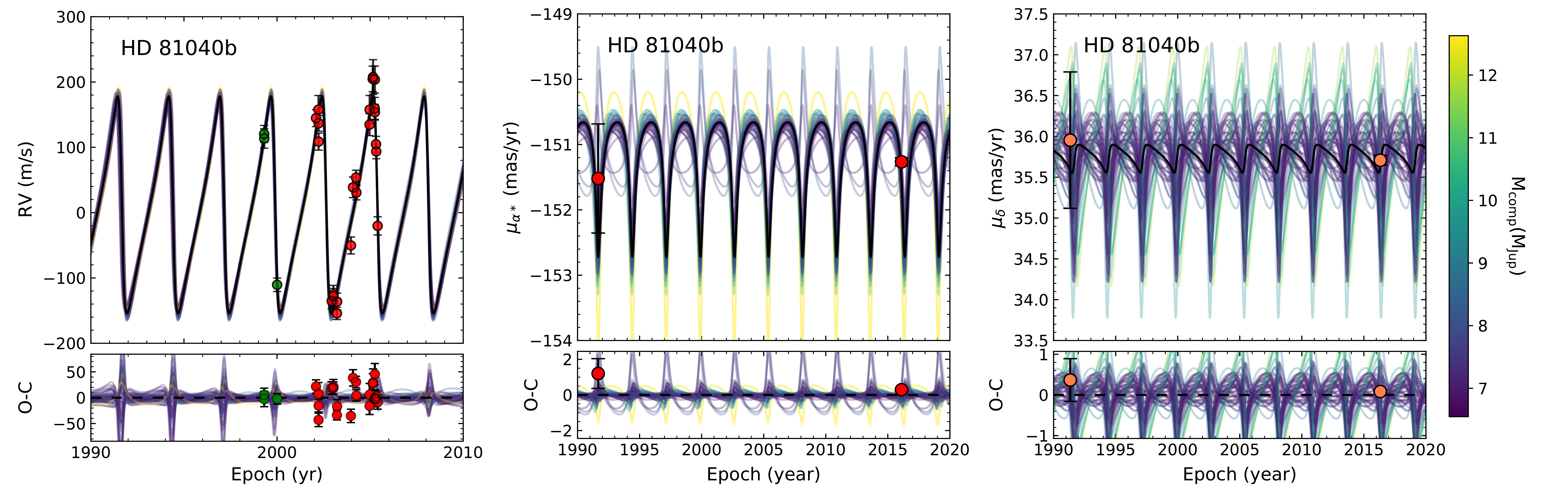}
    \caption{HD~81040~b. Same as Fig.~\ref{fig:HD29021b_RVAST}. Data for RVs come from  HIRES/Keck (green) and ELODIE (red).
\label{fig:HD81040b_RVAST}}
\end{figure*}

\begin{figure*}[htp]
\centering
    \includegraphics[width=\textwidth]{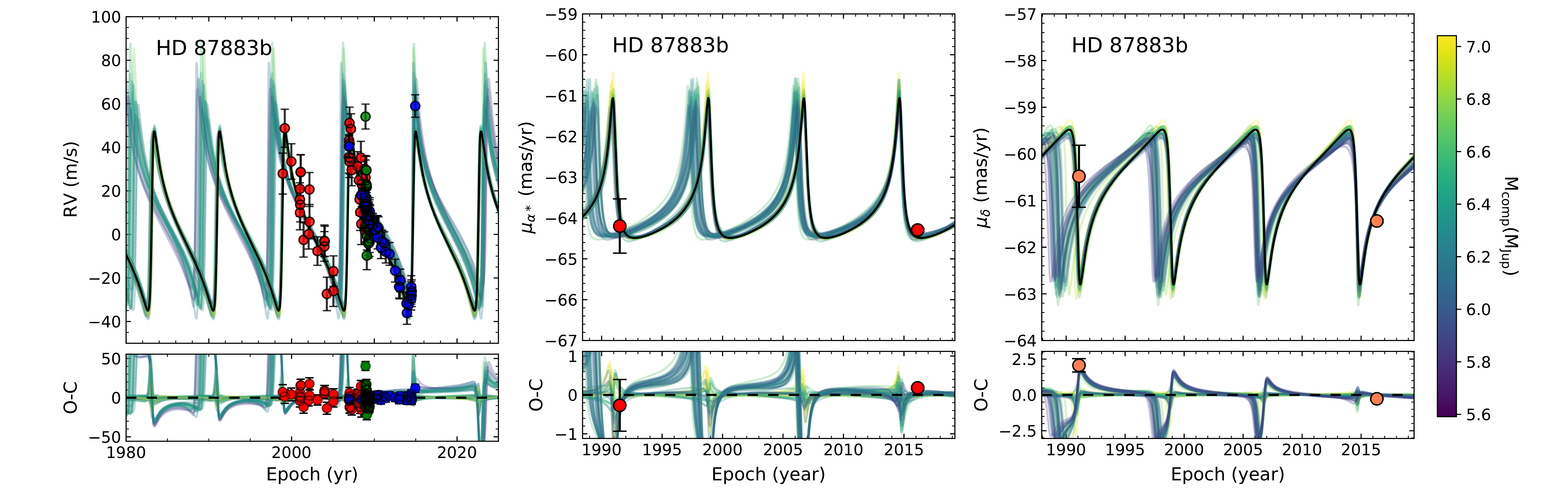}
    \caption{HD~87883~b. Same as Fig.~\ref{fig:HD29021b_RVAST}. Red and blue points with error bars are RVs from Hamilton/Lick published in 2009 and 2017, respectively, and the green points are velocities from HIRES/Keck. 
\label{fig:HD87883b_RVAST}}
\end{figure*}

\begin{figure*}[htp]
\centering
    \includegraphics[width=\textwidth]{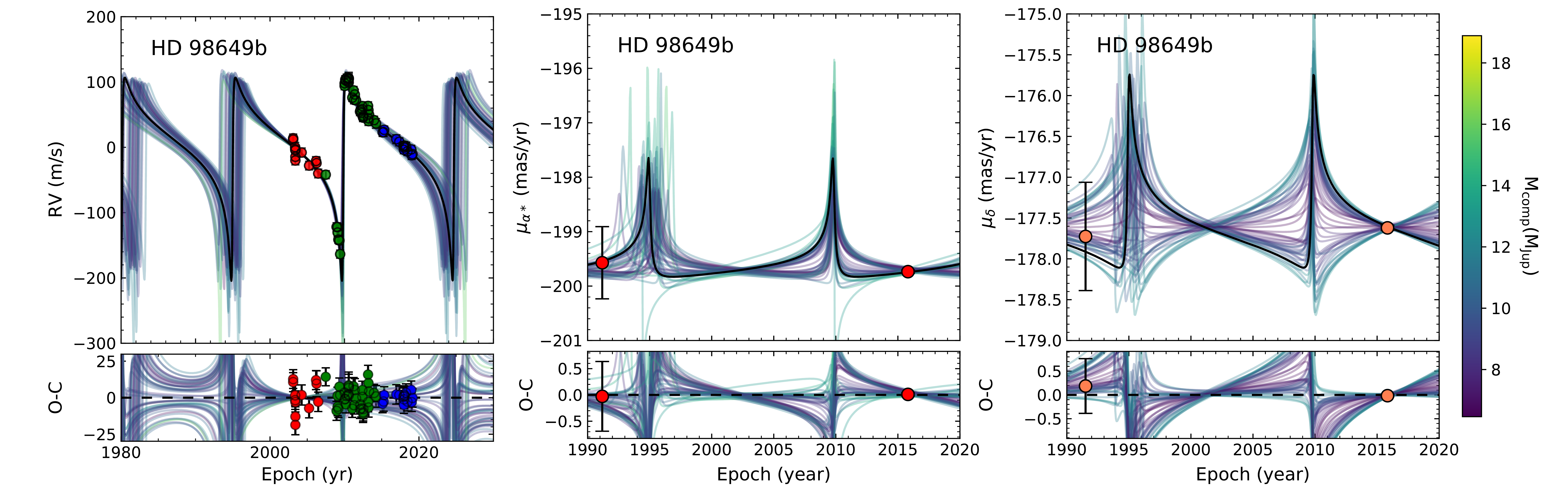}
    \caption{HD~98649~b. Same as Fig.~\ref{fig:HD29021b_RVAST}. The red and green data points in the left panel are RVs from CORALIE-98 and CORALIE-07, respectively.
\label{fig:HD98649b_RVAST}}
\end{figure*}

\begin{figure*}[htp]
\centering
    \includegraphics[width=\textwidth]{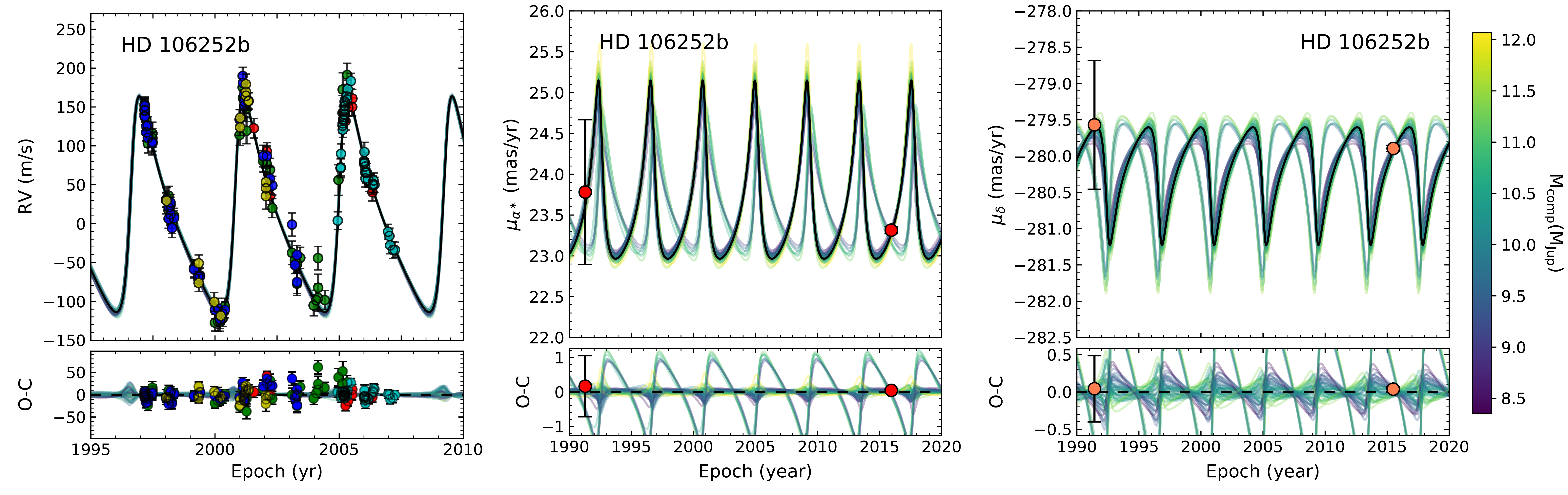}
    \caption{HD~106252~b. Same as Fig.~\ref{fig:HD29021b_RVAST}. The solid circles and error bars in the left most panel represent RVs from ELODIE (blue), Hamilton/Keck (yellow and green), CDES-TS2 (red), and HRS/HET (cyan). 
\label{fig:HD106252b_RVAST}}
\end{figure*}

\begin{figure*}[htp]
\centering
    \includegraphics[width=\textwidth]{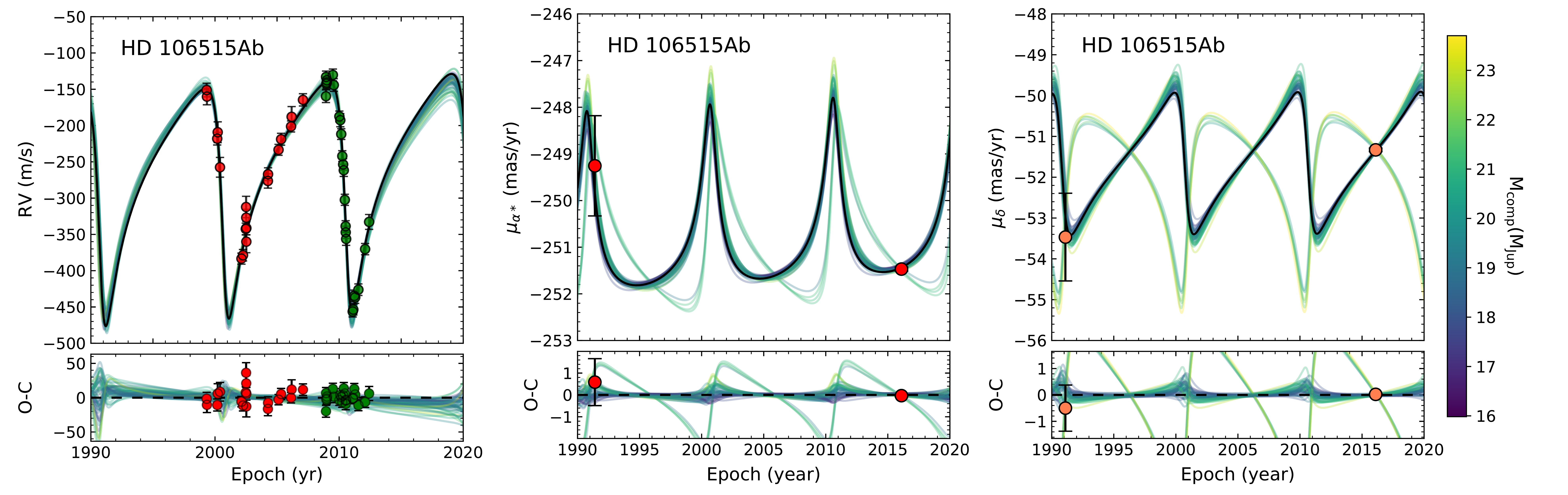}
    \caption{HD~106515~Ab. Same as Fig.~\ref{fig:HD29021b_RVAST}. Red and green RV data show CORALIE-98 and CORALIE-07 measurements, respectively.
\label{fig:HD106515Ab_RVAST}}
\end{figure*}

\begin{figure*}[htp]
\centering
    \includegraphics[width=\textwidth]{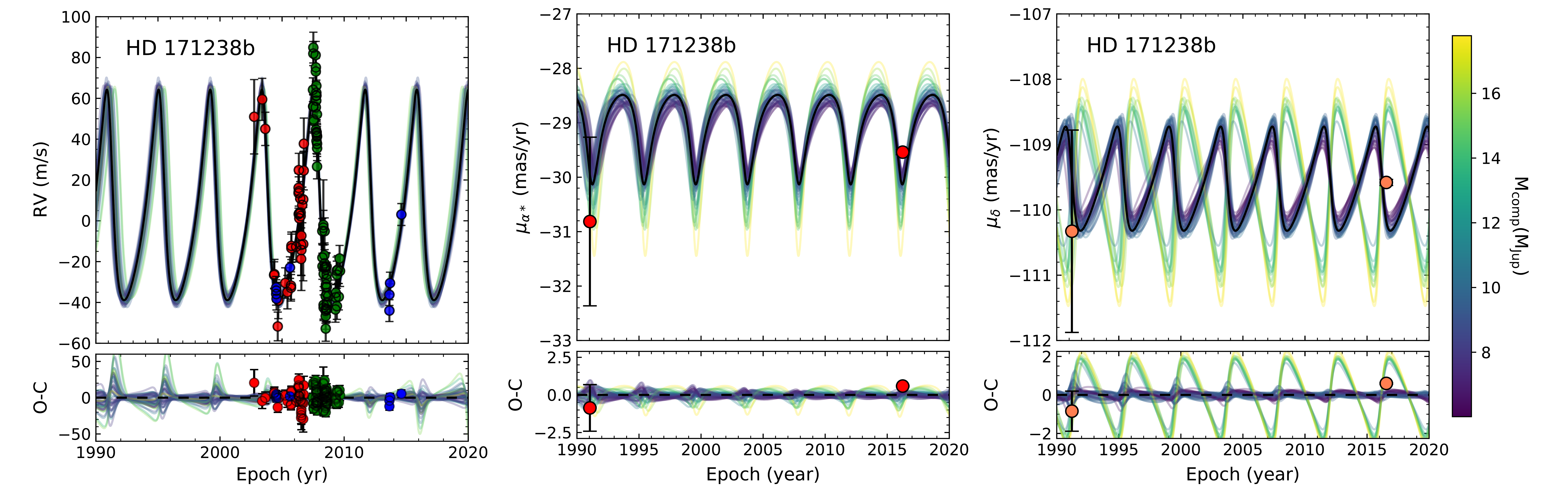}
    \caption{HD~171238~b. Same as Fig.~\ref{fig:HD29021b_RVAST}. In the left panel, the RVs from CORALIE-98 are in red, the velocities from CORALIE-07 are in green, and the blue data points represent velocities from HIRES/Keck published in 2017. 
\label{fig:HD171238b_RVAST}}
\end{figure*}

\begin{figure*}[htp]
\centering
    \includegraphics[width=\textwidth]{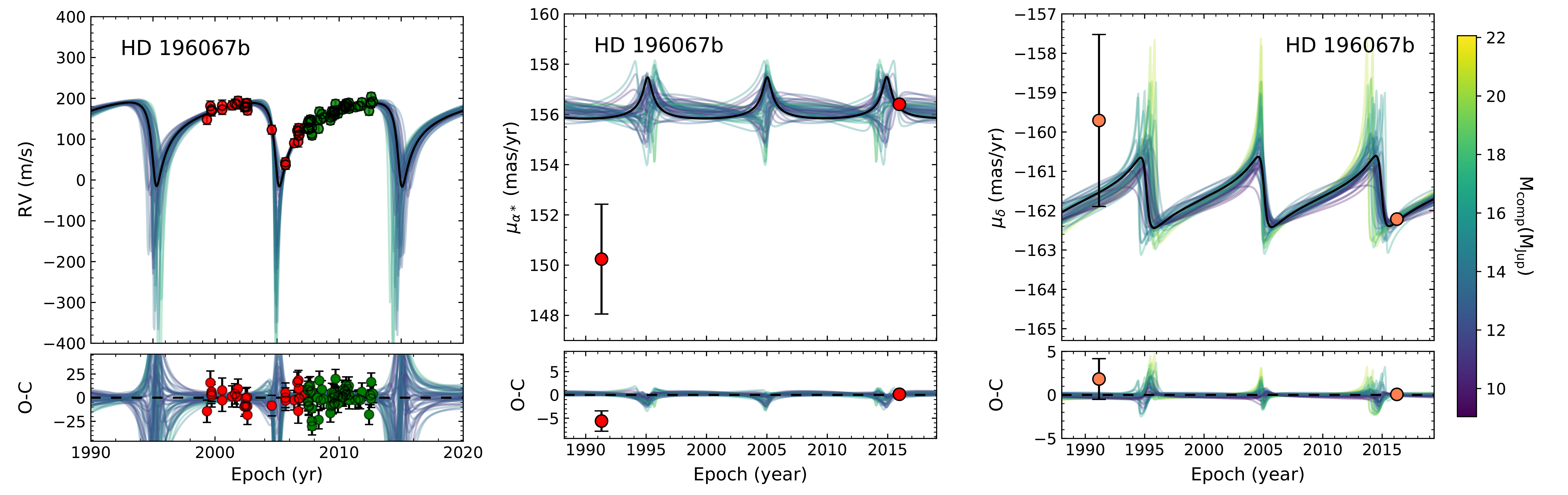}
    \caption{HD~196067~b. Same as Fig.~\ref{fig:HD29021b_RVAST}. The RV measurements are from CORALIE-98 (red) and CORALIE-07 (green).
\label{fig:HD196067b_RVAST}}
\end{figure*}

\begin{figure*}[htp]
\centering
    \includegraphics[width=\textwidth]{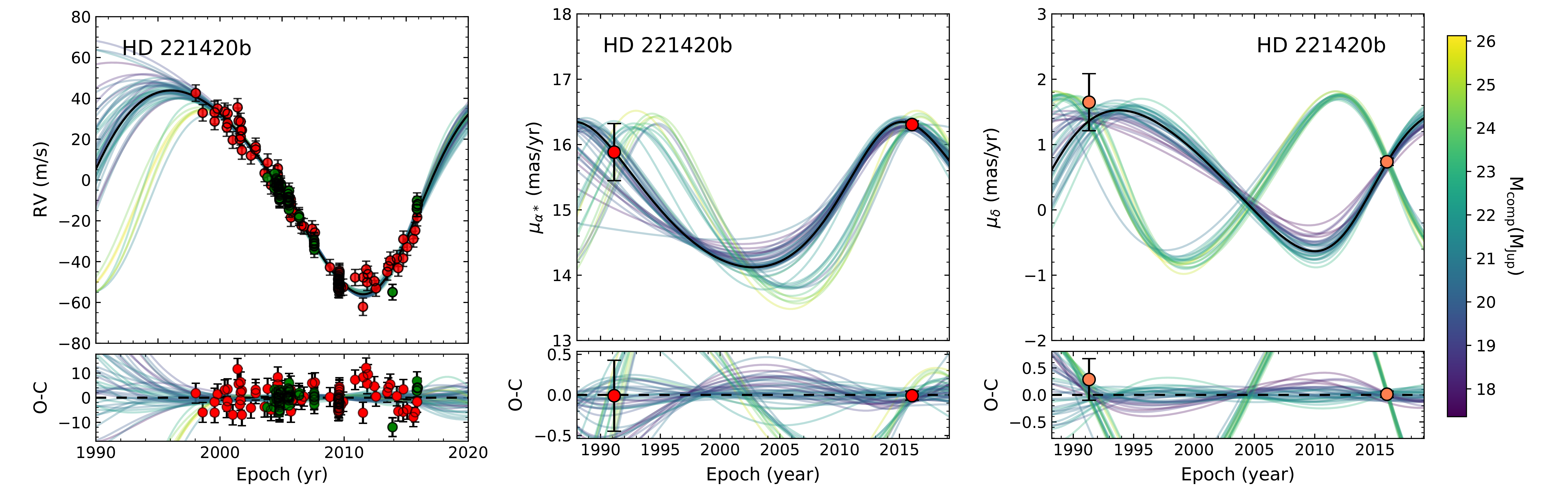}
    \caption{HD~221420~b. Same as Fig.~\ref{fig:HD29021b_RVAST}. In the left panel, the red data points show RVs from AAT, and the HARPS RVs are showcased by green data points. 
\label{fig:HD221420b_RVAST}}
\end{figure*}

\begin{figure*}
    \centering
    \includegraphics[width=\textwidth]{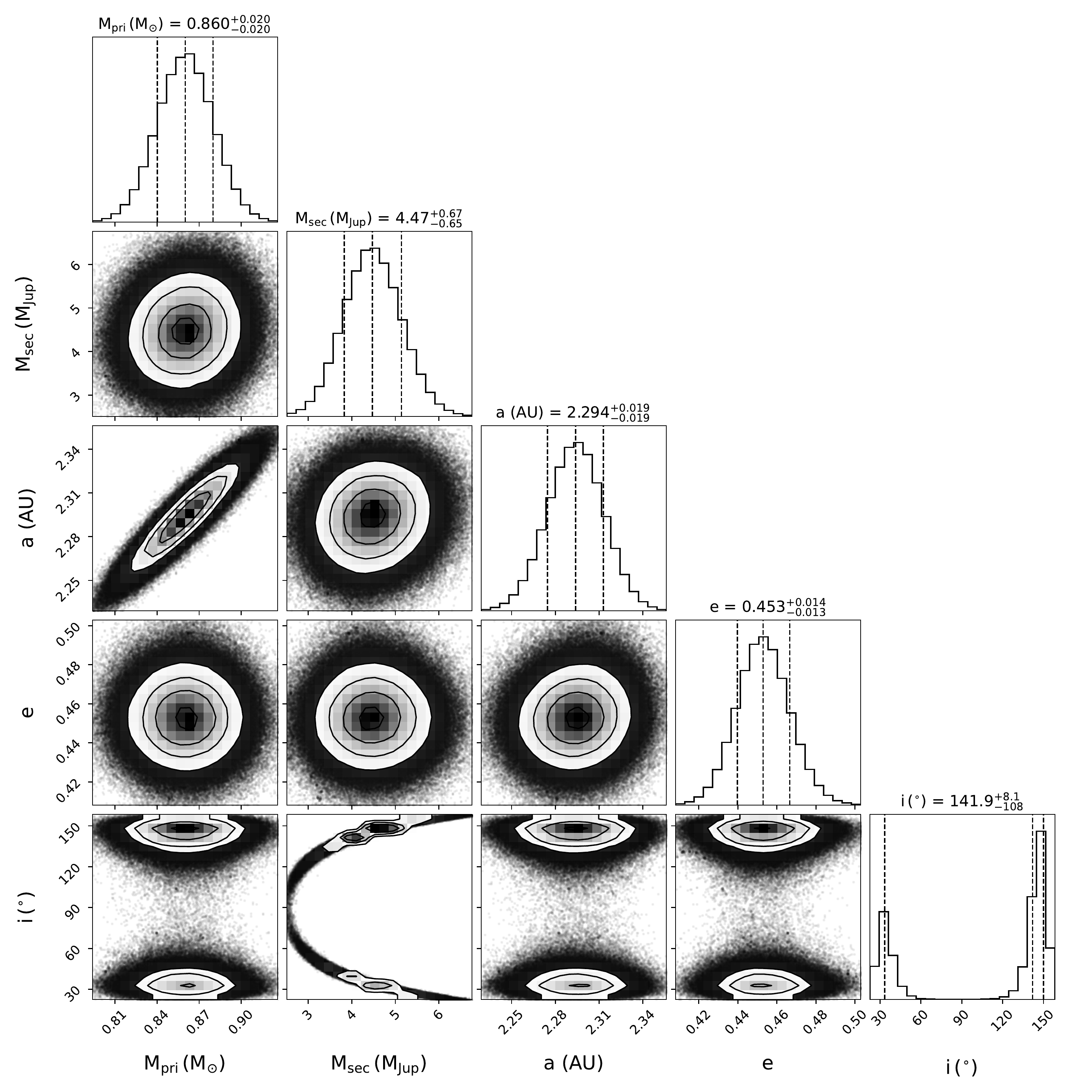}
    \caption{Joint posterior distributions for selected orbital parameters of HD~29021~b. These are the host star's mass ($M_{pri}$), the companion mass ($M_{sec}$), the semi-major axis a, the orbital eccentricity e, and the orbital inclination i. The values and histogram distributions of the posteriors of selected parameters are shown, along with 1 $\sigma$ uncertainties.
\label{fig:HD29021b_corner}}
\end{figure*}

\begin{figure*}
    \centering
    \includegraphics[width=\textwidth]{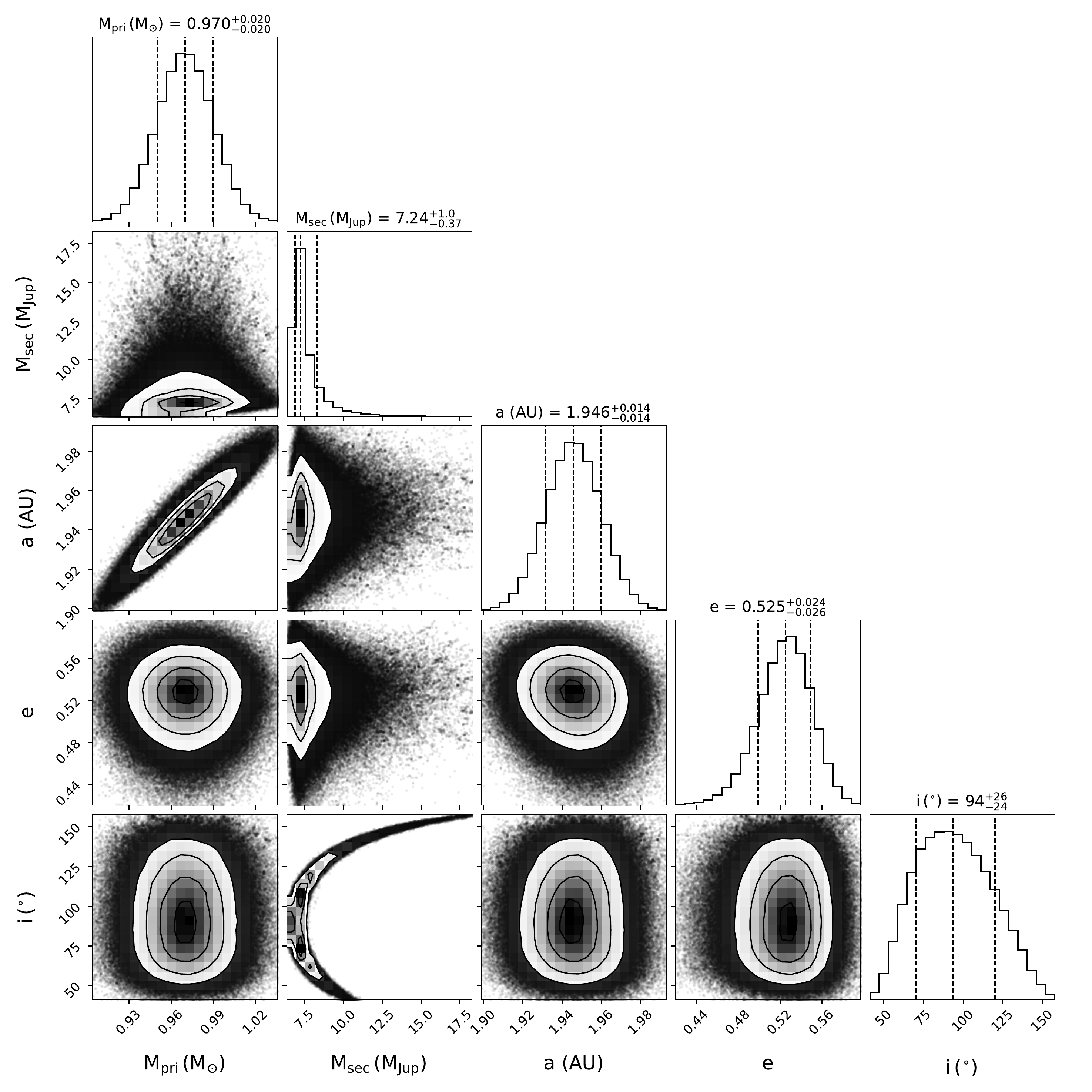}
    \caption{HD~81040~b. Same as Fig.~\ref{fig:HD29021b_corner}.
\label{fig:HD81040b_corner}}
\end{figure*}

\begin{figure*}
    \centering
    \includegraphics[width=\textwidth]{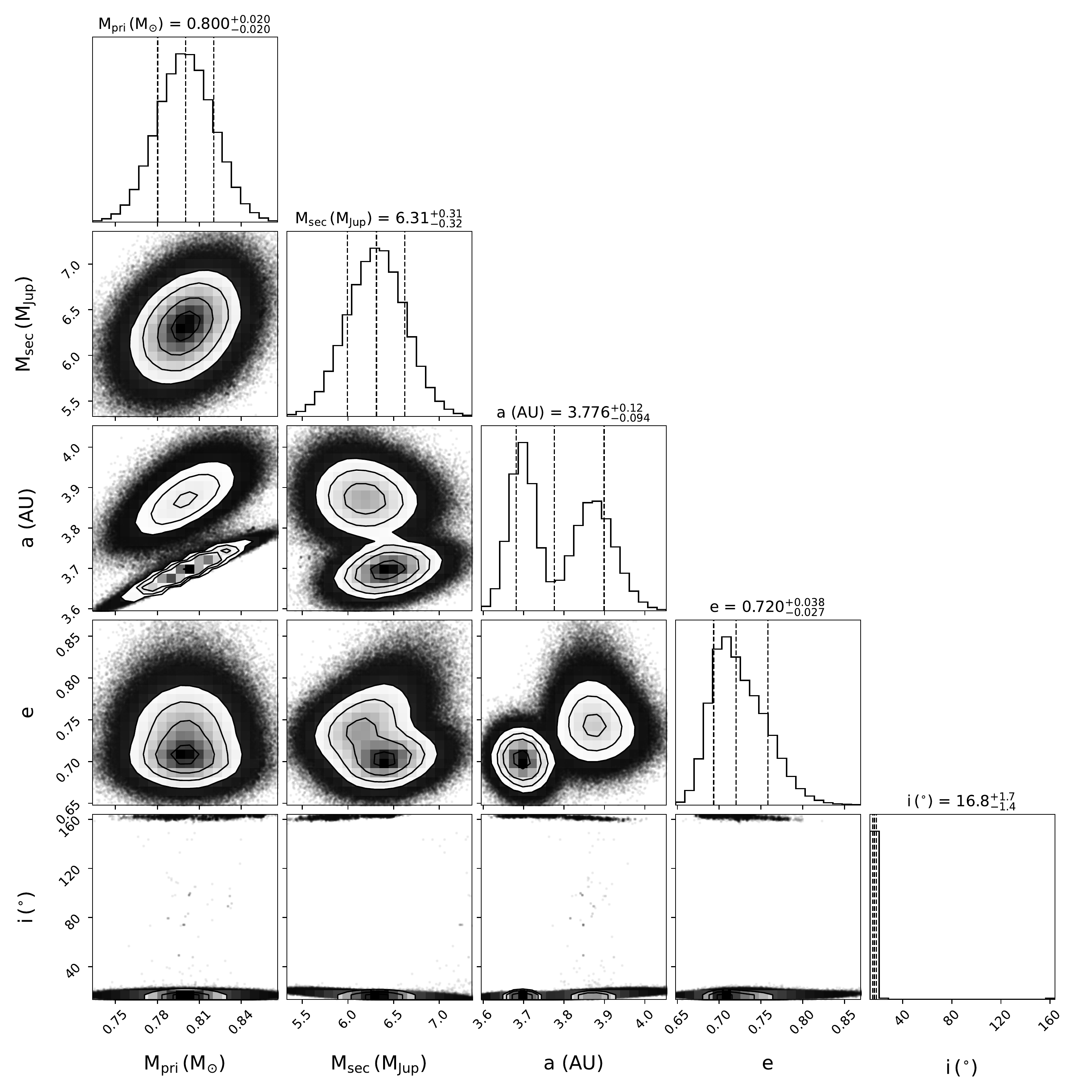}
    \caption{HD~87883~b. Same as Fig.~\ref{fig:HD29021b_corner}.
\label{fig:HD87883b_corner}}
\end{figure*}

\begin{figure*}
    \centering
    \includegraphics[width=\textwidth]{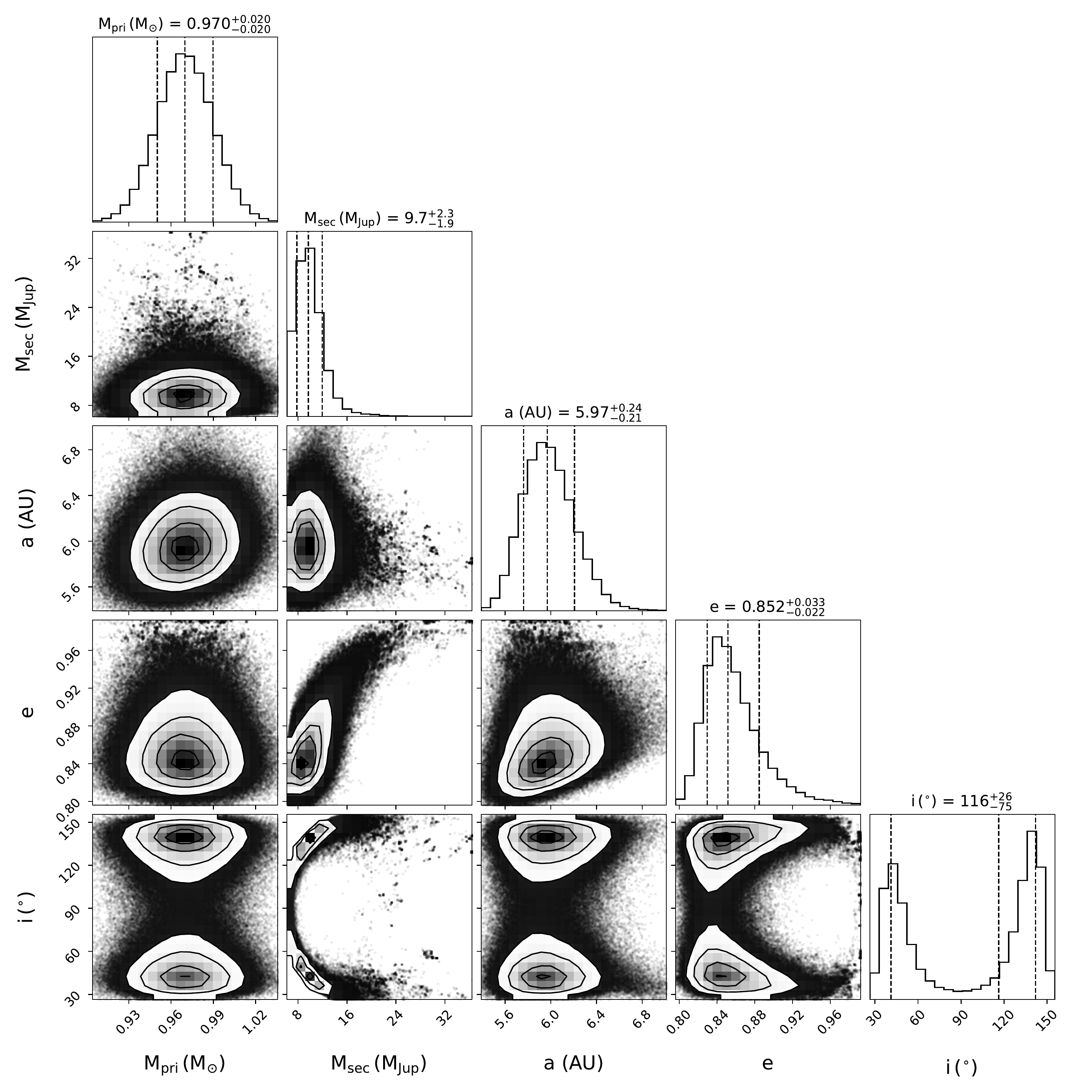}
    \caption{HD~98649~b. Same as Fig.~\ref{fig:HD29021b_corner}.
\label{fig:HD98649b_corner}}
\end{figure*}

\begin{figure*}
    \centering
    \includegraphics[width=\textwidth]{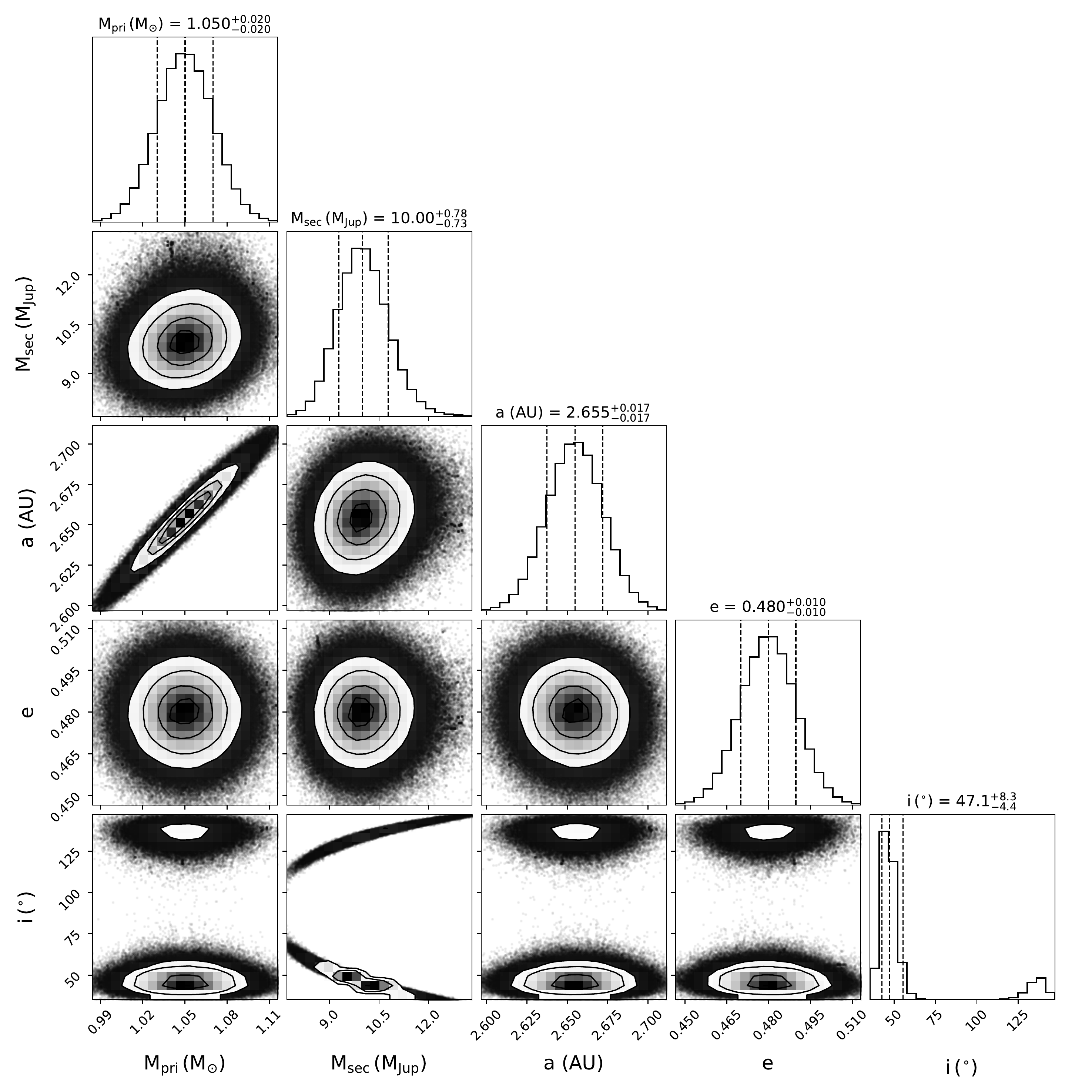}
    \caption{HD~106252~b. Same as Fig.~\ref{fig:HD29021b_corner}.
\label{fig:HD106252b_corner}}
\end{figure*}

\begin{figure*}
    \centering
    \includegraphics[width=\textwidth]{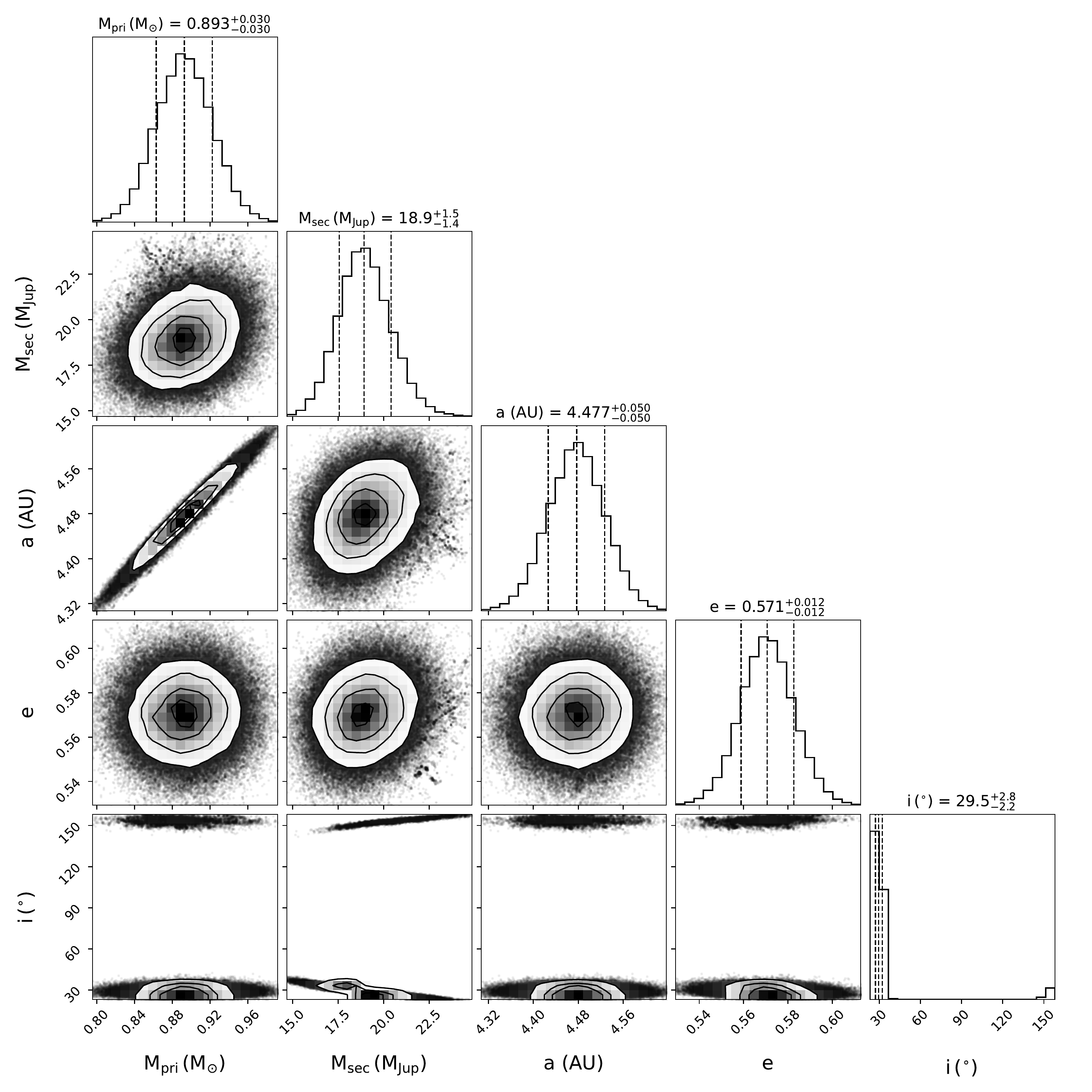}
    \caption{HD~106515~Ab. Same as Fig.~\ref{fig:HD29021b_corner}.
\label{fig:HD106515Ab_corner}}
\end{figure*}

\begin{figure*}
    \centering
    \includegraphics[width=\textwidth]{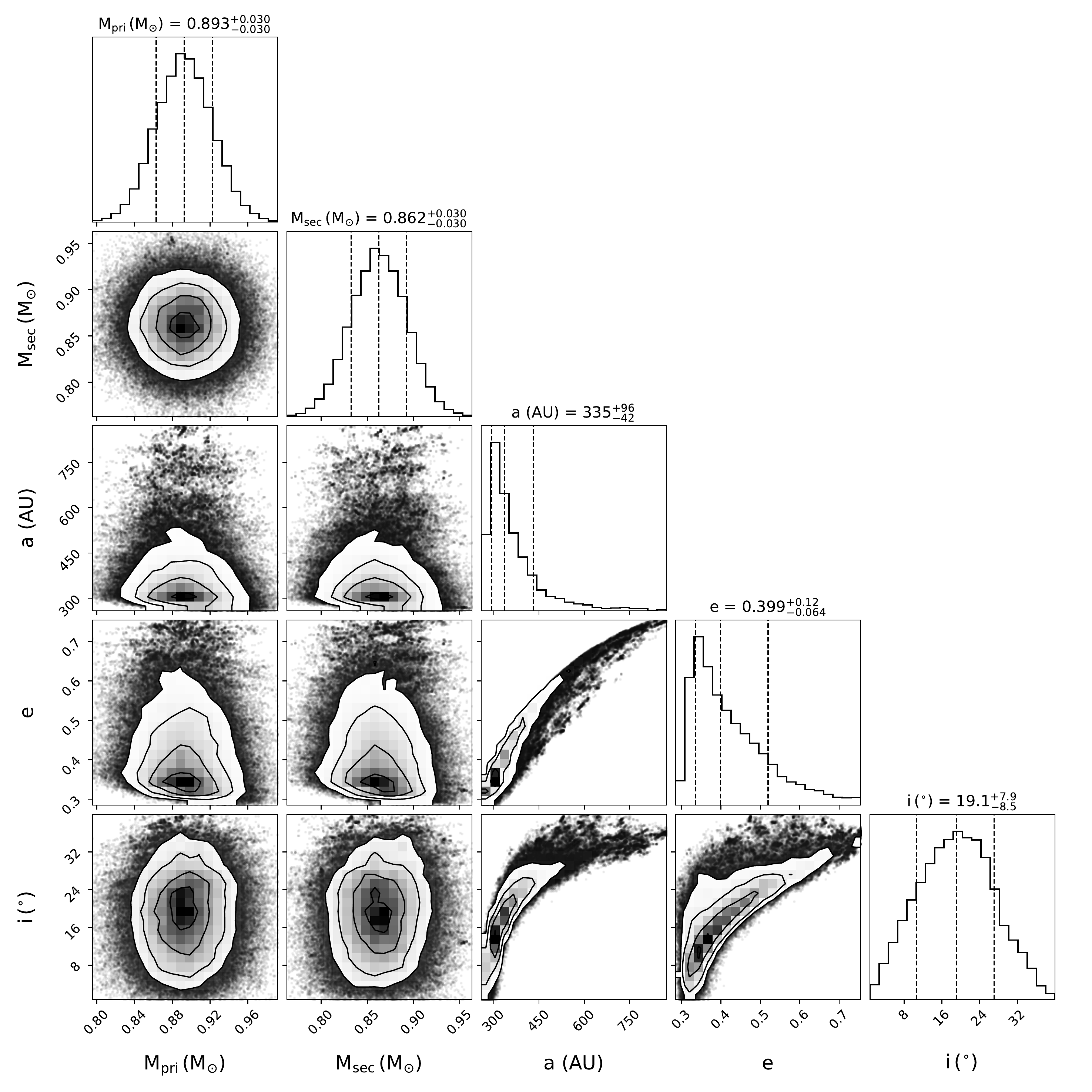}
    \caption{HD~106515~B, the stellar companion to HD~106515~A. Same as Fig.~\ref{fig:HD29021b_corner}.
\label{fig:HD106515B_corner}}
\end{figure*}

\begin{figure*}
    \centering
    \includegraphics[width=\textwidth]{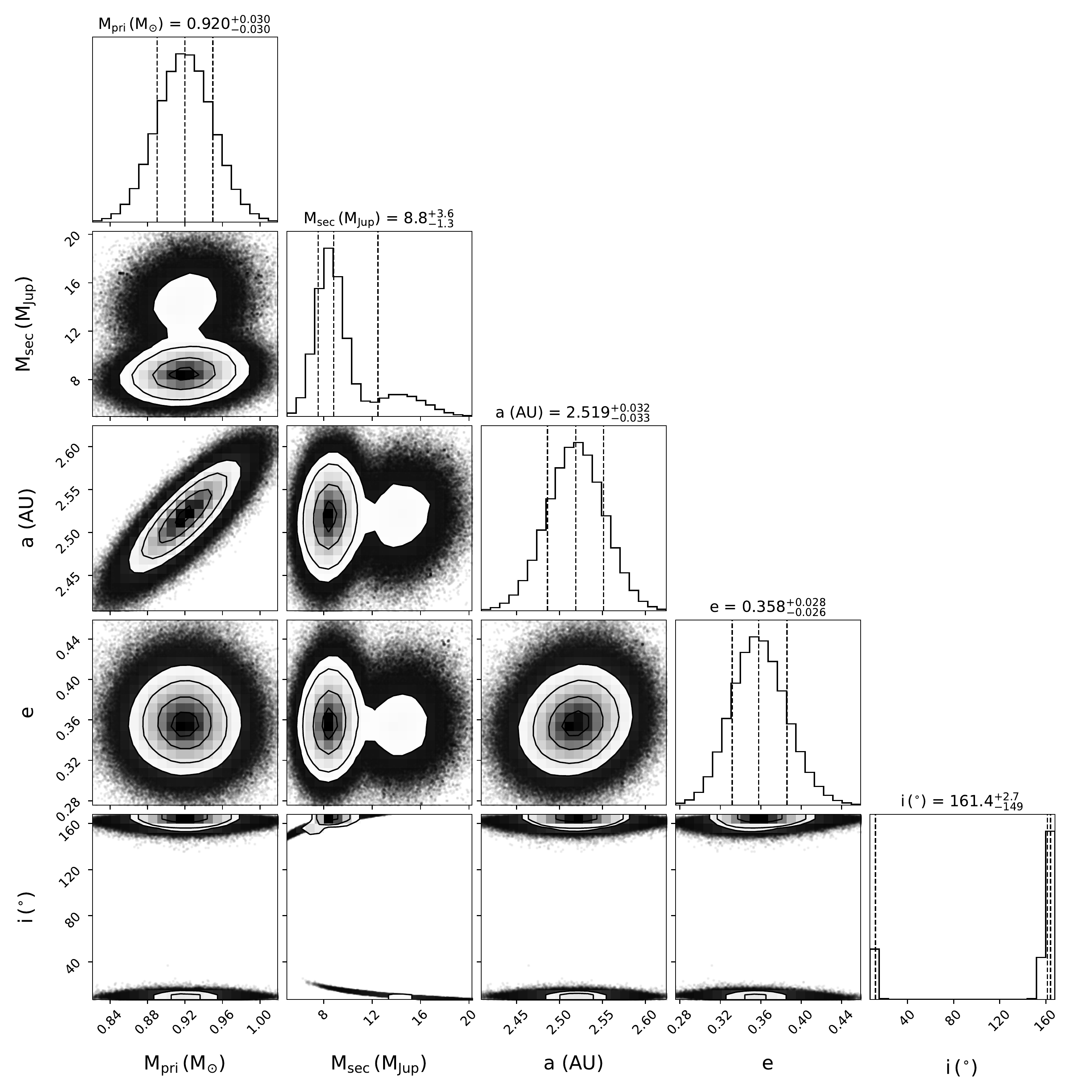}
    \caption{HD~171238~b. Same as Fig.~\ref{fig:HD29021b_corner}.
\label{fig:HD171238b_corner}}
\end{figure*}

\begin{figure*}
    \centering
    \includegraphics[width=\textwidth]{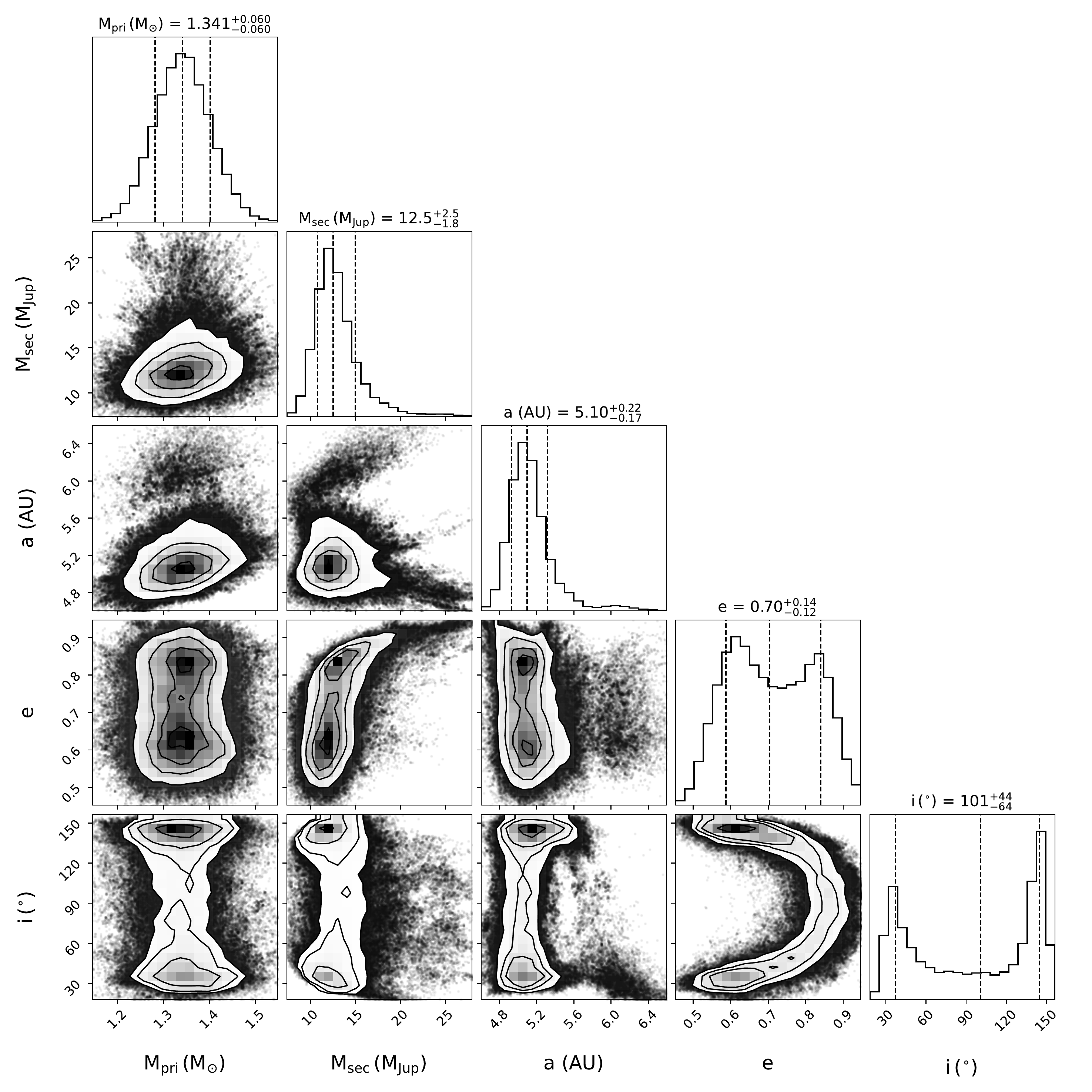}
    \caption{HD~196067~b. Same as Fig.~\ref{fig:HD29021b_corner}.
\label{fig:HD196067b_corner}}
\end{figure*}

\begin{figure*}
    \centering
    \includegraphics[width=\textwidth]{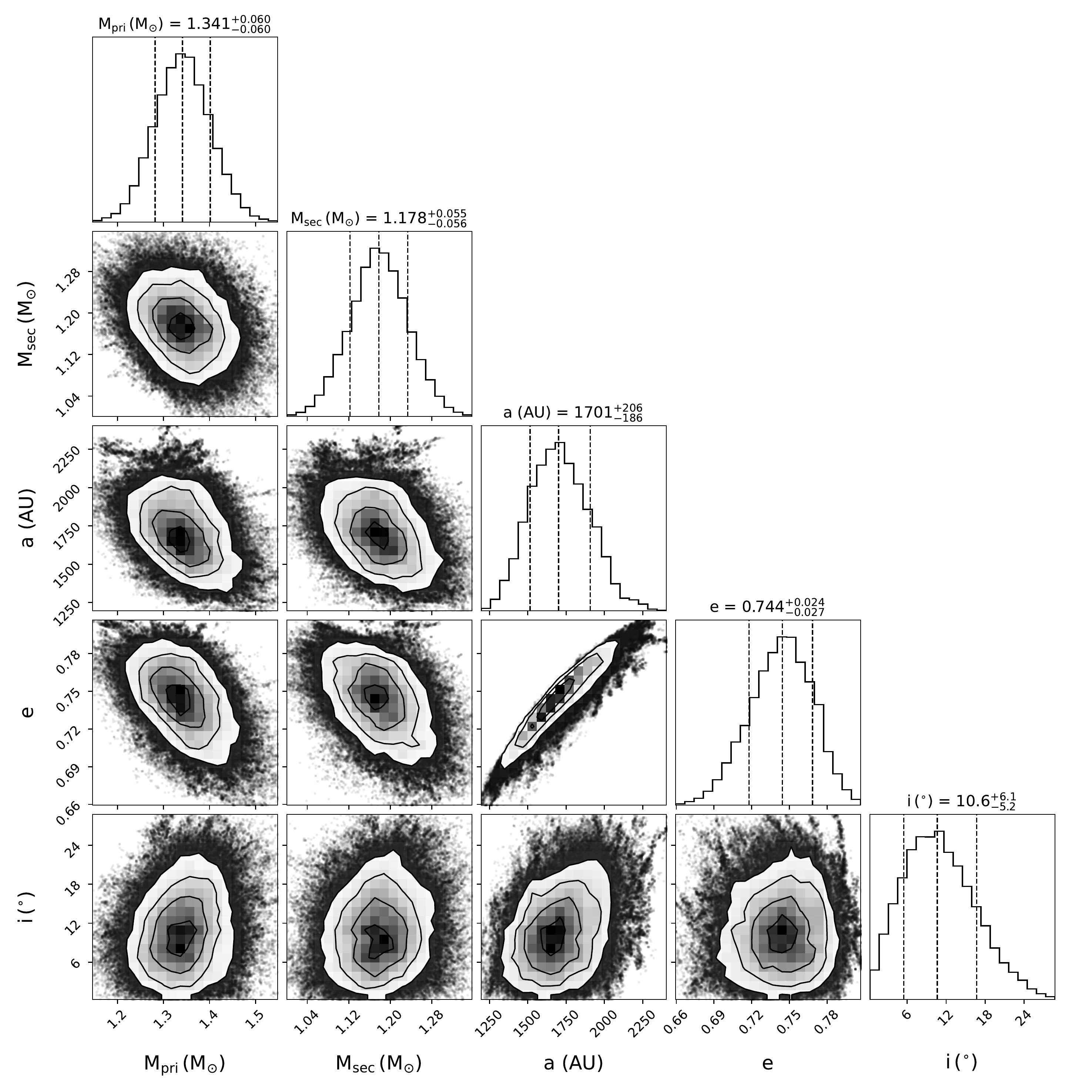}
    \caption{HD~196068. It is bound to HD~196067. Same as Fig.~\ref{fig:HD29021b_corner}.
\label{fig:HD196068_corner}}
\end{figure*}

\begin{figure*}
    \centering
    \includegraphics[width=\textwidth]{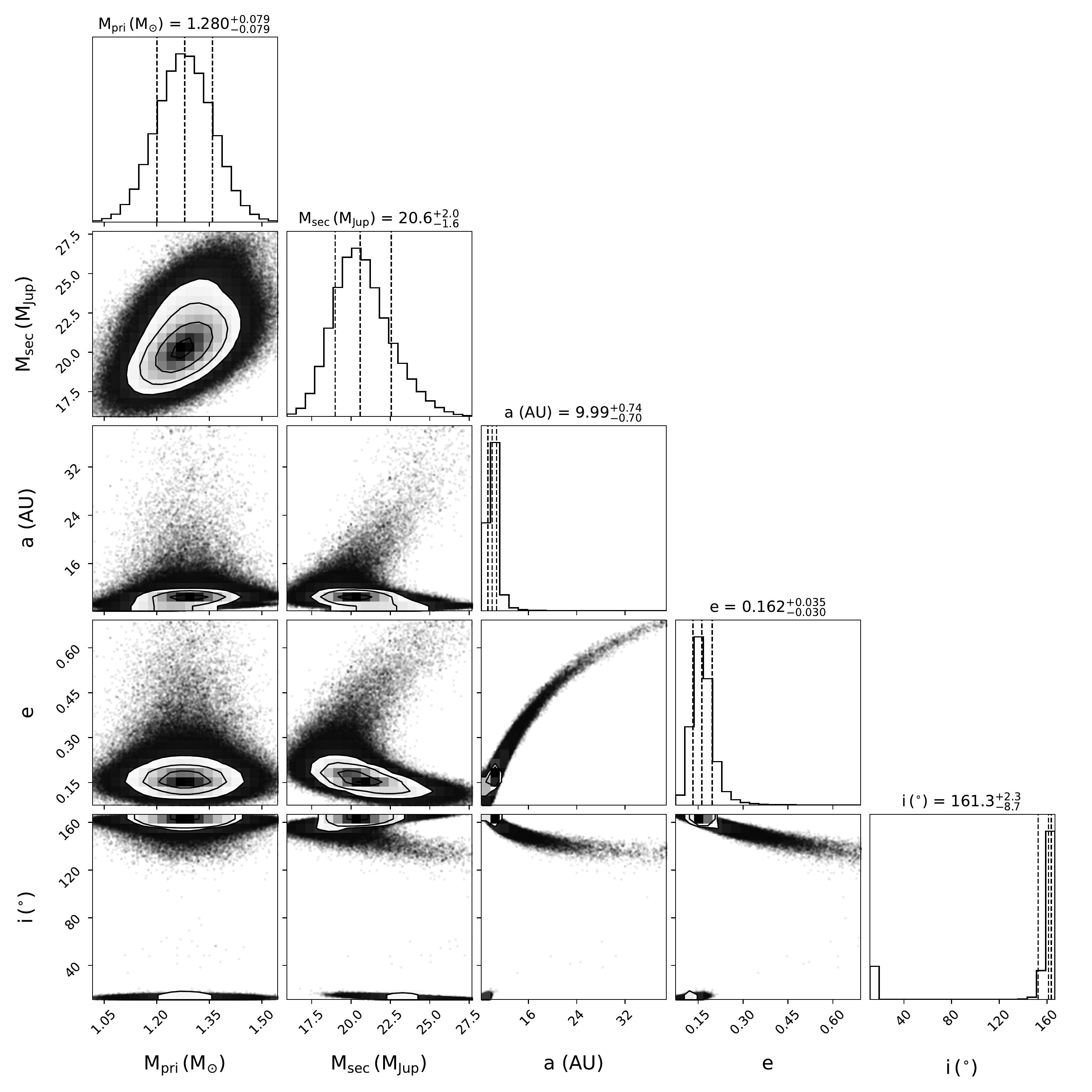}
    \caption{HD~221420~b. Same as Fig.~\ref{fig:HD29021b_corner}.
\label{fig:HD221420b_corner}}
\end{figure*}

\clearpage
\bibliographystyle{apj_eprint}
\bibliography{refs}

\end{document}